\documentstyle[12pt]{article}
\def\hybrid{\topmargin 0pt      \oddsidemargin 0pt
        \headheight 0pt \headsep 0pt
        \textwidth 17.5cm
        \textheight 25cm
        \voffset=-0.7cm
        \hoffset=-0.4cm
       \hoffset=-1.2cm
        \marginparwidth 0.0in
        \parskip 5pt plus 1pt   \jot = 1.5ex}
\catcode`\@=11
\def\marginnote#1{}

\newcount\hour
\newcount\minute
\newtoks\amorpm
\hour=\time\divide\hour by60
\minute=\time{\multiply\hour by60 \global\advance\minute by-\hour}
\edef\standardtime{{\ifnum\hour<12 \global\amorpm={am}%
        \else\global\amorpm={pm}\advance\hour by-12 \fi
        \ifnum\hour=0 \hour=12 \fi
        \number\hour:\ifnum\minute<10 0\fi\number\minute\the\amorpm}}
\edef\militarytime{\number\hour:\ifnum\minute<10 0\fi\number\minute}

\def\draftlabel#1{{\@bsphack\if@filesw {\let\thepage\relax
   \xdef\@gtempa{\write\@auxout{\string
      \newlabel{#1}{{\@currentlabel}{\thepage}}}}}\@gtempa
   \if@nobreak \ifvmode\nobreak\fi\fi\fi\@esphack}
        \gdef\@eqnlabel{#1}}
\def\@eqnlabel{}
\def\@vacuum{}
\def\draftmarginnote#1{\marginpar{\raggedright\scriptsize\tt#1}}

\def\draft{\oddsidemargin -0.1truein
        \def\@oddfoot{\sl preliminary draft \hfil
        \rm\thepage\hfil\sl\today\quad\militarytime}
        \let\@evenfoot\@oddfoot \overfullrule 3pt
        \let\label=\draftlabel
        \let\marginnote=\draftmarginnote
   \def\@eqnnum{{\rm (\theequation)}\rlap{\kern\marginparsep\tt\@eqnlabel}%
\global\let\@eqnlabel\@vacuum}  }

\newdimen\linethick  \linethick=0.4pt
\newdimen\hboxitspace    \hboxitspace=5pt
\newdimen\vboxitspace    \vboxitspace=5pt

\def\fr#1{%
\beq\new
\vcenter{
\hrule height\linethick
           \hbox{\vrule width\linethick
                 \kern\hboxitspace
                 \vbox{\kern\vboxitspace
                       \hbox{$\begin{array}{c}\displaystyle#1
          \end{array}$}%
                       \kern\vboxitspace}%
                 \kern\hboxitspace
                 \vrule width\linethick}%
           \hrule height\linethick}%
\eeq}

\newdimen\Squaresize \Squaresize=14pt
\newdimen\Thickness \Thickness=0.5pt

\def\Square#1{\hbox{\vrule width \Thickness
   \vbox to \Squaresize{\hrule height \Thickness\vss
      \hbox to \Squaresize{\hss#1\hss}
   \vss\hrule height\Thickness}
\unskip\vrule width \Thickness}
\kern-\Thickness}

\def\Vsquare#1{\vbox{\Square{$#1$}}\kern-\Thickness}

\def\numberbysection{\@addtoreset{equation}{section}
        \def\theequation{\thesection.\arabic{equation}}}
\numberbysection

\renewcommand{\theequation}{\thesection.\arabic{equation}}
\newcommand{\l@qq}[2]{\addvspace{2em}
 \hbox to\textwidth{\hspace{1em}\bf #1 \dotfill #2}}

\newcounter{app}

\def\app{\setcounter{equation}{0}
\def\theequation{\Alph{app}.\arabic{equation}}\par
   \addvspace{4ex}
   \@afterindentfalse
  \secdef\@app\@dapp}
\newcommand\@app{\@startsection {app}{1}{0ex}%
                                   {-3.5ex \@plus -1ex \@minus -.2ex}%
                                   {2.3ex \@plus.2ex}%
                                   {\normalfont\Large\bf}}

\def\@dapp#1{%
{\parindent \z@ \raggedright  \bf #1}\par\nobreak}
\def\l@app#1#2{\ifnum \c@tocdepth >\z@
    \addpenalty\@secpenalty
    \addvspace{1.0em \@plus\p@}%
    \setlength\@tempdima{2.5em}%
    \begingroup
      \parindent \z@ \rightskip \@pnumwidth
      \parfillskip -\@pnumwidth
      \leavevmode \bfseries
      \advance\leftskip\@tempdima
      \hskip -\leftskip
      #1\nobreak\hfil \nobreak\hb@xt@\@pnumwidth{\hss #2}\par
    \endgroup\fi}
\newcounter{sapp}[app]

\def\sapp{\def\theequation{\Alph{app}.\arabic{equation}}\par
   \@afterindentfalse
  \secdef\@sapp\@dsapp}
\newcommand\@sapp{\@startsection{sapp}{2}{\z@}%
                                     {-3.25ex\@plus -1ex \@minus -.2ex}%
                                     {1.5ex \@plus .2ex}%
                                     {\normalfont\large\bfseries}}

\def\@dsapp#1{%
{\parindent \z@ \raggedright  \bf #1}\par\nobreak}
\newcommand{\l@sapp}{\@dottedtocline{2}{1.5em}{3em}}

\def\titlepage{\@restonecolfalse\if@twocolumn\@restonecoltrue\onecolumn
     \else \newpage \fi \thispagestyle{empty}\c@page\z@
        \def\thefootnote{\fnsymbol{footnote}} }

\def\endtitlepage{\if@restonecol\twocolumn \else  \fi
        \def\thefootnote{\arabic{footnote}}
        \setcounter{footnote}{0}}  
\relax

\hybrid

\newtoks\@stequation

\def\subequations{\refstepcounter{equation}%
  \edef\@savedequation{\the\c@equation}%
  \@stequation=\expandafter{\theequation}
  \edef\@savedtheequation{\the\@stequation}
  \edef\oldtheequation{\theequation}%
  \setcounter{equation}{0}%
  \def\theequation{\oldtheequation\alph{equation}}}

\def\endsubequations{%
  \setcounter{equation}{\@savedequation}%
  \@stequation=\expandafter{\@savedtheequation}%
  \edef\theequation{\the\@stequation}%
  \global\@ignoretrue}

\makeatletter
\newdimen\normalarrayskip              
\newdimen\minarrayskip                 
\normalarrayskip\baselineskip
\minarrayskip\jot
\newif\ifold             \oldtrue            \def\new{\oldfalse}
\def\arraymode{\ifold\relax\else\displaystyle\fi} 
\def\eqnumphantom{\phantom{(\theequation)}}     
\def\@arrayskip{\ifold\baselineskip\z@\lineskip\z@
     \else
     \baselineskip\minarrayskip\lineskip1\baselineskip\fi}

\def\@arrayclassz{\ifcase \@lastchclass \@acolampacol \or
\@ampacol \or \or \or \@addamp \or
   \@acolampacol \or \@firstampfalse \@acol \fi
\edef\@preamble{\@preamble
  \ifcase \@chnum
     \hfil$\relax\arraymode\@sharp$\hfil
     \or $\relax\arraymode\@sharp$\hfil
     \or \hfil$\relax\arraymode\@sharp$\fi}}

\def\@array[#1]#2{\setbox\@arstrutbox=\hbox{\vrule
     height\arraystretch \ht\strutbox
     depth\arraystretch \dp\strutbox
     width\z@}\@mkpream{#2}\edef\@preamble{\halign \noexpand\@halignto
\bgroup \tabskip\z@ \@arstrut \@preamble \tabskip\z@ \cr}%
\let\@startpbox\@@startpbox \let\@endpbox\@@endpbox
  \if #1t\vtop \else \if#1b\vbox \else \vcenter \fi\fi
  \bgroup \let\par\relax
  \let\@sharp##\let\protect\relax
  \@arrayskip\@preamble}
\def\eqnarray{\stepcounter{equation}%
              \let\@currentlabel=\theequation
              \global\@eqnswtrue
              \global\@eqcnt\z@
              \tabskip\@centering                      
              \let\\=\@eqncr
              $$%
            \halign to \displaywidth  \bgroup
             \eqnumphantom \@eqnsel
      \hskip\@centering                               
    $\displaystyle  \tabskip\z@ {##}$%
    &\global\@eqcnt\@ne \hskip 2\arraycolsep
         $ \displaystyle  \arraymode{##}$\hfil
    &\global\@eqcnt\tw@ \hskip 2\arraycolsep
         $\displaystyle\tabskip\z@{##}$\hfil
         \tabskip\@centering
    &{##}\tabskip\z@\cr}
\makeatother

\def\bea{\begin{eqnarray}}
\def\eea{\end{eqnarray}}

\def\beq{\begin{equation}}
\def\eeq{\end{equation}}
\def\be{\beq\new\begin{array}{c}}
\def\ee{\end{array}\eeq}
\def\bse{\begin{subequations}}                
\def\ese{\end{subequations}}                 %

\begin{document}
\vspace{0.2cm}
\begin{center}
{\LARGE \bf The Gauge String Solution of the} \\
\vspace{0.5cm}{\LARGE \bf $D\geq{3}$ Yang-Mills Loop Equations. } \\
\vspace{0.7cm} {\large\bf Andrey Yu. Dubin}\\
\vspace{0.5cm}{\bf ITEP, B.Cheremushkinskaya 25, Moscow 117259, Russia}\\
\vspace{0.5cm}{{\it e-mail: dubin@heron.itep.ru}}\\
{tel.:~7 095 129 9674~~~~  ;  ~~~~fax:~7 095 883 9601}
 \end{center}
\vspace{0.3cm}

\begin{abstract}

I adapt the Gauge String \cite{Dub3}, representing the strong coupling
expansion in the {\it continuous} $D\geq{3}$ Yang-Mills theory ($YM_{D}$)
with a sufficiently large {\it bare}
coupling constant $\lambda>\lambda_{cr}$ and a fixed ultraviolet cut off
$\Lambda$, to the analysis of the regularized Wilson's loop-averages.
When generalized to describe the {\it fat} (rather than infinitely thin)
flux-tubes, the pattern of thus modified $U(N)$ Gauge
String is proved to be consistent with the chain of the judiciously
{\it regularized} $U(N)$ Loop equations. In particular,
we reveal the dimensional reduction $YM_{D}\rightarrow{YM_{2}}$,
taking place in the extreme $SC$ limit $\lambda\rightarrow{\infty}$, and
compare it with the implications of the $AdS/CFT$ correspondence conjecture.
On the other hand, for the loop-averages
associated to the sufficiently large minimal areas, the proposed stringy
pattern is supposed to be in the one {\it infrared} universality class
(provided the loops are without zig-zag backtrackings) with the {\it novel~}
implementation of the noncritical $D$-dimensional Nambu-Goto string. The
peculiarity is due to the nonstandard $\Lambda^{2}$-scaling,
$\Lambda^{2}=O(\sigma_{ph})$, of the physical string tension $\sigma_{ph}$.
Being well-motivated from the viewpoint of the standard $YM_{4}$ theory with
$\lambda\rightarrow{0}$, this scaling is argued to entail that the considered
modification of the  Nambu-Goto system is in the stringy (rather than in the
branched polymer) regime. In sum, the
confinement in the continuous $D\geq{3}$ $U(N)$ (and, similarly, $SU(N)$)
gauge theory is justified, for the first time, at least when both $N$ and
$\lambda$ are sufficiently large. As a by-product, when continued to
$N=1$, the Gauge String is shown to describe the continuous $U(1)$ gauge
theory enriched with the {\it monopoles} in the phase where the latter are
supposed to be condensed.

\end{abstract}

\begin{center}
\vspace{0.5cm}{Keywords: Yang-Mills, Loop equation, Duality, String,
Strong-coupling expansion} \\
\vspace{0.2cm}{PACS codes 11.15.Pg; 11.15.Me; 12.38.Aw; 12.38.Lg}
\end{center}

\newpage

\section{Introduction.}

An exact stringy reformulation of the continuous four-dimensional Yang-Mills
theory ($YM_{4}$) is one of the fundamental problems in theoretical physics,
resolution of which is supposed to entail wide applications in the realm of
phenomenology. Although a number of deep and impressive ideas have been
invested (see e.g. \cite{Wils}-\cite{GFS},\cite{Polyak2,Witt2}),
the situation in the subject remains far from being settled.
As a step towards the rigorous reformulation, the author has recently
proposed the novel Gauge String representation \cite{Dub3} of certain
$D\geq{3}$ dimensional lattice gauge systems in the large $N$ strong
coupling ({\it SC}) phase. Combining the nonabelian duality transformation
\cite{Dub2} with the Gross-Taylor stringy representation of the
{\it continuous} $YM_{2}$ theory on a $2d$ manifold, we find the exact
reformulation of the considered $D\geq{3}$ $YM_{D}$ systems in terms of the
infinitely thin vortices of the colour-electric $YM$-flux.

In sharp contrast to the previous proposals in this direction, the lattice
construction \cite{Dub3} can be directly employed\footnote{Formally, this is
possible due to the fact that the derived in \cite{Dub3} {\it lattice}
worldsheets's weight (\ref{0.5c}) is invariant, despite the discretization,
under certain {\it continuous} (rather than {\it discrete}) group of the
transformations (homeomorphisms) of the surface $\tilde{M}_{\chi}$. The
extension to the continuous space-time implies to trade the homeomorphisms for
the corresponding diffeomorphisms characteristic for the theories of smooth
strings.} to define the variety of the continuous $D\geq{3}$ Gauge String
models of smooth $YM$ vortices. In turn, one can choose such a model that can
be unambiguously associated to the {\it continuous} $D\geq{3}$ $YM_{D}$
theory with the action
\be
S=\frac{1}{4g^{2}}\int d^{D}x~tr\left( F_{\mu\nu}(x)F_{\mu\nu}(x)\right)~,
\label{1.1}
\ee
where $F_{\mu\nu}\equiv{F^{a}_{\mu\nu}T^{a}_{ij}}$, and
$T^{a}_{ij}$ are the properly normalized generators in the
fundamental representation of a given Lie group to be fixed as $U(N)$ unless
otherwise specified. The considered explicit correspondence with the gauge
theory (\ref{1.1}) is possible owing to the remarkable duality, to be
formalized by eq. (\ref{0.5cxd}).
Certain conglomerates of the Feynman diagrams, comprising the weak-coupling
($WC$) series in a given continuous gauge system, are shown to be in the exact
$WC/SC$ {\it correspondence} with the judiciously associated
variety of the appropriately weighted (piecewise) smooth flux-worldsheets.
In this specific way, certain option of
the smooth Gauge String allows, after the proper regularization (that
will introduce a {\it nonzero} width of the $YM$ vortex),
to implement the $1/N$ {\it strong-coupling} expansion of the regularized
gauge invariant quantities in the theory (\ref{1.1}).

In the present paper, we investigate the regularized pattern of the proposed
$1/N$ $SC$ (as opposed to the conventional $WC$) series further and directly
justify the asserted $YM_{D}/String$ duality employing the power of the Loop
equation \cite{LE/MM,LE/P} associated to the regularized $U(N)$ gauge theory
(\ref{1.1}). To be more explicit, let the system (\ref{1.1}), defined in
the $D\geq{3}$ Euclidean base-space ${\bf R^{D}}$, be regularized at some
ultraviolet ({\it UV}) scale $\Lambda\sim{N^{0}}$. Also, the dimensionless
and $N$-{\it independent} {\it bare} coupling constant 
\be
\lambda=(g^{2}N)\Lambda^{D-4}>\lambda_{cr}(D)
\label{1.1bx}
\ee
is constrained to assume values {\it larger} than certain critical one
$\lambda_{cr}(D)$ (which is expected to be of order of unity)
presumably associated to the large $N$ phase transition.
Once the inequality (\ref{1.1bx}) is fulfilled, the $YM_{D}/String$
duality \cite{Dub3} claims then that the free energy and the Wilson's
loop-averages in the $YM_{D}$ theory (\ref{1.1}) can be reformulated in terms
of the dual microscopic degrees of freedom: respectively the closed and the
open sectors of the proposed smooth Gauge String. {\it Prior} to the $UV$
regularization, the latter is defined through the 'bare' worldsheet's weight
\be
w[\tilde{M}_{\chi}]=N^{\chi}~exp
\left(-\frac{\breve{\lambda}\breve{\Lambda}^{2}}{2}
A[\tilde{M}_{\chi}]\right)~J[\tilde{M}_{\chi}|\breve{\lambda}]~,
\label{0.5c}
\ee
where $A[\tilde{M}_{\chi}]$ denotes the total area of a given
worldsheet $\tilde{M}_{\chi}$, $\breve{\lambda}=
(g^{2}N)\breve{\Lambda}^{D-4}$),
while $N^{\chi}$ is the well-known
't Hooft topological factor (with $\chi$ being the Euler character
of $\tilde{M}_{\chi}$). As for $\breve{\Lambda}$, it is some auxiliary 
parameter which, tending to infinity before the regularization, will be
related (see eq. (\ref{0.1ex})) to the $UV$ cut off $\Lambda$ on a later stage.

In contradistinction to the conventional Nambu-Goto pattern (given by eq.
(\ref{2.5bb})), in eq. (\ref{0.5c}) there is one more factor
$J[\tilde{M}_{\chi}|\breve{\lambda}]$ which is crucial for the
manifest correspondence with the weak-coupling series in the $YM_{D}$ theory
(\ref{1.1}). Being equal to {\it unity} for any nonselfintersecting surface
$\tilde{M}_{\chi}$, this $\breve{\lambda}$-dependent factor  is in addition
sensitive only to the {\it topology}, but {\it not} to the geometry, of
selfintersections of $\tilde{M}_{\chi}$. Another distinction with the
Nambu-Goto Ansatz is that the
worldsheet's selfintersections are allowed to possess certain topological
{\it singularities} absent in the measure of the latter Ansatz.
It is noteworthy, see \cite{Dub3}, that the pattern of {\it both} the
$J[..]$-factor {\it and} the admissible worldsheet's singularities does
depend on the choice of the gauge group and the $YM_{D}$ lagrangian. In
Section 4, it will be shown that, through this somewhat peculiar
representation, one reproduces the contact interactions
between the (self)intersecting (infinitely thin) vortices of the unit
$YM$-flux. These interactions are to be distinguished from the
selfenergy contribution which refers to the sheer Nambu-Goto pattern. 

Next, by construction, eq. (\ref{0.5c}) refers to the dual gauge theory
(\ref{1.1}) considered prior to the $UV$ regularization that, in fact,
is reflected by the {\it vanishing} width of the associated $YM$ vortices.
On the other hand, in order to make contact with the technology of the loop
equations \cite{LE/MM,LE/P}, one is to operate with the {\it regularized}
$YM_{D}$ theory (\ref{1.1}) where, therefore, some {\it quasi-locality} has to
be introduced. In consequence, it turns out that we should deal with the
{\it fat} flux-tubes of a {\it nonzero} width
$\sqrt{<{\bf r^{2}}>}\sim{\Lambda^{-1}}$ described
through certain {\it quasi-local} weight $w_{r}[\tilde{M}]$. The latter weight
is supposed to be deduced via, roughly speaking, a kind of
{\it regularization-dependent} smearing \footnote{The smearing should comply
with the natural requirement
that, modulo admissible rescaling $\lambda\rightarrow{{\lambda}'}$ of
the coupling constant (\ref{1.1bx}), eq. (\ref{0.5c}) represents the
{\it universal} limit of the variety of the differently regularized weights
$w_{r}[\tilde{M}]$. The latter
limit is achieved, possibly with the exception for some {\it measure zero}
subset of the worldsheet's configurations, when the width
$\sqrt{<{\bf r^{2}}>}$ is thinned back to zero.} of
the $\sqrt{<{\bf r^{2}}>}\rightarrow{0}$ pattern (\ref{0.5c}) local on the
worldsheet $\tilde{M}$. As a result, for the economic justification of the
asserted matching (between the smeared variant of the pattern (\ref{0.5c}) and
the regularized Loop equation), it is vital to reduce the
regularization-dependence  of the $YM_{D}$ loop-average $<W_{C}>$ as much as
possible.

Upon a reflection, there are {\it two} limiting regimes (formalized by eqs.
(\ref{0.1eea}) and (\ref{0.1eed}) below) where the dependence of $<W_{C}>$ on
the choice of the regularization indeed can be reduced to the dependence of a
few relevant coupling constants, entering the corresponding implementation
of the Gauge String representation, on the bare coupling (\ref{1.1bx}).
In both cases, we
are dealing with the dominance of the {\it infrared} phenomena: the contours
$C$, being constrained to possess the radius of curvature ${\mathcal{R}}(s)>>
\Lambda^{-1}$ (for $\forall{s}$), should be associated to
sufficiently large values of the minimal area $A_{min}(C)$ of the
saddle-point worldsheet $\tilde{M}_{min}(C)$. Given the latter conditions,
the required reduction is shown to be maintained when the characteristic
amplitude $\sqrt{<{\bf h^{2}}>}$ of the worldsheet's fluctuations is
{\it either} much larger {\it or} much smaller than the flux-tube's width
$\sqrt{<{\bf r^{2}}>}\sim{\Lambda^{-1}}$.

As for the first of the regimes, the precise form of the defining necessary
conditions reads
\be
\frac{<{\bf h^{2}}>}{<{\bf r^{2}}>}\sim
\frac{D-2}{\lambda}\cdot ln[A_{min}(C)\Lambda^{2}]~\longrightarrow{~\infty}~~~~,
~~~~{\mathcal{R}}(s)\Lambda~\longrightarrow{~\infty}~,
\label{0.1eea}
\ee
which (being strengthened by the complementary constraint formulated prior to
eq. (\ref{0.1eel})) entails that the characteristic $YM$ vortices, spanned by
$C$, behave almost as if they are infinitely thin. In this case,
taking advantage of certain hidden simplification inherent in the pattern
(\ref{0.5c}), one can use a particularly simple regularization prescription
for the Gauge String weight. For this purpose, let
us consider the contours with at most {\it point-like} (self)intersections
that, in particular, excludes zig-zag backtrackings of the loops.
Then, the prefered prescription results in the stringy measure which,
prior to the $UV$ regularization at $\Lambda$, is equal to the one in the
conventional Nambu-Goto theory. As for the worldsheet's weight, to reproduce
the regularized pattern of the dual $YM_{D}$ theory (\ref{1.1}), it assumes
the quasi-local
form\footnote{In fact, the pattern (\ref{0.1}) is reminiscent of (but, as it
is discussed in the footnote after eq. (\ref{1.11bbx}), {\it not} equivalent
to) the {\it ad hoc} smearing \cite{LE/MM,ALaw} of the
$m_{0}=0$ option of the Nambu-Goto weight (\ref{2.5bb}). Complementary,
the weight (\ref{0.1}) can be viewed as a specific implementation of the
general confining string Ansatz \cite{PolyakLH,PolyakCS} (see Appendix B for
the explicit comparison) which, in turn, is routed in the abelian Kalb-Ramond
pattern \cite{Kalb&Ramond}. Given the specific choice (\ref{2.5bxd}) of
${\mathcal{G}}(..)$, the $D=4$ pattern
(\ref{0.1}) can be reduced (modulo certain boundary terms) to the {\it rigid}
string but in the phase, see eq. (\ref{0.1eec}) below, which is
{\it different} from the one considered in \cite{ExtrCurv/P,ExtrCurv/K}.}
\be
\tilde{w}_{2}[\tilde{M}_{\chi}]=N^{\chi}~
exp\left(-\frac{\lambda\Lambda^{2}}{4}
\int\limits_{\tilde{M}_{\chi}}\int\limits_{\tilde{M}_{\chi}}
d\sigma_{\mu\nu}({\bf x})d\sigma_{\mu\nu}({\bf y})~
{\Lambda}^{2}{\mathcal{G}}({\Lambda}^{2}({\bf x}-{\bf y})^{2})
\right)~,
\label{0.1}          
\ee
where $d\sigma_{\mu\nu}({\bf x})$ is the standard infinitesimal
area-element (\ref{0.2c}) associated to the surface $\tilde{M}_{\chi}$ of the
Euler character $\chi$. As for
${\Lambda}^{D} {\mathcal{G}}({\Lambda}^{2}{\bf z}^{2})$, being
introduced (see eq. (\ref{0.1bb})) to smear the $D$-dimensional
$\delta_{D}({\bf z})$-function, it is further constrained by the two 
conditions (given by eqs. (\ref{1.3zp}) and (\ref{0.1xem})). We prove that
{\bf thus defined stringy sum provides, in the regime (\ref{0.1eea}), with
the confining solution of the Dyson-Schwinger chain of the judiciously
regularized $D\geq{3}$ $YM_{D}$ loop equations (derived for the system
(\ref{1.1})) to all orders in $1/N$.}

The deep reason, behind the relation between the stringy sums based
respectively on the weights (\ref{0.5c}) and (\ref{0.1}), is the following
{\it infrared universality} taking place in the regime (\ref{0.1eea})
as far as contours without zig-zag backtrackings are concerned.
(The prescription, to implement the mandatory backtracking invariance of the
$YM_{D}$ loop-averages $<W_{C}>$, will be discussed in Section 4.)
To begin with, in this case, {\it the above solution
of the Loop equation is supposed to be in the one infrared universality class
with the unconventional (owing to eq. (\ref{0.1eec})) implementation of the
Nambu-Goto string} which, in
turn, is supposed to be reformulated in the spirit of the
'low-energy' noncritical Polyakov's theory \cite{GFS}. Thus associated
Nambu-Goto theory, presumed to possess the same {\it UV} cut off $\Lambda$,
is endowed with the weight
\be                            
w_{1}[\tilde{M}_{\chi}]=N^{\chi}~
exp\left({-\frac{\bar{\lambda}(\lambda)\cdot\Lambda^{2}}{2}
A[\tilde{M}_{\chi}]}-m_{0}(\lambda)L[\tilde{M}_{\chi}]\right)~,
\label{2.5bb}
\ee
where $L[\tilde{M}]$ is the length\footnote{Actually, as it is clear from
eq. (\ref{2.5bxl}), the simple $L[\tilde{M}_{\chi}]$-dependence of the
subleading boundary-contribution is valid only provided that all the loop's
{\it self}intersections, if present, are {\it point-like} from the low-energy
viewpoint.} of the boundary $\partial{\tilde{M}}$ of
$\tilde{M}$, while $\bar{\lambda}(\lambda)$ and $m_{0}(\lambda)$ are
certain functions of $\lambda$ which depend on the choice of {\it both} the
flux-tube's transverse profile {\it and} the prescription for
the regularization of the string fluctuations.
Then, the remaining step is to take into account
that, being considered as resulting in the zero-width limit
$\lambda\Lambda^{2}<{\bf r^{2}}>\rightarrow{0}$, the associated to
the weight (\ref{0.5c}) stringy sum is equivalent (see Appendix C) to the
Nambu-Goto theory. The latter is endowed with the option of
the weight (\ref{2.5bb}) given by the identifications: $m_{0}=0,~
\bar{\lambda}(\lambda)\Lambda^{2}=\breve{\lambda}\breve{\Lambda}^{2}$.
In particular, the important consequence of the infrared unobservability of the
selfintersection factor $J[..]$, entering eq. (\ref{0.5c}) but missing in eq.
(\ref{2.5bb}), is that the pattern (\ref{2.5bb}) works equally well for a
large {\it variety} of the actions of the dual $YM_{D}$ theories.

Next, despite its familiar appearence, the system (\ref{2.5bb}) is {\it not}
entirely conventional at least for sufficiently large $N\geq{2}$. The reason
is that, within the $1/N$ expansion, the consistent solution of the Loop
equation mandatory results in {\it the physical string tension $\sigma_{ph}$
which is of order of (or, when $\lambda\rightarrow{\infty}$, much larger than)
the {\it UV} cut off $\Lambda$ squared},
\be
\Lambda^{2}=O(\sigma_{ph})~~~~~~~~;~~~~~~~~~
\sigma^{(sc)}_{ph}=\left(\frac{\bar{\lambda}(\lambda)}{2}-\zeta_{D}
\right)\Lambda^{2}~,
\label{0.1eec}
\ee
where, in the $N\rightarrow{\infty}$ semiclassical approximation
$\sigma^{(sc)}_{ph}$ for $\sigma_{ph}$, the $\lambda$-independent constant
$\zeta_{D}$ (given by eqs. (\ref{0.7bcx}),(\ref{0.7bxx})) encodes the
$D\geq{3}$ entropy contribution $\delta \sigma_{ent}=-\zeta_{D}\Lambda^{2}$
due to the regularized transverse string fluctuations. (In particular, eq.
(\ref{0.1eec}) implies that the coupling $\bar{\lambda}(\lambda)$ is
constrained to be sufficiently large.) In other words, the full-fledged
quantum analysis of the stringy sum, endowed with the weight (\ref{2.5bb}),
requires to reformulate
the latter as a somewhat unconventional stringy theory. Possessing the $UV$
cut off $\bar{\Lambda}>>\sqrt{\sigma_{ph}}\sim{\Lambda}$, {\it this theory
should not exhibit propagating degrees of freedom at the 'short distance'
scales $<<{1/\sqrt{\sigma_{ph}}}$, while approaching the proposed Nambu-Goto
pattern in the infrared domain at the scales larger than
$1/\sqrt{\sigma_{ph}}$}.

According to the arguments of Section 7, the $\Lambda^{2}$-scaling
(\ref{0.1eec}) is sufficient to avoid the outgrowth of the microscopic
{\it baby-universes} so that the Gauge String avoids the branched polymer
phase. On the other hand, it is this scaling that hinders
the direct application of the considered construction to the
weak-coupling phase $\lambda\rightarrow{0}$.
Actually, the somewhat odd dependence of $\sigma_{ph}$ on the
{\it UV} cut off $\Lambda$ becomes conceivable if one employs the proper
reinterpretation \cite{Dub3} of the strongly coupled $D=4$ $YM_{4}$
system (\ref{1.1}). In view of the asymptotic freedom, the latter system can
be viewed as a local {\it prototype (or mathematical idealization)} of the
effective low-energy theory of the weakly coupled gauge theory (\ref{1.1}).
In this  perspective, the cut off
$\Lambda$ is to be identified with the confinement-scale which (in the
$D=4$ system (\ref{1.1}) with $\lambda\rightarrow{0}$ and the new $UV$ cut
off $\bar{\Lambda}>>\Lambda$) is supposed to be of order of the lowest
glueball mass. From the results obtained, the considered confinement-scale
can be viewed as the scale where the logarithmic renormgroup flow (of the
running coupling constant) stops. Moreover, given the latter identification,
in the $N\rightarrow{\infty}$ limit the proposed implementation of the
Nambu-Goto theory (endowed with
the weight (\ref{2.5bb})) is presumed to correctly describe the low-energy
dynamics of the $U(N)$ or $SU(N)$ $YM_{4}$ theory (\ref{1.1}) in the
$WC$ phase with $\lambda\rightarrow{0}$. This statement is supported by
the observation of Section 5.2 that the large $N$ limit of the considered
Nambu-Goto sum is supposed to be {\it common} for the infrared description of
{\it any} $D\geq{3}$ $U(\infty)$ or $SU(\infty)$ pure gauge system with an
arbitrary polynomial (in terms of $F_{\mu\nu}$) lagrangian providing with the
$N^{2}$-scaling of the free energy.

In view of the latter universality, it is
desirable to reconcile the reduction of the regularization-dependence with
the observability of the particular choice of the associated $U(N)$ action
(\ref{1.1}). On the side of the dual string theory, it is tantamount to
the fact that the {\it deviation} of the selfintersection $J[..]$-factor (of
eq. (\ref{0.5c})) from unity is observable. This is indeed possible to
achieve in the regime when the extreme {\it SC} limit
$\lambda\rightarrow{\infty}$ is performed {\it before} the large
$[A_{min}(C)\Lambda^{2}]$ limit (and, possibly, even before the large $N$
limit):
\be
\frac{ln[A_{min}(C)\Lambda^{2}]}{\lambda}~\longrightarrow{~0}~~~,
~~~[A_{min}(C)\Lambda^{2}]~\longrightarrow{~\infty}~~~,~~~
{\mathcal{R}}(s)\Lambda~\longrightarrow{~\infty}~,
\label{0.1eed}
\ee
while the radius of curvature ${\mathcal{R}}(s)$ is also kept to be
sufficiently large (and the condition stated prior to eq. (\ref{0.1eel}) is
also assumed).
The major simplification of the regime (\ref{0.1eed}) is that the
characteristic amplitude $\sqrt{<{\bf h^{2}}>}$ of the fluctuations
is {\it much smaller} than the width $\sqrt{<{\bf r^{2}}>}
\sim{\Lambda^{-1}}$ of $YM$ vortex. Furthermore, the leading asymptotics of
the average $<W_{C}>$ is represented by the contribution of the saddle-point 
vortex corresponding
to the minumal area worldsheet(s) $\tilde{M}_{min}(C)$ with the
characteristic radius ${\mathcal{R}}(\gamma)$ of curvature being (at any point
$\gamma\equiv{(\gamma_{1},\gamma_{2})}$ of the surface) much larger
than the flux-tube width:
\be
{\mathcal{R}}(\gamma)\Lambda~\longrightarrow{~\infty}~.
\label{0.1eeb}
\ee
Then, the land-mark of the regime (\ref{0.1eed}) is that
the 'minimal area' worldsheet(s) are
correctly described directly by the weight (\ref{0.5c}) so that the specifics
of the $YM_{D}$ action (\ref{1.1}) is indeed observable.

In compliance with the prediction of \cite{Dub3}, this can be formalized by
the following concise prescription\footnote{The prescription (\ref{0.5bxx})
will be deduced, {\it directly} from the loop equations, in a separate paper.
Here, we restrict our analysis only to a preliminary justification given
by eq. (\ref{2.12bl}).} (to be compared with the implications of the
$AdS/CFT$ correspondence conjecture in Section 9). For simplicity, let the
saddle-point (minimal area) worldsheets $\tilde{M}_{min}(C)$ possess the
support $T_{min}=T_{min}(C)$ which, for simplicity, is presumed to be unique
for a given $C$. Additionally, keeping the conditions (\ref{0.1eed})
fulfilled, let both the coupling constant $\lambda$ and
$({\mathcal{R}}(s)\Lambda)^{2}$ be much larger than
$N^{2}$. Then, summing up the leading (with respect to the $1/\lambda$
expansion) subseries in $1/N^{2}$, one obtains that {\it the
pattern of the $D\geq{3}$ averages $<W_{C}>|_{YM_{D}}$ is reduced,
\be
<W_{C}>|_{YM_{D}}~\longrightarrow{~<W_{C}>|_{YM_{2}(T_{min})}}~~~~~~if
~~~~~~\lambda,~({\mathcal{R}}(s)\Lambda)^{2}~~>>N^{2}~,
\label{0.5bxx}
\ee
to the one $<W_{C}>|_{YM_{2}(T_{min})}$ in the continuous $D=2$
$YM_{2}$ theory (\ref{1.1}).} Provided $T_{min}(C)$ can be embedded into a
$2d$ manifold, the latter $YM_{2}$ system is conventionally defined on
$T_{min}(C)$ as on the base-space; otherwise a slightly more subtle
construction \cite{Dub3} is to be utilized. (When both $\lambda$ and
$({\mathcal{R}}(s)\Lambda)^{2}$ are large but $\leq{N^{2}}$, in
the r.h. side of eq. (\ref{0.5bxx}) one is to retain only the leading $O(N)$
order of the $1/N$ expansion of the $YM_{2}(T_{min})$ average.) The $YM_{2}$
coupling constant $g_{YM_{2}}$ is related to the original
$D\geq{3}$ $YM_{D}$ constant $g_{YM_{D}}$ via the rescaling 
\be
g^{2}_{YM_{2}}=\xi~g^{2}_{YM_{D}}~\Lambda^{D-2}=
\xi~\frac{\lambda}{N}~\Lambda^{2}
\label{0.4bc}
\ee
that can be viewed (modulo the auxiliary
$\xi$-factor\footnote{In eq. (\ref{0.1ew}), the parameter $\xi$ will be
related to the smearing function (\ref{0.1bb}) reflecting the ambiguity
of the $UV$ regularization.} which can be fixed to be unity) to the
{\it dimensional reduction} $YM_{D}\rightarrow{YM_{2}}$.
In particular, eq. (\ref{0.5bxx}) implies that the leading asymptotics of the
physical string tension $\sigma_{ph}(R)$ (associated to the average
$<W^{R}_{C'}>$ of the Wilson loop in any {\it (anti)chiral} representation
$R$ of the $U(N)$ group) is proportional, see eq. (\ref{2.12bb}),
to the eigenvalue $C_{2}(R)$ of the second $U(N)$ Casimir operator. As a
result, the averaged force between the collinear elementary $YM$
vortices is {\it repulsion}. Borrowing the classification from the pattern of
the (dual) abelian superconductor, the (associated to the gauge theory
(\ref{1.1})) Gauge String corresponds therefore to the type-II superconductor.

Finally, the considered in our paper stringy solution (\ref{0.1}) of the Loop
equation can
be formally {\it continued} into the weak-coupling $\lambda\rightarrow{+0}$
phase as well. Yet, the scaling (\ref{0.1eec}) is evidently in conflict with
the implications of the standard perturbative computations valid for $D=3$ and
$4$ in the limit $\lambda\rightarrow{+0}$. Therefore, in the large $N$
weak-coupling limit, this solution in $D=3,4$ is presumed to describe the
{\it metastable} pattern of the microscopic excitations which are less
energetically favourable than the ordinary perturbative gluons. Nevertheless,
provided all the phase transitions (taking place in the process of decreasing
$\lambda$ from infinity to zero) are {\it not} of deconfining nature, the
analysed {\it SC} dynamics foreshadows confinement in the {\it WC} phase of
the theory (\ref{1.1}) when $D=3,4$.

\subsection{Organization of the paper.}

In Section 2, we first show that the Loop equation (\ref{1.11b}) reduces, for
the contours {\it without} nontrivial selfintersections, to the 
linear equation (\ref{1.11bfb}) (which, for a subclass of the solutions, can
be further reduced to the transparent form of eq. (\ref{1.11bbs})) formulated
directly for the worldsheet weight $w_{2}[\tilde{M}(C)]$. Being restricted to
the canonical space of
the area-functionals, the general solution $w_{2}[\tilde{M}(C)]$ of eq.
(\ref{1.11bfb}) is given by the product of the particular solution (\ref{0.1})
and the generic admissible {\it zero mode} $w^{(0)}_{2}[\tilde{M}(C)]$ of the
the Loop operator $\hat{\mathcal{L}}_{\nu}$.
Also, we briefly sketch how the $U(N)$ gauge theory's correlators can be
reconstructed on the stringy side and make a brief detour to the
regime (\ref{0.1eed}).
Complementary, it is emphasized that the $N=1$ option of
the weight (\ref{0.1}) constitutes, for arbitrary {\it selfintersecting}
contours, the {\it general} $SC$
solution of the full Dyson-Schwinger chain of the loop equations associated to
the $D\geq{2}$ $U(1)$ gauge theory\footnote{In $D\geq{3}$, the
latter abelian theory is enriched with the monopoles (to be treated similarly
to the Wu-Yang formalism \cite{Wu&Yang}) which are supposed to be condensed.} (\ref{1.1}) which is
regularized according to the prescription (\ref{0.1bb}). In Section 3,
on the example of 'short
distance' expansion of the exponent of the weight (\ref{0.1}), it is
explained how the limit of the infinitely thin flux-tube can be used for the
analysis of the Gauge String dynamics (of the fat $YM$ vortices) in the two
distinguished regimes (\ref{0.1eea}) and (\ref{0.1eed}).
In addition, we comment on the existence of the
metastable stringy phase in the $WC$ regime $\lambda\rightarrow{0}$.

Next, to properly utilize the solution of eq. (\ref{1.11bfb})
in the context of the $1/N$ $SC$ expansion for $N\geq{2}$, in Section 4 the
stringy sum endowed with the weight (\ref{0.1}) is compared with the
representation corresponding to the weight
(\ref{0.5c}). It is shown that the difference between the 'zero vortex width'
limit (\ref{2.5bxa}) of the weight (\ref{0.1}) and the weight (\ref{0.5c})
reduces to the one associated to certain {\it contact} interactions (of the
elementary $YM$ vortices) which are totally
{\it unobservable} within the corresponding stringy sums.
On the other hand, the quasi-contact interactions, built into
the quasi-local pattern (\ref{0.1}), are found to be of the abelian nature that has
important implications for the regime (\ref{0.1eed}). Also,
the two complementary prescription, to implement the invariance (absent in
the conventional Nambu-Goto string) of the averages $<W_{C}>$ under the
backtrackings of the contour $C$, are introduced.

In Section 5, it is proved that, in the regime (\ref{0.1eea})/(\ref{0.1eec}),
both the stringy sum endowed with the weight (\ref{0.1}) and the Gauge String
representation based on the (smeared option of the) weight (\ref{0.5c}) are
in the infrared universality class of the Nambu-Goto theory with the weight
(\ref{2.5bb}). In Section 6, the extreme $SC$ limit (\ref{0.1eed}) is briefly
discussed and an interesting infrared reduction of the number of the
independent low-energy stringy excitations (corresponding to the
nonelementary fat $YM$ vortex) is revealed in
this regime. In particular, the $U(N)$ Casimir $C_{2}(R)$-scaling of the
$SC$ asymptotics of the string tension $\sigma_{ph}(R)$ is deduced
for $\forall{R\in{Y^{(N)}_{n}}}$.

In the dynamical regime (\ref{0.1eea}), the novel implications of the
$\Lambda^{2}$-scaling (\ref{0.1eec}) of $\sigma_{ph}$ are
clarified in Section 7. Section 8 contains the proof that the conditions
(\ref{0.1eea})/(\ref{0.1eec}) ensure that the sum over the $YM$ vortices
assigned with the weight (\ref{0.1}) indeed provides with the solution of
the large $N$ Loop equation (\ref{1.11b}) (and, more generally, of the full
Dyson-Schwinger chain of the loop equations). We demonstrate
that the unconventional scaling (\ref{0.1eec}) of $\sigma_{ph}$ implies the
inapplicability of the arguments \cite{MigdElf} concerning
certain mismatch between the structure of the r.h. side of the Loop equation
(\ref{1.11b}) and the pattern of the Nambu-Goto/Polyakov's string. Also,
we adapt the gauge invariant regularization \cite{MigdRep} of eq.
(\ref{1.11b}) in such a way that makes it consistent with the pattern
(\ref{0.1}) of the solution of the reduced Loop equation (\ref{1.11bfb}).
The conclusion provides with the summary of our results and emphasizes the
central direction of the further research. A brief comment on the
interrelation with the approach, based on the so-called $AdS/CFT$
correspondence, is given as well.
Finally, the Appendices serve to decrease the amount of technical details in
the main text.

\section{The solutions of the Loop equation on $\Upsilon_{0}$.}

It is reasonable to expect that the (regularized implementation of the)
$SC/WC$ correspondence \cite{Dub3}, discussed in the begining of the
introduction, is extended to the entire (regularized) $YM_{D}$ functional
integral. In other words, the latter is supposed to be
reformulated as the one over the smooth worldsheets with the
string action predetermined by a smeared counterpart of the weight
(\ref{0.5c}). To justify this expectation, we start with the
representation of the continuous $YM_{D}$ theory (\ref{0.1}) in the space of
the correlators of the Wilson loops 
\be
W_{C}=\frac{1}{N}Tr\Bigg[~{\mathcal{P}}~exp\left(i\oint_{C} dx_{\mu}A_{\mu}(x)
\right)\Bigg]~~~~~,~~~~~A_{\mu}(x)\equiv{A^{a}_{\mu}(x)T^{a}_{ij}}~,
\label{1.2}
\ee
parametrized by the contours $C$ which are represented by the trajectory
${\bf x}(s) \equiv{x_{\mu}(s)}$ in the Euclidean base-space ${\bf R^{D}}$.
The averages of the (products of the) loops (\ref{1.2}) are known to satisfy
the set of the Dyson-Schwinger equations which, in the limit
$N\rightarrow{\infty}$ (so that $<W_{C}>_{\infty}\equiv{<W_{C}>
|_{N\rightarrow{\infty}}}$), can be reduced to the
single Loop equation \cite{LE/MM,LE/P}
\be
\hat{\mathcal{L}}_{\nu}({\bf x}(s))<W_{C}>_{\infty}=\tilde{g}^{2}
\oint\limits_{C} dy_{\nu}(s')~\delta_{D}({\bf y}(s')-{\bf x}(s))
<W_{C_{xy}}>_{\infty}<W_{C_{yx}}>_{\infty},
\label{1.11b}
\ee
where $\tilde{g}^{2}=g^{2}N=\lambda\Lambda^{4-D}$, and
$\hat{\mathcal{L}}_{\nu}({\bf x})$ is the Loop operator specified
by eq. (\ref{1.6bf}) below. As for the loop
$C$, owing to the constraint imposed by the
$D$-dimensional $\delta_{D}(..)$-function, the r.h. side of eq.
(\ref{1.11b}) vanishes unless $C$ has
a selfintersection at a point ${\bf x}(s)={\bf y}(s')$. In other words,
$C\equiv{C_{xx}}=C_{xy}C_{yx}$ is decomposable into the two subloops $C_{xy}$
and $C_{yx}$, with $C_{xy}$ (or $C_{yx}$) collapsing to a point in the case
of the trivial selfintersection.

The eq. (\ref{1.11b}) is notorious for being generically
{\it nonlinear}, and the 'the devil' resides in the effects due to the
nontrivial selfintersections of the loop $C$. Furthermore, a priori, one may
expect
that in eq. (\ref{1.11b}) {\it not} any 'natural' (from the weak-coupling
phase viewpoint) regularization of the $\delta_{D}({\bf x}-{\bf y})$-function
results in a {\it tractable} regularization of the presumed stringy solution
of the Loop equation in the $SC$ phase.
On the contrary, in the sector $\Upsilon_{0}$ (of the full loop space
$\Upsilon$) comprised of the contours $C$ which do {\it not} (nontrivially)
selfintersect, eq. (\ref{1.11b}) considerably simplifies. In particular, there
exists a simple gauge invariant regularization consistent with the transparent
pattern of the regularized stringy solution on $\Upsilon_{0}$.
To take advantage of the latter simplification, the idea\footnote{This can
be compared to the qualitative analysis of \cite{ALaw,MigdRep} where it is
argued that the area-law asymtotics of $<W_{C}>$ is not in conflict
with the Loop equation (\ref{1.11b}) in the $WC$ limit
$\lambda\rightarrow{0}$.} is to decompose the
solution of the Loop equation (\ref{1.11b}) into the two steps. First, one is
to find a subclass of the regularized solutions of the considered
$\Upsilon_{0}$-reduction of
the Loop equation that is the subject of the present section. Then, one is
to analyse under what circumstances thus obtained solutions
correctly reproduce the loop-averages for nontrivially {\it selfintersecting}
contours. In other words, under what conditions there is such a judicious
regularization of the full eq. (\ref{1.11b}) that makes the latter equation
consistent with the considered $\Upsilon_{0}$-solutions.

For any nonintersecting loop $C\in{\Upsilon_{0}}$, the
r.h. side of eq. (\ref{1.11b}) receives a nonzero contribution only from the
{\it trivial} selfintersection point ${\bf x}(s)={\bf y}(s')$ with ${s'=s}$
so that one can put $<W_{C_{xy}}>=<W_{C_{xx}}>$ while $<W_{C_{yx}}>=1$.
Actually, the same simplification takes place for the finite $N$ extension
(given by the $n=1$ option of eq. (\ref{1.11bz}) in Appendix D) of eq.
(\ref{1.11b}) provided by the substitution
\be
\left(\lim_{N\rightarrow{\infty}}
<W_{C_{xy}}><W_{C_{yx}}>\right)~\longrightarrow{~<W_{C_{xy}}W_{C_{yx}}>}~.
\label{0.9zzg}
\ee
It reduces the latter finite $N$ equation to the {\it linear}
one\footnote{When $N=1$, the average
$<W_{C}>$ satisfies eq. (\ref{0.9za}) for an {\it arbitrary} selfintersecting
contour.}
\be                                                      
\hat{\mathcal{L}}_{\nu}({\bf x}(s))<W_{C}>=\tilde{g}^{2}<W_{C}>
\oint\limits_{C} dy_{\nu}(s')~\delta_{D}({\bf y}(s')-{\bf x}(s))~~~~,~~~~
C\in{\Upsilon_{0}}~,
\label{0.9za}
\ee
where $\tilde{g}^{2}=g^{2}N\sim{N^{0}}$, and one can implement (consistently
with the manifest gauge invariance) the smearing prescription
\be
\delta_{D}({\bf x}-{\bf y})~\longrightarrow{~
{\Lambda}^{D}{\mathcal{G}}({\Lambda}^{2}({\bf x}-{\bf y})^{2})}
~~~~~;~~~~~\int d^{D}z~{\mathcal{G}}({\bf z}^{2})=1~.
\label{0.1bb}
\ee
In eq. (\ref{0.1bb}), the sufficiently smooth function
${\mathcal{G}}({\bf z}^{2})$ (so that all its $n\geq{0}$ moments
(\ref{0.3bbx}) are
presumed to be well-defined quantities) satisifies the natural
normalization-condition  that will be complemented by the
two additional constraints (\ref{1.3zp}) and (\ref{0.1xem}).

Next, a priori, one can search for the solution $<W_{C}>$ of the reduced eq.
(\ref{0.9za}) in the form of the sum $\sum w_{2}[\tilde{M}(C)]$ over the
worldsheets $\tilde{M}(C)$, immersed into
${\bf R^{D}}$, which are weighted by some factor $w_{2}[\tilde{M}(C)]$.
(More precise and formal definition of the measure in the latter sum will be
given in Section 4.) To handle eq. (\ref{0.9za}), the first step is to plunge
the latter stringy Ansatz and then perform the {\it interchange} of the
relative order of the differentiation by $\hat{\mathcal{L}}_{\nu}$ and
summation over the worldsheets $\tilde{M}$ that amounts to
\be
\sum_{\tilde{M}} {}^{(r)} \left( \hat{\mathcal{L}}_{\nu}({\bf x}(s))
w_{2}[\tilde{M}(C)]-\tilde{g}^{2}~w_{2}[\tilde{M}(C)]
\oint\limits_{C} dy_{\nu}(s')~\delta_{D}({\bf y}(s')-{\bf x}(s))\right)=0~,
\label{1.6bmb}
\ee
where the sum's superscipt $(r)$ recalls that the transverse string's
fluctuations are to be regularized at the scale $\Lambda$.
The point is to search for such solution $w_{2}[\tilde{M}(C)]$ that
the expression in the big round parentheses of eq.
(\ref{1.6bmb}) does vanish for {\it any} $\tilde{M}(C)$ involved into the
sum. To this aim, let us first take into account
that the Loop operator $\hat{\mathcal{L}}_{\nu}$ is, in
fact, the {\it first} (rather than second) order operator \cite{MigdRep}
complying therefore with the {\it Leibnitz} rule\footnote{
As it will be clear, the solution
(\ref{0.1bbe})/(\ref{0.1}) of the regularized eq. (\ref{1.11bb}) is indeed
consistent with the conditions selecting the latter type of the
$C$-functionals.} As a result, eq. (\ref{0.9za}) can be transformed into
the linear equation (to be regularized according to eq. (\ref{0.1bb}))
\be
\hat{\mathcal{L}}_{\nu}({\bf x}(s))~
ln\left(w_{2}[\tilde{M}(C)]\right)=\tilde{g}^{2}
\oint\limits_{C} dy_{\nu}(s')~\delta_{D}({\bf y}(s')-{\bf x}(s))
\label{1.11bfb}
\ee
which, operating directly with the worldsheet's weight $w_{2}[\tilde{M}(C)]$,
is suitable for our further analysis. Remark also that, in accord with the
general structure of
the Dyson-Schwinger equations, eq. (\ref{1.11bfb}) merely determines the
{\it discontinuity} that, being represented by its r.h. side,
is exhibited by the total {\it YM}-flux 
('localized' within the $YM$ vortex) which terminates at the Wilson loop
source.

\subsection{The stringy form of Loop Equation for $C\in{\Upsilon_{0}}$.}

Due to the first-order nature of 
$\hat{\mathcal{L}}_{\nu}$, the general solution $w_{2}[\tilde{M}(C)]$ of eq.
(\ref{1.11bfb}) assumes the form
\be
w_{2}[\tilde{M}(C)]=\tilde{w}_{2}[\tilde{M}(C)]~w^{(0)}_{2}[\tilde{M}(C)]
~~~~~~~;~~~~~~~
\hat{\mathcal{L}}_{\nu}({\bf x}(s))~ln(w^{(0)}_{2}[\tilde{M}(C)])=0,
\label{1.9add}
\ee
where $\tilde{w}_{2}[..]$ is any particular solution
of eq. (\ref{1.11bfb}) (to be identified with the one of eq. (\ref{0.1})),
while $w^{(0)}_{2}[\tilde{M}(C)]$ is formally allowed to be an arbitrary
$N$-{\it independent zero mode}\footnote{More precisely,
$ln(w^{(0)}_{2}[..])$ is presumed
to support the proper cluster decomposition (to be introduced in Appendix B)
similar to the one discussed in \cite{MigdRep}.} of the
Loop operator. The latter
can be defined \cite{MigdRep} as the combination
\be
\hat{\mathcal{L}}_{\nu}({\bf x}(s))=
\partial_{\mu}^{{\bf x}(s)}~
\frac{\delta}{\delta \sigma_{\mu\nu}({\bf x}(s))}
\label{1.6bf}
\ee
where $\delta/\delta\sigma_{\mu\nu}({\bf x}(s))$ and
$\partial_{\mu}^{{\bf x}(s)}$ denote respectively the Mandelstam
area-derivative and the path-derivative (see Appendix B which recalls their
definitions).

To proceed further, it is helpful to rewrite the r.h. side of eq.
(\ref{1.11bfb}) as the surface integral over $\tilde{M}(C)$. This
can be done making use of the simple identity
\be
\oint\limits_{C=\partial\tilde{M}(C)}dy_{\nu}~
{\mathcal{K}}(({\bf y}-{\bf x})^{2})=
-\partial_{\mu}^{{\bf x}(s)}
\int\limits_{\tilde{M}(C)} d\sigma_{\mu\nu}({\bf y})~
{\mathcal{K}}(({\bf y}-{\bf x})^{2})~,
\label{1.11bfs}
\ee
where ${\mathcal{K}}(({\bf y}-{\bf x})^{2})$ is a generic smooth function
which regularizes, according to eq. (\ref{0.1bb}), the
$\delta_{D}(..)$-function of eq. (\ref{1.11bfb}). 
(In the derivation of eq. (\ref{1.11bfs}), one is to employ that in this case
the path-derivative $\partial_{\mu}^{{\bf x}(s)}
\equiv{\delta/\delta x_{\mu}(s)}$ can be substituted by the
ordinary derivative ${\partial}/{\partial x_{\mu}}$ provided
$|\delta x_{\mu}(s)|$ is infinitesimal compared to the decay-length of
${\mathcal{K}}(({\bf y}-{\bf x})^{2})$. Also, the
abelian Stokes theorem should be used.) Comparing the r.h. side of eq.
(\ref{1.11bfs}) with the pattern (\ref{1.6bf}) of $\hat{\mathcal{L}}_{\nu}$,
one concludes that a particular solution $\tilde{w}_{2}[..]$ of eq.
(\ref{1.11bfb}) may be found from the equation
\be
\frac{\delta~ln\left(\tilde{w}_{2}[\tilde{M}(C)]\right)}
{\delta\sigma_{\mu\nu}({\bf x}(s))}=
-\frac{\tilde{g}^{2}}{2}~(\delta_{\mu\alpha}\delta_{\nu\beta}-
\delta_{\mu\beta}\delta_{\nu\alpha})~
\int\limits_{\tilde{M}(C)} d\sigma_{\alpha\beta}({\bf z})~
\delta_{D}({\bf x}-{\bf z})~.
\label{1.11bbs}
\ee
Therefore, applying the area-derivative
$\delta/\delta\sigma_{\rho\sigma}({\bf y}(s'))$ to both
sides of eq. (\ref{1.11bbs}), we arrive at the relatively simple equation
which, prior to the regularization (\ref{0.1bb}) of its r.h. side, reads
formally
\be
\frac{2~\delta^{2}~ln\left(\tilde{w}_{2}[\tilde{M}(C)]\right)}
{\delta\sigma_{\mu\nu}({\bf x}(s))~\delta\sigma_{\rho\chi}({\bf y}(s'))}=
-(\delta_{\mu\rho}\delta_{\nu\chi}-\delta_{\mu\chi}\delta_{\nu\rho})~
\tilde{g}^{2}~\delta_{D}({\bf x}-{\bf y})~.
\label{1.11bb}
\ee

Next, judging from the pattern of eq. (\ref{1.11bb}), it is suggestive that
this equation would be further simplified if the
Mandelstam area-derivatives were traded for the ordinary
functional area-derivatives (preliminary restricted to the boundary $C$)
\be
\frac{\delta}{\delta \sigma_{\mu\nu}({\bf x}(s))}~\longrightarrow~
\frac{\delta_{f}}{\delta p_{\mu\nu}({\bf x}(\gamma))}
\Bigg|_{{\bf x}(\gamma)={\bf x}(s)\in{C}}
\label{1.9adb}
\ee
with respect to the standard infinitesimal area-element
\be
d\sigma_{\mu\nu}({\bf x}(\gamma))=
p_{\mu\nu}(\gamma)~d^{2}\gamma~~~~~~~,~~~~~~~
p_{\mu\nu}(\gamma)=\varepsilon^{ab}~\frac{\partial x_{\mu}(\gamma)}
{\partial\gamma^{a}}
\frac{\partial x_{\nu}(\gamma)}{\partial\gamma^{b}}~,
\label{0.2c}
\ee
where the coordinates $x_{\mu}(\gamma)\equiv{x_{\mu}(\gamma_{1},\gamma_{2})}$
define the position
of a given worldsheet $\tilde{M}$ in the base-space ${\bf R^{D}}$.
To justify the substitution (\ref{1.9adb}), the simplest and natural option
is to constrain that the solution $-ln(w_{2}/N^{\chi})$ of eq.
(\ref{1.11bb}) belongs to the space ${\bf \Psi}$ of the
area-functionals\footnote{The futher analysis of the full $U(N)$ Loop
equation (\ref{1.11b}) reveals that the restriction to the 
${\bf \Psi}$-space (\ref{1.9adc}) is indeed adequate at least in the regime
(\ref{0.1eea})/(\ref{0.1eec}).}
\be
{\bf \Psi}~:~~~\sum_{n\geq{2}}~\int\limits_{\tilde{M}}...
\int\limits_{\tilde{M}}d\sigma_{\mu_{1}\nu_{1}}(\gamma^{(1)})...
d\sigma_{\mu_{n}\nu_{n}}(\gamma^{(n)})~
{\bf \mathcal{S}}^{(n)}_{\{\mu_{k}\nu_{k}\}}
(\{{\bf x}(\gamma^{(i)})-{\bf x}(\gamma^{(j)})\})~,
\label{1.9adc}
\ee
where ${\bf \mathcal{S}}^{(n)}_{\{\mu_{k}\nu_{k}\}}(..)
\equiv{{\mathcal{S}}^{(n)}_{\mu_{1}\nu_{1}..\mu_{n}\nu_{n}}(..)}$ is an
arbitrary translationally invariant $2n$-tensor (with respect to the $O(D)$
group of the Euclidean rotations) depending on the relative coordinates
${\bf x}(\gamma^{(i)})-{\bf x}(\gamma^{(j)})$ which are allowed to take
values in the full base-space ${\bf R^{D}}$ rather than only on $\tilde{M}$.
(Eq. (\ref{1.9adb}) implies that the area-derivative's variation
$|\delta \sigma_{\mu\nu}({\bf x}(s))|\rightarrow{0}$ is infinitesimal
compared to the decay-length of any of the kernals
${\bf \mathcal{S}}^{(n)}_{\{\mu_{k}\nu_{k}\}}(..)$.)

Furthermore, on the space (\ref{1.9adc}), in eq. (\ref{1.9adb}) the
restriction of the functional derivatives to the boundary $C$ can be safely
omitted. In sum, we arrive at the following reduction of 
eq. (\ref{1.11bb}):             
\be
\frac{2~\delta^{2}_{f}~ln\left(\tilde{w}_{2}[\tilde{M}(C)]\right)}
{\delta p_{\mu\nu}({\bf x}(\gamma))~
\delta p_{\rho\sigma}({\bf y}(\gamma'))}=
-(\delta_{\mu\rho}\delta_{\nu\sigma}-
\delta_{\mu\sigma}\delta_{\nu\rho})~\lambda{\Lambda}^{4}~
{\mathcal{G}}({\Lambda}^{2}({\bf x}-{\bf y})^{2})~,
\label{1.11bbx}
\ee
where $\lambda$ is defined in eq. (\ref{1.1bx}), and we have implemented the
smearing (\ref{0.1bb}) of the $\delta_{D}$-function.
Then, modulo the overall constant
(excluded, in fact, through the matching with the full eq. (\ref{1.11b})),
the pattern\footnote{In \cite{Ol&Pet}, the $\chi=1$ option of the pattern
(\ref{0.1}) was discussed, with $\tilde{M}$ being {\it fixed} as the minimal
area-worldsheet, as the ansatz for the solution $<W_{C}>$ of the 
loop-equation considered on a limited {\it subspace} of the loop space.
In \cite{LE/MM,ALaw}, a 'cousin' of the
pattern (\ref{0.1}) was considered: the normalization (\ref{0.1bb}) (required
for the consistency with the regularized option of eq. (\ref{0.9za})) of the
$(D-2)$th moment (\ref{0.3bbx}) of ${\mathcal{G}}({\bf z}^{2})$ is
{\it absent} while
the alternative constraint, corresponding to $\xi=1$ in eq. (\ref{0.3b}), is
imposed on the {\it zeroth} moment.}
 (\ref{0.1}) is the most general solution of eq. (\ref{1.11bbx})
when we presume one more constraint motivated by the $YM_{D}/String$ duality
\cite{Dub3}. Namely, we require the $exp[a\chi]$-dependence of the weight on
the Euler characteristics $\chi$ of the worldsheet $\tilde{M}$. As it is
predetermined by the relations (\ref{0.9zzi}) of Appendix D, in the
regime (\ref{0.1eea})/(\ref{0.1eec}), the consistency of the
pattern (\ref{0.1}) with the full Loop equation (\ref{1.11b}) requires to
identify $a=ln(N)$, where $N$ parametrizes the associated $U(N)$ group.

\subsection{A brief digression to the extreme $SC$ asymptotics
(\ref{0.1eed}).}

At this step it is appropriate to remark that, when the first of the
conditions (\ref{0.1eea}) is violated, our previous analysis can be utilized
at least in the following two situations. Namely, {\it either}
in the $N=1$ case of the $U(1)$ pure gauge theory (with monopoles, see eq.
(\ref{1.1bb}), in $D\geq{3}$) {\it or}, for $N\geq{2}$, in the regime
(\ref{0.1eed}) when the saddle-point flux-tube is further {\it restricted} to
be represented by an elementary $YM$ vortex with {\it unobservable} (modulo
exponentially suppressed contribution going beyond the short-distance
expansion (\ref{0.3bbq})) selfoverlapping. In particular, in the extreme $SC$
limit (\ref{0.1eed}), the ${\bf \Psi}$-space (\ref{1.9adc}) can be employed
to deduce the leading asymptotics (entering e.g. eq. (\ref{0.5bxx})) of the
average $<W_{C'}^{R}>$ in any given (anti)chiral representation
$R\in{Y^{(N)}_{n}}$ of $U(N)$. The short-cut way to recover
this asymptotics is to consider the analogue of the reduced Loop equation
(\ref{0.9za}) but formulated for $<W_{C'}^{R}>$ rather than for
$<W_{C}>\equiv{<W_{C}^{f}>}$.
Repeating the previous steps, one arrives at the modification of eq.
(\ref{1.11bb}) provided by the substitution
\be
g^{2}N~\longrightarrow{~g^{2}C_{2}(R)}~~~~~,~~~~~~
\lambda~\longrightarrow{~\lambda C_{2}(R)/N}~~~~~~~;~~~~~~~R\in{Y^{(N)}_{n}}~,
\label{2.12xc}
\ee
where $C_{2}(R)$ is the (associated to $R$) eigenvalue (\ref{2.12bv}) of the
second $U(N)$ Casimir operator.
Thus modified equation evidently admits the solution similar to
eq. (\ref{0.1}) that can be utilized to write down the following asymptotics
of the average
\be
\frac{<W^{R}_{C'}>}{dimR}\longrightarrow
exp\left(-\frac{\lambda\Lambda^{2}C_{2}(R)}{4N}
\int\limits_{T_{min}}\int\limits_{T_{min}}
d\sigma_{\mu\nu}({\bf x})d\sigma_{\mu\nu}({\bf y})
~{\Lambda}^{2}{\mathcal{G}}({\Lambda}^{2}({\bf x}-{\bf y})^{2})\right),
\label{2.12bl}
\ee
where $dimR$ is the dimension of the (anti)chiral $U(N)$ representation $R$
in question, and $T_{min}\equiv{T_{min}(C')}$ is the support (of the relevant
conglomerate, see eq. (\ref{0.5cxd})) of the minimal area worldsheets.
A more refined analysis, to be presented in a separate paper,
reveals that in the extreme $SC$ limit (\ref{0.1eed}) the asymptotics
(\ref{2.12bl}) is indeed consistent with
the full chain of the loop equations, at least provided one
more condition is imposed. The support $T_{min}(C')$
should be further constrained to possess
topology of such a smooth disc (which is, by construction, devoid of
selfintersections) that the short-distance expansion (\ref{0.3bbq}) of the
exponent of eq. (\ref{2.12bl}) disregards\footnote{More explicitly, at any
point of $T_{min}(C')$, the line in the normal direction either does not have
the second intersection with $T_{min}(C')$ or the second intersection takes
place at a distance $>>\Lambda^{-1}$.}  in the latter equation only some
exponentially suppressed 'nonperturbative' contribution irrelevant in the
considered $SC$ limit.

\subsection{Zero modes of the Loop operator $\hat{\mathcal{L}}_{\nu}$.}

Given the particular solution (\ref{0.1}) of eq. (\ref{1.11bfb}), 
the construction of the general solution (\ref{1.9add}) requires to find the
admissible class of the zero modes of the Loop operator
(\ref{1.6bf}). Being as previously constrained to belong to the
${\bf \Psi}$-space (\ref{1.9adc}), the general $N$-independent solution
$w^{(0)}_{2}[\tilde{M}(C)]$ of the zero-mode
equation (\ref{1.9add}) is obtained in Appendix B. There, we also make
contact with the confining string Ansatz \cite{PolyakCS} employing the
Kalb-Ramond representation \cite{Kalb&Ramond} of the area-functionals
(\ref{1.9adc}). As for the present subsection, we merely present the
simplest variety of $w^{(0)}_{2}[\tilde{M}(C)]$ and comment on its
interpretation.

The selected variety can be deduced from the general
pattern (\ref{1.9adc}) retaining only the bilocal $n=2$ contribution where
${\bf \mathcal{S}}^{(2)}_{\{\mu_{k}\nu_{k}\}}({\bf x}-{\bf y})$ is reduced to
${\Lambda}^{4}{\mathcal{M}}_{\mu\nu;\rho\sigma}({\Lambda}({\bf x}-{\bf y}))$
with
\be
{\mathcal{M}}_{\mu\nu;\rho\sigma}({\bf z})=
[\delta_{\mu\rho}\delta_{\nu\sigma}-\delta_{\mu\sigma}\delta_{\nu\rho}]
\partial_{\lambda}\partial^{\lambda}{\mathcal{V}}({\bf z}^{2})-
[\partial_{\mu}\partial_{\rho}\delta_{\nu\sigma}-...]
{\mathcal{V}}({\bf z}^{2})~,
\label{0.1xx}          
\ee
where $\partial_{\lambda}\equiv{\partial}/{\partial z_{\lambda}}$, and
${\mathcal{V}}({\bf z}^{2})$ is an arbitrary smooth function
vanishing sufficiently fast at infinity ${\bf z}^{2}\rightarrow{\infty}$.
As for the second square braket in eq. (\ref{0.1xx}), it stands for  
$[\partial_{\mu}\partial_{\rho}\delta_{\nu\sigma}-
\partial_{\mu}\partial_{\sigma}\delta_{\nu\rho}+
\partial_{\nu}\partial_{\sigma}\delta_{\mu\rho}-
\partial_{\nu}\partial_{\rho}\delta_{\mu\sigma}]$ so that
${\mathcal{M}}_{\mu\nu;\rho\sigma}({\bf z})$ is totally
antisymmetric with respect to the two pairs (i.e.
$\{\mu\nu\}$ and $\{\rho\sigma\}$) of the tensor indices.
The defining property of the tensor (\ref{0.1xx}) reads as
$\partial_{\mu}{\mathcal{M}}_{\mu\nu;\rho\sigma}({\bf z})=0$
which, as it is shown in Appendix B, can be easily generalized to construct
the more general zero-mode solutions corresponding
to the $n\geq{3}$ terms of the area-functional (\ref{1.9adc}).

Finally, it is straightforward to verify that the modification (\ref{1.9add})
of the particular solution (\ref{0.1}), due to the inclusion of the
admissible
zero-modes, does {\it not} alter the general conclusions of Sections 3 and 5
concerning the stringy dynamics in the regime (\ref{0.1eea})/(\ref{0.1eec}).
(In particular, the general pattern of the short-distance expansion
(\ref{0.3bbq}) arises for any area-functional (\ref{1.9adc}).) To be more
specific, let us consider the bilocal mode
$w^{(0)}_{2}[\tilde{M}(C)]$ corresponding to eq. (\ref{0.1xx}).
On the one hand, the contribution associated to the first term of the latter
equation can be viewed as the additive 'renormalization',
${\mathcal{G}}\rightarrow{{\mathcal{G}}+8\partial_{\lambda}
\partial^{\lambda}{\mathcal{V}}({\bf z}^{2})/\lambda}$, of the smearing
function in eq. (\ref{0.1}). Furthermore, the normalization (\ref{0.1bb}) of
the $(D-2)$th moment (\ref{0.3bbx}) of ${\mathcal{G}}$ is not changed by this
modification, while the remaining moments of ${\mathcal{G}}$ a priori are not
fixed anyway. On the other hand, the double surface-integral of the second
term ($\sim{{\mathcal{V}}({\bf z}^{2})}$) of eq. (\ref{0.1xx}) can be
transformed, via the abelian Stokes theorem, into the double contour-integral
over the boundary $C$. In consequence, this subleading perimeter-type
contribution is of secondary importance for the indentification of the
infrared universality class (predetermined, in the leading order, by the
dynamics of the worldsheet's interior) to which the analysed stringy system
(\ref{0.1}) belongs.

\subsection{The $N=1$ versus $N\geq{2}$ cases.}

Judging from the previous discussion, there still remains uncertainty if the
solution (\ref{1.9add})/(\ref{0.1}) of eq. (\ref{1.11bfb}) can be employed to
reproduce the $1/N$ {\it expansion} in the $SC$ phase of the $U(N)$ gauge
theory (\ref{1.1}). The problem is foreshadowed by the fact that 
the purported 't Hooft topological weight $N^{\chi}$ can {\it not} be
determined from the reduced eq. (\ref{0.9za}) merely because the latter
$\Upsilon_{0}$-equation, being $N$-{\it independent} (once $\tilde{g}^{2}$
is fixed independently of $N$), evidently {\it decouples} into the set of
equations each associated to a given Euler character $\chi$ of the
worldsheets $\tilde{M}_{\chi}$ involved into the relevant stringy sum
(\ref{1.6bmb}) to be formalized by eq. (\ref{0.1bbe}).
Upon a reflection, the derivation of the $N^{\chi}$-factor requires to
check the consistency of the pattern (\ref{0.1bbe})/(\ref{0.1}) with the
whole Dyson-Schwinger chain of the $U(N)$ loop equations considered within
the framework of the $1/N$ series. In particular, this analysis will reveal
that, even for $N\geq{2}$, the Ansatz (\ref{0.1bbe})/(\ref{0.1}) (and its
generalization (\ref{1.9add})) is indeed consistent with the latter chain in
the regime (\ref{0.1eea})/(\ref{0.1eec}), provided the additional constraint
(\ref{1.3zp}) on the normalized smearing function (\ref{0.1bb}) is satisfied.
On the contrary, in the extreme $SC$ limit (\ref{0.1eed}), the considered
$N\geq{2}$ Anzatz works only on a limited sector
of the loop space (including a subclass of nonselfintersecting loops).

Before we begin to discuss the $N\geq{2}$ issues, it is instructive
to make a brief detour to the $N=1$ case where the situation is
substantially different. The reason is that, when $N=1$, the reduced Loop
equation (\ref{0.9za}) is valid for an {\it arbitrary} (selfintersecting)
loop $C$. Furthermore, for a generic contour, the $N=1$ option of the Ansatz 
(\ref{0.1bbe})/(\ref{0.1}) is consistent with the chain of the $U(1)$ loop
equations (and, in fact, with the abelian Bianchi identities $
\epsilon_{\mu\nu\rho\lambda}\partial_{\rho}^{{\bf x}(s)}
\frac{\delta}{\delta \sigma_{\mu\nu}({\bf x}(s))}<W_{C}>=0$ as it can
be shown employing, e.g., the arguments from \cite{PolyakCS}). On the
other hand, the subtlety arises because in $D\geq{3}$ the considered variety
(\ref{1.9add}) of the solutions is associated, see Appendix B, to the
abelian gauge theory enriched with the monopoles\footnote{Similar continuous
systems have been recently considered in \cite{PolyakCS}.} which are supposed
to be condensed. Actually, the presence of the Dirac-like monopoles
(so that the magnetic current has the support on the union of the
$(D-3)$-dimensional hypersurfaces) could be anticipated in advance.
Indeed, the $N=1$ pattern (\ref{0.1bbe})/(\ref{0.1}) can be viewed as the
continuum counterpart of the representation \cite{Gop&Mack} (arising after
the so-called abelian duality transformation) in the {\it lattice} pure $U(1)$
gauge theory with the Heat-Kernal action.

Similarly to the latter lattice theory, it is reasonable to expect that the
$N=1$ variant of the Gauge String \cite{Dub3} provides with the representation
of the $SC$ series in the following {\it compactified} variant of the
ordinary continuous $U(1)$ system (\ref{1.1}). Selecting for simplicity the
$D=4$ case, in the $N=1$ action (\ref{1.1}) the field-strength tensor
$F_{\mu\nu}=\partial_{\mu}\wedge A_{\nu}({\bf x})$
is to be 'minimally' extended,
\be
F_{\mu\nu}({\bf x})~\longrightarrow{~
\tilde{B}_{\mu\nu}({\bf x})=
F_{\mu\nu}({\bf x})+\frac{1}{2}\epsilon_{\mu\nu\rho\sigma}
\frac{\partial}{\partial x_{\rho}}\wedge
\int\limits_{\bf R^{4}}d^{4}y
<{\bf x}|\frac{1}{\partial_{\lambda}\partial^{\lambda}}|{\bf y}>
k_{\sigma}({\bf y})}~,
\label{1.1bc}
\ee
to incorporate the contribution generated by the {\it magnetic}
current $k_{\sigma}({\bf y})$, to be averaged over, corresponding to the
elementary magnetic charge $2\pi$. (The singular Dirac string is supposed to
be eliminated through the Wu-Yang construction \cite{Wu&Yang}.)
In the sector of $n$ monopole's
loops $C_{q}$, the current conventionally reads:
$k_{\sigma}({\bf y})=\sum^{n}_{q=1}2\pi\int_{C_{q}} dz_{\sigma}(s_{q})
\delta_{4}({\bf z}(s_{q})-{\bf y})$, where one is sum over all immersions
${\breve{C}_{q}}\rightarrow{C_{q}}$ into ${\bf R^{4}}$.
Then, according to Appendices B and D, the $N=1$ Ansatz
(\ref{0.1bbe})/(\ref{0.1}) is supposed to be the general solution (of the set
of the abelian loop equations) which reproduces the {\it SC} expansion of the
arbitrary loop-observables
in the regularized $D\geq{3}$ $U(1)$ system\footnote{For $D=2$, the result
(\ref{0.1}) can be alternatively obtained applying the standard
Feynman diagrammatics to the reduced variant of the $U(1)$ system
(\ref{1.1bb}) {\it without} monopoles.} with the action
\be
\sum_{n=0}^{\infty}\sum_{\{C_{q}|n\}}\frac{1}{n!}~
exp\Bigg[{-\frac{1}{4g^{2}}\int d^{4}x~d^{4}y
\left(\tilde{B}_{\mu\nu}({\bf x})~{\Lambda}^{D}<{\bf x}|~
1/\tilde{\mathcal{G}}(\partial_{\mu}\partial^{\mu}/
\Lambda^{2})~|{\bf y}> \tilde{B}_{\mu\nu}({\bf y})\right)}\Bigg],
\label{1.1bb}
\ee
(where $\tilde{\mathcal{G}}(p_{\mu}p^{\mu})$ is the four-dimensional
Fourier image of the smearing function ${\mathcal{G}}(z_{\mu}z^{\mu})$)
corresponding to the compactified (via the substitution (\ref{1.1bc}))
counterpart of the $U(1)$ action (\ref{1.1}) regularized in compliance with
eq. (\ref{0.1bb}).
(Note that, akin the lattice case, one is to perform the canonical averaging
over the $n$-loop ensembles $\{C_{q}|n\}$ with $n=1,2,...$.)
On the other hand, apart from the regime (\ref{0.1eea}) where the
infrared universality takes place, the nontrivial zero-modes
$w^{(0)}_{2}[\tilde{M}(C)]\neq{1}$ (being
constrained to meet the requirement of the proper cluster
decomposition, see Appendix B) refer in $D\geq{3}$ to the continuous $U(1)$
gauge theories
other than the one of eq. (\ref{1.1bb}): the 'compactification' is
represented by the {\it combination} of the minimal extension (\ref{1.1bc}) with
some further $F_{\mu\nu}$-{\it independent} modification
(see e.g. the expression in the large round brakets of eq. (\ref{1.6zxc}) in
Appendix B) of the abelain action.

\subsection{The gauge theory's correlators on the stringy side.}

Let us presume that the smearing is consistent with the Bianchi identities
(which indeed can be manifestly justified at least in the regimes
(\ref{0.1eea}) and (\ref{0.1eed})) reformulated, according to eq.
(\ref{1.6bs}) below, in
terms of the loop-operators \cite{MigdRep}. Then, one can
build up the regularized $YM_{D}$ correlators from the properly associated
states of the open\footnote{As usual, the correspondence with the closed
string states is maintained when the 'holes', associated to the boundary
loops, collapse to the points.} Gauge String visualized in terms of the fat
rather than infinitely thin flux-tubes. For this purpose, we are to employ the
canonical reinterpretation \cite{MigdRep} of the area- and path-derivatives,
combined into the Loop operator (\ref{1.6bf}), in terms of the corresponding
$YM_{D}$ structures (see eq. (\ref{1.6bs}) below). To this aim, recall
first that, provided some (gauge invariant) regularization of the
$\delta_{D}$-function, the representation (\ref{1.11b}) of the Loop equation
refers to the regularized, at some $UV$ scale $\Lambda$, but {\it not} to the
renormalized $YM_{D}$ theory (\ref{1.1}). The consistency of
the regularized eq. (\ref{1.11b}) requires that
$\Lambda^{2}\cdot|\delta \sigma_{\mu\nu}({\bf x}(s))|\rightarrow{0},~
\Lambda\cdot|\delta x_{\mu}(s)|\rightarrow{0}$,
where the variations $\delta \sigma_{\mu\nu}({\bf x}(s))$ and
$\delta x_{\mu}(s)$ refer, see Appendix B, to the corresponding derivatives
in the Loop operator.

The key-point is that, employing thus introduced derivatives, from the
multiloop averages $<\prod_{p} W_{C_{p}}>$ one can reconstruct, at least in
principle, the stringy representation for (the infrared asymptotics of) a
generic field-strength correlator, with some path-ordered exponents of the
gauge field possibly
included, in the regularized $U(N)$ gauge theory (\ref{1.1}).
This can be done because, when the path-derivatives act on
$<\prod_{p} W_{C_{p}}>$ {\it after} the area-derivatives, they altogether can
insert into the trace of each ordered exponent (\ref{1.2}) respectively the
field-strength tensor and the conventional covariant derivative,
\be
\frac{\delta}{\delta \sigma_{\mu\nu}({\bf x}(s))}~\longrightarrow{~
iF_{\mu\nu}^{a}({\bf x}(s))}T^{a}_{kj}~~~~~,~~~~~
\partial_{\mu}^{{\bf x}(s)}~\longrightarrow
{\hat{D}_{\mu}^{a}[A({\bf x}(s))]T^{a}_{kj}}~,
\label{1.6bs}
\ee
where the colour $k.j$-indices are to be contracted with the corresponding
indices of the path-ordered exponents (originally combined into the set of
the loops $W_{C_{p}}$) 'ingoing' and 'outgoing' from a particular point
${\bf x}(s_{p})$ of a given boundary $C_{p}$. Contracting all the involved
loops $C_{p}$ to the points, in this way one is supposed to refine the
amplitudes which originally appeared in the framework of the so-called dual
resonance models stepping out at dawn of the hadron's physics.

\section{The fat vs. infinitely thin $YM$ vortices.}

Prior to the analysis of the fat flux-tubes' dynamics, it is helpful to
consider first the mathematical idealization (e.g., of
the solution (\ref{0.1})) corresponding to the infinitely thin $YM$
vortex described, therefore, through the weight which is {\it local}
(like the one of eq. (\ref{0.5c}) or (\ref{2.5bxa}) below) on the worldsheet.
The reason is twofold. On the one hand, in the extreme $SC$ limit
(\ref{0.1eed}), the condition (\ref{0.1eeb}) will be shown to ensure that the
considered idealization provides with the adequate description of the
(conglomerate of the) saddle-point worldsheets entering the r.h. side of the
asymptotics (\ref{0.5bxx}) of $<W_{C}>$. On the other hand,
in the regime (\ref{0.1eea})/(\ref{0.1eec}), this local deformation of the
Gauge String weight is expected to provide with the short-cut way to the
identification of the universality class (stated in the Introduction)
corresponding to the original fat $YM$ vortex.

As for the latter identification, according to Section 5, one
can take advantage of the following evident freedom in the definition of the
smearing of the Nambu-Goto ($NG$) weight (\ref{2.5bb}). {\it In the
considered
regime, any two stringy systems (formulated in terms of possibly fat
$YM$ vortices) are supposed to belong to the same infrared universality class
if and only if their local deformations are equivalent, modulo possible
rescaling of the coupling constants, to the Nambu-Goto theory}. The stated
selection-rule is aimed to take into account that there are infinitely many
(see Appendix C for a general discussion) seemingly different but in fact
equivalent formulations of the considered $NG$ theory which are all
admissible as the starting point for the subsequent smearing of the weight.
The reformulations in question are generated merely by arbitrary nonsingular
modifications (retaining the locality on the worldsheet) of the
original $NG$ weight (\ref{2.5bb}) on any {\it measure zero} subspace of
the space spanned by relevant worldsheet's configurations. In this
perspective, the Nambu-Goto Ansatz can be reinterpreted as the
{\it simplest} implementation of the stringy theory which alternatively can
be defined, in the apparently more complicated way, either through the weight
(\ref{0.5c}) or via the local deformation (given by eq. (\ref{2.5bxa})) of
the weight (\ref{0.1}).

\subsection{The 'brute force' local limit of the quasi-local weight
(\ref{0.1}).}

In order to obtain the required local deformation of the pattern (\ref{0.1}),
we have to omit temporarily the normalization condition
(\ref{0.1bb}) and allow that the smearing function is
{\it analytically continued},
\be
{\Lambda}^{2}{\mathcal{G}}({\Lambda}^{2}({\bf x}-{\bf y})^{2})
\longrightarrow{\xi~\delta^{w}_{2}({\bf x}(\gamma)-{\bf y}(\gamma'))}=
\xi~\frac{\delta_{2}(\gamma-\gamma')}{\sqrt{p^{2}_{\mu\nu}(\gamma)/2}}~,
\label{0.3b}          
\ee
to merge with the 2-dimensional delta-function
$\delta^{w}_{2}({\bf x}-{\bf y})$
on the {\it worldsheet} $\tilde{M}$ so that ${\bf x},{\bf y}\in{\tilde{M}}$.
(The area-element $p_{\mu\nu}(\gamma)$, entering eq. (\ref{0.3b}), has been
introduced in eq. (\ref{0.2c}).)
In this case, the quasi-local weight (\ref{0.1}) reduces to its local limit
\be
w_{3}[\tilde{M}_{\chi}]=
N^{\chi}~exp\left(-\frac{\bar{\lambda}\Lambda^{2}}{2}
\sum_{q}~(n_{q})^{2}~\bar{A}_{q}\right)~~~~~~~;~~~~~~~
\bar{\lambda}=\xi\lambda~,
\label{2.5bxa}
\ee
where the integer number $n_{q}$ is to be identified with the number of the
(oriented) sheets covering the corresponding $q$th
elementary domain $T_{q}$ (of the area $A[T_{q}]
\equiv{\bar{A}_{q}}=\int d^{2}\gamma_{q}\sqrt{p^{2}_{\mu\nu}(\gamma_{q})/2}$)
of the support $T$ of $\tilde{M}_{\chi}$ so that
$\sum_{q} |n_{q}| \bar{A}_{q}=A[\tilde{M}_{\chi}]$.

Upon a reflection, the pattern (\ref{2.5bxa}) implies that the net effect of
the contact interactions (to be discussed in Section 4 in more details)
between the associated elementary $YM$ vortices  is represented by the
{\it ratio}
\be
\varrho[\tilde{M}_{\chi}]=\frac{w_{3}[\tilde{M}_{\chi}]}
{~~~~w_{1}[\tilde{M}_{\chi}]|_{m_{0}=0}}=
exp\left(-\frac{\bar{\lambda}\Lambda^{2}}{2}\sum_{q}~[~(n^{2}_{q}-|n_{q}|)~
\bar{A}_{q}~]\right)
\label{2.5bxb}
\ee
of the pattern (\ref{2.5bxa}) and the
$m_{0}=0,~\bar{\lambda}(\lambda)=\xi\lambda$ option of the
Nambu-Goto weight (\ref{2.5bb}), both assigned to a given worldsheet
$\tilde{M}_{\chi}$ corresponding to a particular $\{n_{q}\}$-covering. 
Therefore, for $\prod_{q}n_{q}\neq{0}$, $\varrho[\tilde{M}_{\chi}]$ is
{\it not} equal to unity only for the multi-sheet covering of $T$, i.e. there
must be at least one
$q$ so that $|n_{q}|\geq{2}$ which can be visualized as the corresponding
{\it 2-dimensional selfintersection} (with the support on $T_{q}$)
of the worldsheet $\tilde{M}_{\chi}$. In consequence, for any
worldsheet $\tilde{M}_{\chi}$ {\it without} selfintersections on some $2d$
submanifolds, the pattern (\ref{2.5bxa}) reduces to the sheer $m_{0}=0$
option of the Nambu-Goto weight (\ref{2.5bb}) corresponding to the bare
string tension
\be
\sigma_{0}=\frac{\xi\lambda\Lambda^{2}}{2}~~~~~~~~;~~~~~~~~
\xi=\int d^{2}z~{\mathcal{G}}({\bf z}^{2})\sim{1}~.
\label{0.1ew}
\ee
furthermore, one can argue that (at least when $\lambda\sim{1}$) the factor
$\xi$, reflecting the $D\geq{3}$ regularization ambiguity, is
supposed to be {\it of order of unity} once the solution (\ref{0.1}) describes
stable, rather than metastable, stringy excitations. (As we will
see, the same ambiguity is built into the mechanism \cite{Dub3} of the
$YM_{D}/String$ duality formalized by eq. (\ref{0.5cxd}) of Section 4.)

Next, to appropriately utilize the Ansatz (\ref{0.1}), it is helpful first to
compare its local limit (\ref{2.5bxa}) with the pattern corresponding
to the worldsheet's weight (\ref{0.5c}). Postponing the full analysis till
the next section, here we state the two
$D\geq{3}$ equivalence-relations\footnote{The first of the relations was
implicitly assumed in the context of the heuristic 'regularization'
\cite{LE/MM,ALaw} of the area-functional $A[\tilde{M}]$.}
proved in Appendix C. In fact, these relations present a particular example
of certain general freedom in the definition of a given variety of stringy
theories (describing {\it infinitely thin} flux-tubes) that provide with one
and the same set of the physical
observables. The statement is that {\it both the stringy sum (see eq.
(\ref{0.1bbe})) endowed with the local weight (\ref{2.5bxa}) and the
idealized implementation (formalized by eq. (\ref{2.5}) of Section 4) of the
Gauge String, corresponding to the direct application of the 'bare' weight
(\ref{0.5c}), in $D\geq{3}$ represent one and the same simpler system in
disguise.} Irrespectively of the value of $\sigma_{ph}/\Lambda^{2}>0$, the
latter system is the $m_{0}=0$ reduction
of the good old Nambu-Goto Ansatz associated to eq. (\ref{2.5bb}).

Then, in the regime
(\ref{0.1eea})/(\ref{0.1eec}), the latter equivalence-relations allow to
design the economic prescription for the smearing of the 'bare' Gauge String
weight (\ref{0.5c}). Namely, one is allowed to perform the smearing
regularization\footnote{Owing to the normalization (\ref{0.1bb}) (to be
reinterpreted from the viewpoint of the field-theoretic regularization of the
gauge theory), the latter expansion has to be formulated in terms of the
{\it fat}, rather than infinitely thin, $YM$ vortices of the width which is
$\sim{\Lambda^{-1}}$ provided the condition (\ref{0.1ew}) is fulfilled.} of
the worldsheet's weight {\it after} the selfintersection factor $J[..]$ in
eq. (\ref{0.5c}) is omitted (and neglecting all the irregular topologies
absent in the Nambu-Goto measure). In this way, the pattern (\ref{0.5c}) is
traded for the one of eq. (\ref{0.1}). As for the the physical interpretation
of the latter two theorems, it is transparent:
{\it when the corresponding elementary $YM$ vortices are infinitely thin,
their contact interactions (responsible for the discrepancy with the
Nambu-Goto pattern) are completely unobservable within the associated
$D\geq{3}$ stringy sums}. In particular, the first theorem merely formalizes
the fact that the deviation
(\ref{2.5bxb}) between the $m_{0}=0$ option of the Nambu-Goto weight
(\ref{2.5bb}) and the corresponding pattern (\ref{2.5bxa}) takes place only
for those worldsheet's configurations which are of {\it measure zero} in the
$D\geq{3}$ immersion space ${\mathcal{I}}(M,{\bf R^{D}})$ relevant for both
of the stringy representations.

\subsection{The short-distance expansion of the quasi-local weight.}

For the analysis of the next two sections, we will need the description of the
quasilocal weight (given either by eq. (\ref{0.1}) or by a smeared counterpart
of eq. (\ref{0.5c})) through certain short-distance expansion which starts
with  the leading term corresponding to the previously discussed local
deformation of the weight in question. But before we handle this issue, it is
appropriate to illuminate
the two related points. The first is the physical reason why the
regularization of the $YM_{D}$ theory (\ref{1.1}) {\it mandatory} requires
to work with the {\it smeared} worldsheet's weight assigned to the {\it fat}
flux-tube. In consequence, the second question arises: what is the precise
interpretation of the weight (\ref{0.5c}), formally associated to the
infinitely thin $YM$ vortex, in the context of the stringy represetation
of the regularized gauge theory (\ref{1.1}).

For this purpose, we begin with the observation that the choice of
the flux-tube profile (defined, in the case of the bilocal pattern
(\ref{0.1}), through the kernal ${\mathcal{G}}({\bf z}^{2})$) can be
reinterpreted as the particular prescription for the regularization of the
worldsheet's weight. The subtlety is that {\it not} every regularization
of the weight can be mapped onto the corresponding regularization of the dual
gauge theory (\ref{1.1}). The point is that, within the Loop equation
(\ref{1.11b}), the kernal ${\mathcal{G}}({\bf z}^{2})$ satisfies the
constraint\footnote{On the field-theoretic side, the meaning of the latter
normalization is transparent. Taking the $N=1$ case
(\ref{1.1bb}) as the simplest example, one immediately concludes: the
considered condition (\ref{0.1bb}) ensures that the residue, of the
$1/(p_{\mu})^{2}$-pole in the {\it regularized} gauge field propagator
(defined within the framework of the standard weak-coupling expantion), is
conventionally fixed to be unity modulo the standard prefactor.}
(\ref{0.1bb}) which is {\it not}
necessary from the viewpoint of the stringy representation alone.
In particular, presuming that $\xi$ meets the condition (\ref{0.1ew}), the
$D\geq{3}$ analytical continuation (\ref{0.3b}) {\it disagrees}
with the normalization (\ref{0.1bb}) of the $(D-2)$th moment of
${\mathcal{G}}({\bf z}^{2})$. In turn, it matches with the fact that the
Nambu-Goto pattern (\ref{2.5bb}) by itself is in conflict (see e.g.
\cite{MigdRep,Polyak2}) with the definition of Stoks functional \cite{MigdRep}
which makes the action of the Loop operator $\hat{\mathcal{L}}_{\nu}$
ill-defined.

Taking into account the above discussion, one may expect (and this will be
confirmed in Section 4) that the weight (\ref{0.5c}) refers to the formal
implementation of the $1/N$ $SC$ series in the 'bare' $YM_{D}$ system
(\ref{1.1}) prior to the full-fledged $UV$ regularization of the gauge
theory. This point can be most transparently justified at the example of the
$U(1)$ gauge theory (\ref{1.1bb}). When the smearing
function (\ref{0.1bb}) approaches the original
$\delta_{D}({\bf x}-{\bf y})$-function (that can be translated into the
infinite thinning of the $YM$ vortices), the
corresponding $N=1$ option of the weight (\ref{0.1}) evidently becomes
divergent. The point is that the {\it leading divergency} is evidently
associated to the ($N=1$ variant of the) local limit (\ref{2.5bxa}) of the
quasi-local pattern (\ref{0.1}) so that in eq. (\ref{2.5bxa})
one is to identify
\be
\breve{\Lambda}^{D-2}=\xi\Lambda^{D-2}~\longrightarrow{~\infty}~.
\label{0.1ex}
\ee
Similarly, presuming the identification (\ref{0.1ex}),
the weight (\ref{0.5c}) is supposed to refer to the
'leading divergency' (of some quasi-local weight $w_{r}[\tilde{M}]$)
exhibited in the considered local limit when the 'bare' $YM_{D}$ theory is
formally recaptured.

On the other hand, the 'bare' weight (given, e.g., by eq. (\ref{2.5bxa}) or
by the $\breve{\Lambda}^{D-2}=\xi\Lambda^{D-2}$ option of eq. (\ref{0.5c}))
can be properly utilized for the desription of the fat $YM$-vortex as well.
Indeed, it can be viewed as the 'low-energy' limit (taking place when the
characteristic worldsheet's curvature satisfies the constraint (\ref{0.1eeb}))
of the corresponding variety of the quasi-local weights assigned to the
flux-tubes of nonzero width. More generally, the 'bare' weight can be viewed
as the leading term in the 'short-distance' expansion of the
{\it quasi-local} interaction between the surface elements (\ref{0.2c}) in
the area-functionals akin to the one of eq. (\ref{1.9adc}).
To be more specific, we consider the latter expansion on the example
of the bilocal exponent of the weight (\ref{0.1}) that reduces to
the pattern (\ref{2.5bxa}) in the 'brute force' local limit (\ref{0.3b}).
The expansion runs (see e.g. \cite{PolyakLH,PolyakCS},
\cite{Orland,AES} and Appendix A) in terms of the scalar operators which,
roughly speaking, refer to an effective theory of the two-dimensional gravity
defined on the worldsheet $\tilde{M}$ as on the base-space.

The simplest situation arises in the case of the worldsheets
$\tilde{M}$ {\it without}
selfintersections\footnote{The lines/points of selfintersections of
$\tilde{M}$ are assigned with some extra factors. Also, in addition to the
bulk contribution (\ref{0.3bbq}), the expansion of the exponent of eq.
(\ref{0.1}) includes the boundary contribution which, in the leading order, is
reduced to the perimeter-term (\ref{2.5bxl}) generalizing the one in eq.
(\ref{2.5bb}).} so that the formal series start with the Nambu-Goto
pattern (\ref{2.5bb}). In particular, the contribution, associated to the
interior of $\tilde{M}$, can be symbolically written in the form
\be
\frac{\lambda\Lambda^{2}}{2}\sum_{p\geq{0}}~K_{2p}[{\mathcal{G}}]
\sum_{k=1}^{l(2p)}~\frac{H^{(k)}_{2p}}{\Lambda^{2p}}~
\int_{\tilde{M}} d^{2}\gamma~{\mathcal{Q}}^{(k)}_{2p}(\gamma)~,
\label{0.3bbq}
\ee
where the operators\footnote{The operators ${\mathcal{Q}}^{(k)}_{p}$
are characterized by their canonical dimension $[{\mathcal{Q}}^{(k)}_{p}]=p$
which is evaluated postulating that $[x_{\mu}]=[\gamma]=-1$.
The extra label $k$ is introduced to distinguish, for $p\geq{1}$, between
$l(p)\geq{1}$ different operators of the same dimension $p$.}
${\mathcal{Q}}^{(k)}_{2p}(\gamma)$ are composed of the
{\it local} tensors of the (intrinsic and extrinsic) curvature and torsion
(and their derivatives) associated to a given point $\gamma$ of $\tilde{M}$.
As for $H^{(k)}_{2p}$, it is some $\tilde{M}$- and ${\mathcal{G}}$-independent
numerical constant so that $l(0)=1$ and $H^{(1)}_{0}=1$. Finally, we have to
pay for the {\it microscopic} nature of the (fat) $YM$ vortex by the explicit
{\it regularization-dependence} of its  {\it bare} 'effective action'
(\ref{0.3bbq}). Namely, the overall coefficient, in front of the operators of
a given dimension $2p$, depends on the transverse profile of the flux-tube
through the $2p$th moment 
\be
K_{n}[{\mathcal{G}}]=
\int_{{\mathcal{P}}} d^{2}z~({\bf z}^{2})^{\frac{n}{2}}~
{\mathcal{G}}({\bf z}^{2})~~~~~,~~~~~n\geq{0}~,
\label{0.3bbx}          
\ee
of the smearing function (\ref{0.1bb}) (and the integration in eq.
(\ref{0.3bbx}) runs over the infinite $2d$ plane ${\mathcal{P}}$, with
${\bf z}^{2}$ being already measured in {\it dimensionless} units.)

Returning to the substitution (\ref{0.3b}), for $\xi\sim{1}$ it becomes exact
only when the smearing function (\ref{0.1bb}) is such that
\be
K_{0}=\xi~~~~~~;~~~~~~K_{n}\longrightarrow{0}~~~~~if~~~~~n\geq{1}~.
\label{0.3cb}
\ee
Similarly to the prescription of \cite{Dub3}, it corresponds
to the infinitely small flux-tube width $<{\bf r}^{2}>$ which is to be 
estimated as
\be
<{\bf r}^{2}>~\sim{~\Lambda^{-1}\cdot
supr\left(\sqrt[n]{<({\bf z}^{2})^{\frac{n}{2}}>}~\right)}
~~~~~~;~~~~~~<({\bf z}^{2})^{\frac{n}{2}}>=K_{n}/K_{0}~,
\label{0.3csb}
\ee
where $supr(..)$ denotes the supremum with respect to $n\geq{1}$. Then, the
closest to eq. (\ref{0.3cb}) $D\geq{3}$ smearing, consistent with the
normalization condition (\ref{0.1bb}), reads
\be
K_{0}=\xi~~~,~~~K_{D-2}=\frac{2V_{2}}{D\cdot V_{D}}~~~~~~~;~~~~~~~
K_{n}\longrightarrow{0}~~,~~
n\neq{0,~(D-2)}~,
\label{2.5bxd}
\ee
where $V_{D}$ is the volume (e.g. $V_{2}=\pi$) of the $D$-dimensional ball
possesing the unit radius. Therefore, when $D$ is even, the prescription
(\ref{2.5bxd})
turns to zero the coefficients (\ref{0.3bbq}) in front of all the operators 
${\mathcal{Q}}^{(k)}_{2p}$ except for the ones of the dimension $2p=0$ (i.e.
the area term $A[\tilde{M}]$) and of the dimension $2p=(D-2)$. Complementary,
it implies that, when the dimension $D$ is odd, the additional
(to $A[\tilde{M}]$) unsuppressed operators are associated to the
boundary of the surface $\tilde{M}$. Consequently,
once the conditions (\ref{0.1eea}) are valid for such $D$, in the interior of
$\tilde{M}$ the Gauge String weight can be traded (after the identification
(\ref{0.1ex})) for the sheer $m_{0}=0$ option of the Nambu-Goto pattern
(\ref{2.5bb}) with $\bar{\lambda}(\lambda)=\xi\lambda$.

As for $D=4$, apart from the full-derivative term localized at the boundary,
the only nontrivial operator of dimension $(D-2)=2$ yields the well-known
extrinsic curvature term so that the prescription (\ref{2.5bxd}) converts the
exponent of eq. (\ref{0.1}) into the pattern reminiscent, modulo certain
boundary terms,  of the so-called
{\it rigid} string \cite{ExtrCurv/P,ExtrCurv/K}. (As it is clear from
Appendix A, the parameter $\xi$ in eq. (\ref{2.5bxd}) is
constrained to be larger than certain critical value when $D=4$ case.) The
important difference with the latter proposal is that the required
$\Lambda^{2}$-scaling (\ref{0.1eec}) of $\sigma_{ph}$ entails the dynamical
regime distinct from the conventional regime presumed in \cite{ExtrCurv/P}. To
say the least, eq.
(\ref{0.1eec}) can be shown to imply the {\it absence} of the logarithmic flow
of the running coupling constant in front of the extrinsic curvature term.

\subsection{Existence of the metastable stringy phase in the
$\lambda\rightarrow{0}$ limit.}

So far, we have never explicitly used that the coupling constant (\ref{1.1bx})
is sufficiently large. The necessary condition, constraining $\lambda$, stems
from the natural requirement that the physical string tension
$\sigma_{ph}\sim{\Lambda^{2}}$ (given, in the semiclassical approximation, by
eq. (\ref{0.1eec})) is {\it positive}. When the condition (\ref{0.1ew}) on
the auxiliary parameter $\xi$ (entering the bare string tension $\sigma_{0}$)
is fulfilled, the constraint $\sigma_{ph}>0$ indeed implies that $\lambda$ can
not approach zero. On the other hand, the freedom in the choice of the
$YM_{D}$ regularization can be employed to formally continue the considered
stringy sum, endowed with the weight
(\ref{1.9add}), into the weak-coupling regime $\lambda\rightarrow{0}$.
To explain this continuation understand its status in the simplest setting,
we restrict our attention to the regime (\ref{0.1eea}) presuming that the
weight (\ref{0.1}) is regularized according to the prescription
(\ref{2.5bxd}).

The key-observation is that the ambiguity, in the choice of $\xi$,
allows to keep $\bar{\lambda}=\xi\lambda$ sufficiently large even when
$\lambda\rightarrow{+0}$ at the expense of the scaling\footnote{Furthermore,
owing to eqs. (\ref{0.3bbq}) and (\ref{2.5bxd}), in this case the quasilocal
weight (\ref{0.1}) approaches (in the interior of $\tilde{M}$) 
the sheer Nambu-Goto pattern (\ref{2.5bb}) with $\bar{\lambda}(\lambda)=
\xi\lambda$.} $\xi\sim{1/\lambda}\rightarrow{\infty}$. Taking into account
the estimate (\ref{0.7bcx}) of the entropy constant $\xi_{D}$,
the latter observation can be employed to ensure that the physical string
tension (\ref{0.1eec}) (remaining positive even when
$\lambda\rightarrow{0}$) still complies with the
$\Lambda^{2}$-scaling (\ref{0.1eec}) that is mandatory for the consistency of
the solution (\ref{0.1}) with the full Loop equation (\ref{1.11b}).
Unfortunately, being in the apparent conflict with the implications of the
standard weak-coupling perturbative analysis (valid for $D=3,4$), thus constructed
stringy representation provides with the {\it metastable} weak-coupling
solution at least in $D=3$ and $D=4$. In turn, it matches with the
conjecture of \cite{Dub3} that certain large $N$ phase
transition\footnote{The precise values of both $\lambda_{cr}(D)$ and
$\zeta_{D}$ depend on the details of the regularization.} at
$\lambda=\lambda_{cr}(D)\geq{\bar{\lambda}^{-1}(2\zeta_{D})}$
may exclude the neighborhood of $\bar{\lambda}(\lambda)
\rightarrow{2\zeta_{D}}$
(where eq. (\ref{0.1eec}) is violated) from the domain of validity of the
considered $YM_{D}/String$ duality. Let us also remark that, in the abelian
case as well, one may expect that certain finite $N$ phase transition
(or crossover) at some $g^{2}=g^{2}_{cr}$ makes the $N=1$ strong-coupling
expansion (\ref{0.1})/(\ref{0.1bbe}) the {\it unfaithful} representation of
the $U(1)$ gauge theory (\ref{1.1bb}).

\section{Gauge vs. Nambu-Goto string.}

In the previous section, we have shown that the bridge, between the
stringy sums based respectively on the weight (\ref{0.1}) and on the
smeared counterpart of the weight (\ref{0.5c}), is formed by the two
equivalence-theorems. The theorems state that the deformations of the
latter sums, corresponding to the 'brute force' thinning of the flux-tube's
width to zero, are both equivalent to the Nambu-Goto theory provided the
appropriate identification of the relevant coupling constants.
The aim of the present Section is to provide with the necessary details
on the second of the above stringy sums. The central role in our discussion
is played by the concept of the $YM_{D}/String$ duality \cite{Dub3}
formalized by the identity (\ref{0.5cxd}) which can be
used as the {\it generating function} for the 'bare' Gauge String's weight
(\ref{0.5c}) associated to the infinitely thin $YM$ vortex. Complementary,
the duality-relation (\ref{0.5cxd}) will be employed in Section 6 in order
to deduce the $SC$ asymptotics (\ref{0.5bxx}) implementing the dimensional
reduction $YM_{D}\rightarrow{YM_{2}}$.

\subsection{A formal definition of the generic stringy Ansatz.}

To begin with, taking as an example the stringy sum endowed with the weight
(\ref{2.5bb}) or (\ref{1.9add})/(\ref{0.1}), let us introduce a more precise
and abstract formulation of the latter systems which is presumably applicable
to the generic stringy Ansatz. As for the sum associated to the weight
(\ref{1.9add})/(\ref{0.1}), it can be
symbolically written in the following form
\be
N^{b}<\prod_{k=1}^{b}W_{C_{k}}>=\int\limits^{\partial \bar{\vartheta}=
\cup_{k}C_{k}}_{\bar{\vartheta}
\in{{\mathcal{I}}(M,{\bf R^{D}})}}{\mathcal{D}}\bar{\vartheta}~
w_{2}[\tilde{M}(\{C_{k}\})]~\Bigg|_{\tilde{M}=\bar{\vartheta}(M)}~.
\label{0.1bbe}
\ee
As the average (\ref{0.1bbe}) is {\it not} irreducible for
$b\geq{2}$, one is to sum over the surfaces $\tilde{M}(\{C_{k}\})$
possessing $1\leq{p}\leq{b}$ {\it open} connected components with the
boundaries associated to an arbitrary (ordered) partition $\{b_{j}\}$ of the
set of the $b=\sum_{j=1}^{p}b_{j}$ contours $C_{k}$  into {\it nonempty}
subsets. As for the functional measure $\sum_{\tilde{M}}\rightarrow
{\int {\mathcal{D}}\bar{\vartheta}}$, it refers to the standard representation
\cite{StabMap} of a given worldsheet $\tilde{M}(\{C_{k}\})\equiv{\tilde{M}}$
as the {\it image}\footnote{The somewhat
unconventional description (\ref{0.4}) of the measure is helpful to make
contact with the formalisms of \cite{Gr&Tayl,Dub3}. Not less important, it
will allow for the economic implementation of the backtracking invariance of
the average $<W_{C}>$ briefly sketched in the end of
this section.} $\tilde{M}=\vartheta(M)$ of the corresponding map
\be
\vartheta~:~~~~~~~M(\{\breve{C}_{k}\})~\longrightarrow{~\bf R^{D}}
~~~~~~~~;~~~~~~~~\vartheta(\breve{C}_{k})=C_{k}~,
\label{0.4}
\ee
of an {\it oriented} (not necessarily connected) $2d$ manifold
$M(\{\breve{C}_{k}\})\equiv{M}$, with a given number $b$ of the
boundaries $\breve{C}_{k}$ (or without them at all), into the Euclidean
base-space ${\bf R^{D}}$. As a result, {\it prior} to the regularization, the
generic measure is determined by the specification of the space
${\mathcal{X}}$ of the admissible mappings (\ref{0.4}) (i.e. of the admissible
topologies of $\tilde{M}$) while the area $A[\tilde{M}]$ of $\tilde{M}$ is
postulated to be averaged over.
(As before, the frequences of the transverse fluctuations
of the worldsheet are to be somehow regularized at the $UV$ scale $\Lambda$
that trades ${\mathcal{D}}\bar{\vartheta}$ for its regularized counterpart
${\mathcal{D}}_{r}\bar{\vartheta}$. To simplify the notations the subscript
$r$ will be skipped.)

Next, returning to the case of the Ansatz (\ref{2.5bb}) or
(\ref{1.9add})/(\ref{0.1}), let us temporarily presume that the contours
$C_{k}$ are {\it without} backtrackings. Then, once the conditions
(\ref{0.1eea}) are satisfied, the relevant set ${\mathcal{X}}$ of the maps
(\ref{0.4}) can be consistently reduced to the space
${\mathcal{I}}(M,{\bf R^{D}})$ of the {\it immersions} $\bar{\vartheta}$ of
the manifold $M$ into ${\bf R^{D}}$. The latter type of the
mappings is characterized by the fact that any worldsheet
$\tilde{M}\in{{\mathcal{I}}(M,{\bf R^{D}})}$ locally looks like a domain of
the 2-dimensional Euclidean space ${\bf R^{2}}$. Therefore,
having fixed a partition $\{b_{j};~j=1,...,p\}$, a given topology of an open
surface $\tilde{M}\in{{\mathcal{I}}(M,{\bf R^{D}})}$ in eq. (\ref{0.1bbe}) is
specified  by the $p$ positive integers $h_{j}$ (with $h=\sum_{j=1}^{p} h_{j}$)
standing for the number of handles of $j$th open connected component of the
worldsheet $\tilde{M}_{\chi}$. In turn, the total Euler character $\chi$ of
$\tilde{M}_{\chi}$ (or, equally, of $M_{\chi}$) is related to the set of the
quantum numbers $\{h_{j},~b_{j}\}$ through the standard identity
\be
\chi=\left(\sum_{j=1}^{p}2-2h_{j}-b_{j}\right)=2p-2h-b~~~~~~~~;~~~~~~~~
b=\sum_{j=1}^{p}b_{j}~~~,~~~1\leq{p}\leq{b}~,
\label{0.9zzk}
\ee
which ensures that $\chi=\sum_{j=1}^{p} \chi_{j}$ is the sum of the Euler
characters $\chi_{j}$ associated to the connected components of
$\tilde{M}_{\chi}$.

Finally, to obtain the conventional
representation \cite{GFS} of the stringy sum (\ref{0.1bbe})/(\ref{2.5bb}),
one is to represent both the action (\ref{2.5bb}) and the measure (i.e. the
space ${\mathcal{I}}(M,{\bf R^{D}})$) via the {\it equivalence class} of the
functions
\be
\{~x_{\mu}(\gamma_{1},\gamma_{2})~~|~~
x_{\mu}(\gamma_{1},\gamma_{2})
\cong{x_{\mu}(f_{1}(\gamma),f_{2}(\gamma))}~\}~,
\label{0.4zh}
\ee
where the identifications are performed with respect to the group of the
area-preserving
diffeomorphisms induced by the reparametrization functions
$f_{a}(\gamma)$. Actually, to faithfully reproduce the immersion-space
${\mathcal{I}}(M,{\bf R^{D}})$, the smooth functions in eq. (\ref{0.4zh})
should satisfy the two complementary conditions. Namely, 
for all admissible $\gamma$, the {\it rank} $r(\gamma)$ of the
$D\times 2$ matrix $\partial x_{\mu}(\gamma)/\partial \gamma_{a}$ is
equal to {\it two},  while the Jacobian
\be
{\mathcal{J}}(\gamma)=
\frac{\partial(f_{1}(\gamma),f_{2}(\gamma))}
{\partial(\gamma_{1},\gamma_{2})}
=det\Bigg[{
\frac{\partial f_{a}(\gamma)}
{\partial \gamma_{b}}}\Bigg]>0
\label{0.4zhh}
\ee
of any reparametrization (\ref{0.4zh}) is constrained to be {\it positive}.

\subsection{The basics of the $YM_{D}/String$ duality.}

\subsubsection{The idealized Gauge String sum in the limit of the zero vortex
width.}

Before we introduce the basic duality-relation (\ref{0.5cxd}),
it is appropriate to present the precise formulation of the idealized
Gauge String sum, endowed with the $\breve{\lambda}\breve{\Lambda}^{2}=
\bar{\lambda}{\Lambda}^{2}$ option (resulting after the identification
(\ref{0.1ex})) of the 'bare' weight (\ref{0.5c}). The representation in
question, arising when the formal limit\footnote{The dimensional reduction
(\ref{0.5bxx}), that takes place in the extreme $SC$ regime when the limits
$\lambda,({\mathcal{R}}(s)\Lambda)\rightarrow{\infty}$ are performed
{\it before} any other limits, can be formally deduced retaining from eq.
(\ref{2.5}) only the saddle-point contribution associated to the conglomerate
of the minimal-area worldsheets.} 
\be
<{\bf r}^{2}>~\longrightarrow{~0}
\label{2.5zxz}
\ee
of the zero flux-tube's width (\ref{0.3csb}) is performed {\it prior} to any
other relevant limit, reads
\be
N^{b}<\prod_{k=1}^{b}W_{C_{k}}>=\sum_{\chi}N^{\chi}
\int\limits^{\partial \vartheta_{\chi}=\cup_{k}C_{k}}_{
\vartheta_{\chi}\in{{\Delta}_{\chi}(M,{\bf R^{D}})}}
{\mathcal{D}}\vartheta_{\chi}~J[\vartheta_{\chi}|\bar{\lambda}]~
exp\left({-\frac{\bar{\lambda}\Lambda^{2}}{2}A[\vartheta_{\chi}]}\right),
\label{2.5}
\ee
where the worldsheets $\tilde{M}_{\chi}=\vartheta_{\chi}(M)
\equiv{\vartheta_{\chi}}$, possessing a given total Euler character $\chi$,
result from the mappings (\ref{0.4}) comprised into certain space
$\Delta(M,{\bf R^{D}})=\cup_{\chi}~\Delta_{\chi}(M,{\bf R^{D}})$.
(The functional integral $\int {\mathcal{D}}\vartheta_{\chi}=
\sum_{\{h_{j},b_{j}\}} \int {\mathcal{D}}\vartheta_{\{h_{j},b_{j}\}}$ implies
the discrete sum over the relevant set $\{h_{j},~b_{j}\}$, constrained by eq.
(\ref{0.9zzk}), which enumerates different topological sectors.)
Without loss of the physical data, all the contours $C_{k}$ are presumed
to be devoid of backtrackings; the prescription to implement the backtracking
invariance of $<W_{C}>$ is discussed in Appendix E. Also, as previously,
an effective {\it UV} cut off $\Lambda$ should be somehow
imposed on the frequences of the transverse fluctuations of the worldsheets
$\tilde{M}_{\chi}$.

According to Appendix C, the stringy sum (\ref{2.5}) is
equivalent\footnote{The advantage of the representation (\ref{2.5}) is that, contrary to the
other two reformulations, it allows to make contact with the $1/N$
{\it weak-coupling} expansion in the $YM_{D}$ theory (\ref{1.1}).}, in
$D\geq{3}$, to the $m_{0}=0,~\bar{\lambda}(\lambda)=\xi\lambda$ option of
the Nambu-Goto Ansatz (\ref{0.1bbe})/(\ref{2.5bb}) which,
in turn. is indistinguishable from the local deformation
(\ref{0.1bbe})/(\ref{2.5bxa}) of the system (\ref{0.1bbe})/(\ref{0.1}).
Let us briefly explain why it is the case. To begin with, the conventional
immersion space
${\mathcal{I}}(M,{\bf R^{D}})$ (entering the sum (\ref{0.1bbe})) can be viewed
as the {\it dense} subspace of the $\Delta(M,{\bf R^{D}})$-space relevant for
eq. (\ref{2.5}). Furthermore, the selfintersection-factor is considerably
reduced when the worldsheet $\tilde{M}$ belongs to the latter
immersion-subspace:
\be
J[\vartheta(M)|\bar{\lambda}]=\frac{1}{|{\mathcal{C}}_{\vartheta}|}~~~~~~~
if~~~~~~~\tilde{M}=\vartheta(M)~\in{~{\mathcal{I}}(M,{\bf R^{D}})}~,
\label{0.5ccx}
\ee
where $|{\mathcal{C}}_{\vartheta}|$ is the number
of discrete transformations leaving $\tilde{M}$ invariant, i.e. the number
of the topologically inequivalent automorphisms $\kappa$ of the
immersion-map (\ref{0.4}) which represents $\tilde{M}$:
$\vartheta\circ \kappa=\vartheta$. Then, the equivalence-theorem (stated in
the begining of this section) merely reflects the fact that, for $D\geq{3}$,
there is a dense subspace of ${\mathcal{I}}(M,{\bf R^{D}})$ where
$|{\mathcal{C}}_{\vartheta}|=1$. Summarizing, in $D\geq{3}$, the unobservable
difference between the $N\geq{2}$ stringy sums (\ref{2.5}) and
(\ref{0.1bbe})/(\ref{2.5bxa}) refers exclusively to the distinction in the
contact interactions between the elementary $YM$ vortices of zero width.

\subsubsection{The $SC/WC$ correspondence.}

Now we are in a position to formulate the duality-relation which allows to
make contact between the representation (\ref{2.5}) and the $1/N$
weak-coupling series in the associated gauge theory (\ref{1.1}). Also, in
Section 6, this relation will be crucial when we deduce
the $SC$ asymptotics (\ref{0.5bxx}) of $<W_{C}>$.

For this purpose, let us consider the $T$-restriction
$\Delta_{T}(M,{\bf R^{D}})$ of $\Delta(M,{\bf R^{D}})$ that consists of all
the maps $\vartheta_{T}$ which
result in the (piecewise) smooth worldsheets $\tilde{M}(\{C_{k}\})$ with the
support ${Sup[\vartheta_{T}]}$ (in ${\bf R^{D}}$) belonging to a given $2d$
cell-complex\footnote{Due to possible
selfintersections of $\tilde{M}(\{C_{k}\})$, the support $T(\{C_{k}\})$ may
have topology of the $2d$-skeleton of a $D$-dimensional lattice (embedded
into ${\bf R^{D}}$) rather than simply of a $2d$ manifold. (Also, owing to the
irrelevance \cite{Dub3} of the worldsheet backtrackings, $T(\{C_{k}\})$ is
restricted to contain {\it no} backtracking $2d$ submanifolds.)
On any given $T(\{C_{k}\})$, one can
choose the graph ${\mathcal{T}}_{T}$ where more than two outgoing $2d$
submanifolds of $T(\{C_{k}\})$ merge. Cutting along ${\mathcal{T}}_{T}$,
we make $T(\{C_{k}\})$ into a disjoint union of the elementary domains
$T_{p}(\{C_{k}\})$ (entering e.g. eq. (\ref{2.5bxa})) which possess topology
of $2d$ manifolds with a boundary. By
construction of \cite{Dub3}, the support of a particular worldsheet
$\vartheta_{T}$ must be a nonempty union of some of the domains
$T_{p}(\{C_{k}\})$.} $T(\{C_{k}\})$ so that
$\partial T(\{C_{k}\})\in{\cup_{k}C_{k}}$. Also, let
$w[\vartheta(M)|\bar{\lambda}]$ denote the modification of the weight
(\ref{0.5c}) arising after the identification (\ref{0.1ex}). Then, the central
implication of
the $YM_{D}/String$ duality is that the corresponding $T$-restriction of the
idealized $D\geq{3}$ Gauge String sum (\ref{2.5}) represents the $1/N$
$SC$ expansion of the (multi)loop average
\be
N^{b}<\prod_{k}W_{C_{k}}>\Bigg|_{YM_{2}(T)}=
\int\limits_{\vartheta_{T}\in{\Delta_{T}(M,{\bf R^{D}})}}
d\vartheta_{T}~w[\vartheta_{T}|\bar{\lambda}]~+~O(e^{-|a|\cdot N})~,
\label{0.5cxd}
\ee
evaluated in the {\it two}- (rather than $D$-) dimensional $YM_{2}$ system
(\ref{1.1})/(\ref{0.4bc}) defined on the curved taget-space
$T(\{C_{k}\})\equiv{T}$ as on the base-space. (In the r.h. side of eq.
(\ref{0.5cxd}), the exponentially suppressed residual term refers to the
contribution which is essentially 'nonperturbative' with respect to the
considered $1/N$ $SC$ series. For example, in the $SU(N)$ case, the latter
contribution includes the effects associated to the $SU(N)$ string-junctions.)

By construction, the relation (\ref{0.5cxd}) can be used as the generating function
which formalizes the algorithm \cite{Dub3} to determine, for a generic
nonabelian group, the pattern both of the relevant space
$\Delta(M,{\bf R^{D}})$ of the mappings (\ref{0.4}) and of the explicit form
of the bare worldsheet's weight $w[\vartheta(M)]$. On the other hand, eq.
(\ref{0.5cxd}) maintains the required correspondence between the
{\it two} distict $1/N$ expansions in the $YM_{D}$ theory (\ref{1.1}). To
explain the point, one is observe first the weaker variant of this
correspondence between the amplitudes, associated to the conglomerates of the Feynman
diagrams of the $1/N$ {\it WC} series (written for the $YM_{2}$ averages in
the l.h. side of eq. (\ref{0.5cxd})), and the amplitudes constituted by the
$1/N$ $SC$ series organized (in the r.h. side of eq.
(\ref{0.5cxd})) as the string-like representation. The final step is to
notice that the latter {\it WC} conglomerates can be reinterpreted as certain
subclasses of the Feynman diagrams in the {\it bare} $D$-dimensional $YM_{D}$
theory (\ref{1.1}) {\it prior} to the $UV$ regularization of the latter.
More precisely, one is to rewrite first the perturbative
propagators as the path integrals over the gluonic tragectories
assigned with the colour indices.
(To circumvent gauge-fixing, tricky to explicitly match between the
$T$-restriction's and standard formulations, one is to introduce an
infinitesimally small mass term for the gauge field and then perform the
considered comparison.) For a given support $T$, the relevant subclass of the
diagrams includes those fishnet configurations which comply with the twofold
constraint. The gluonic trajectories should be localized to $T$.  while
{\it each} colour $a$-component of the associated gluonic
field-strength tensor  $F^{a}_{\mu\nu}({\bf z})$ (for any given
${\bf z}\in{\it T}$) as the Euclidean $O(D)$-tensor should belong to the
tangent space of $T$ at ${\bf z}$.

In particular, the $\xi\Lambda^{D-2}=\breve{\Lambda}^{D-2}$ implementation of
the relation (\ref{0.4bc}), between the $D$- and $2$-dimensional coupling
constants, emerges in the following way. The considered $WC/SC$ resummation
implies that, to each factor $g^{2}_{YM_{D}}$ of the amplitude associated to
any given $WC$ Feynman diagram, one is to
assign the auxiliary factor $\breve{\Lambda}^{D-2}$ which is to account 
for the {\it different} dimensionality of the microscopic excitations
entering respectively the $WC$ and $SC$ series in the $YM_{D}$ system
(\ref{1.1}). Indeed, $(D-2)$ is the
dimensionality of the subspace orthogonal to the tangent space associated to
the support $T$ (spanned by the selected dual variety of the gluonic
trajectories) of the corresponding conglomerate of the $2d$ flux-worldsheets.
Summarizing, thus introduced {\it SC/WC}
correspondence yields the refinement of the heuristic 't Hooft arguments
\cite{'t Hooft1} (see also \cite{LE/MM,MigdRep}) concerning possible relation
of the topological $1/N$ resummation (of the $WC$ expansion) to the existence
of the string-like representation for the matrix theories like (\ref{1.1}).

The subtle point is that, as the $WC$
series refer to the $YM_{D}$ theory before the $UV$ regularization, the
(apparently arbitrary) auxiliary parameter $\breve{\Lambda}$ can be initially
interpreted in the sense of the formal limiting procedure (\ref{0.1ex}).
Complementary, the representation (\ref{2.5}) is to viewed as the
{\it analytical continuation} (\ref{2.5zxz}) of the stringy sum based on
certain quasi-local smearing $w_{r}[\tilde{M}]$ of the weight (\ref{0.5c}).
This continuation is supposed to be performed extending the procedure which
has been used to show that, in the local limit (\ref{0.3b}) of the zero
flux-tube's width, the Ansatz
(\ref{0.1bbe})/(\ref{0.1}) is reduced to the system
(\ref{0.1bbe})/(\ref{2.5bxa}) describing the infinitely thin vortices.
Actually, to deduce the smearing (consistent with the regularization (\ref{0.1bb}) of
the reduced Loop equation (\ref{0.9za})) of the worldsheet weight
(\ref{0.5c}), one is to trade the average in the l.h. side of eq.
(\ref{0.5cxd}) for its 'regularized' counterpart. The simple example of this
prescription, implementing the proper smearing of the pattern (\ref{0.3csd})
below, has been introduced in eq. (\ref{2.12bl}).
In particular, it provides with the complementary explanation of the reason
why the $2d$ coupling constant $g_{YM_{2}}$ (entering, e.g., the $SC$
asymptotics (\ref{0.5bxx}) of $<W_{C}>$) is related to the original
$D$-dimensional one $g_{YM_{D}}$ via the rescaling (\ref{0.4bc}).

\subsubsection{The $YM$ vortices as seen within the $1/N$ $SC$ expansion.}

At this step, it is appropriate to comment on the precise relation between
the sum over the worldsheets $\tilde{M}$ in the r.h. side of eq.
(\ref{0.5cxd}) and the superposition of the configurations visualized as
(infinitely thin) flux-tubes. To identify the latter configurations, we
recall first the following symbolic representation
\be
Z(\{C_{k}\})<\prod_{k}W_{C_{k}}>\Bigg|_{YM_{2}(T)}=\sum_{\{R_{q}\}}
F(\{R_{q}\})~
exp\left(-\frac{\bar{\lambda}\Lambda^{2}}{2}
\sum_{q}~C_{2}(R_{q})~\bar{A}_{q}\right)
\label{2.5bxh}
\ee
corresponding to the standard {\it character expansion} \cite{Dr&Zub} in the
Heat-Kernal lattice gauge system (on $T(\{C_{k}\})$) that, in \cite{Dub3}, is
shown to reproduce the average in the l.h. side of eq. (\ref{0.5cxd}).
In eq. (\ref{2.5bxh}), $F(\{R_{q}\})$ is some $\{\bar{A}_{q}\}$-independent
function, while  $Z(\{C_{k}\})|_{YM_{2}(T)}$ is the partition
function\footnote{This partition function is to be evaluated
with the {\it free} boundary conditions for the $YM_{2}(T)$ gauge fields on
the boundary $\partial T(\{C_{k}\})$ of the base-space
$T\equiv{T(\{C_{k}\})}$.} associated to $<\prod_{k}W_{C_{k}}>|_{YM_{2}(T)}$.
As for $\bar{A}_{q}$, it is the area of the $q$th elementary domain
$T_{q}(\{C_{k}\})$ introduced in the footnote prior to eq. (\ref{0.5cxd}) so
that $\sum_{q}\bar{A}_{q}=\bar{A}$ is the total area of $T(\{C_{k}\})$. Each
domain $T_{q}(\{C_{k}\})$ is assigned with some representation
$R_{q}$ (entering the eigenvalue (\ref{2.12bv}) of the second Casimir
operator $C_{2}(R_{q})$) of the Lie group in
question, and the sum in eq. (\ref{2.5bxh}) runs over all admissible
$\{R_{q}\}$-assignements satisfying the (nonabelian) {\it fusion-rule}
algebra. In particular, a specifically transparent example is provided by
the loop average 
\be
<W^{R}_{C'}>|_{YM_{2}(T)}=(dimR)^{\chi}~
exp\left(-\frac{\bar{\lambda}A[T]}{2N}~C_{2}(R)\right)~,
\label{0.3csd}
\ee
in a given representation $R$ of the $U(N)$ group, and the support
$T$ (with $\partial T=C'$) is constrained to be a curved disc
of the area $A[T]$ and with $h$ handles so that $\chi=1-2h$.

Upon a reflection, a given flux-tube configuration is to be parametrized
(as it is implied in the interpretation of the ratio (\ref{2.5bxb})) by a
single  admissible $\{R_{q}\}$-assignement that refers to the corresponding
term in the r.h. side of eq. (\ref{2.5bxh}). In consequence, once the
l.h. side of eq. (\ref{0.5cxd}) does {\it not} comply with the sheer
Nambu-Goto pattern, the expansion in the r.h. side of eq. (\ref{0.5cxd})
is related to the character/flux-tube expansion (\ref{2.5bxh}) via a
nontrivial resummation. The latter resummation, which {\it serves to
represent the contact interactions between the elementary $YM$ vortices
in stringy terms}, is the clue for the proper understanding of the
unconventional features of the (idealized) Gauge String 
pattern (\ref{2.5}). One observes that, apart from the simplest situations
(which, roughly speaking, refer to surfaces $\tilde{M}$ selfintersecting
{\it at most} on a set of isolated points), there is {\it no} a direct
identification of the weight (\ref{0.5c}) with the amplitude assigned to
a particular flux-tube (of zero width). To say the least, a given $YM$
vortex may contribute to all orders of the $1/N$ series. More precisely,
the duality-relation (\ref{0.5cxd}) implies that, generically, certain
superpositions of the flux-tubes (of regular topology) can be effectively
reproduced through the corresponding stringy sums over the surfaces (including
the ones of singular topology, {\it not} included into the immersion space
${\mathcal{I}}(M,{\bf R^{D}})$) endowed with the weight (\ref{0.5c}).

\subsection{The case of the $D\geq{3}$ $U(1)$ gauge theory with monopoles.}

The so far revealed matching, between the $N\geq{1}$ patterns
(\ref{0.1bbe})/(\ref{2.5bxa}) and (\ref{2.5}), is rooted in the
unobservability (within the corresponding $D\geq{3}$ stringy sums) of the
difference in the associated contact interactions between the infinitely thin
(elementary) flux-tubes. Actually, the matching becomes substantially more
transparent in the $N=1$ case. It may be expected from the fact that,
irrespective of whether the conditions (\ref{0.1eea})/(\ref{0.1eec})) are
satisfied, the $N=1$ option of the Ansatz (\ref{0.1bbe})/(\ref{0.1}) provides
with the solution of the $U(1)$ loop equations for generic Wilson loops.
For a preliminary orientation, observe first that
the l.h. side (\ref{2.5bxh}) of the $U(1)$
relation (\ref{0.5cxd}) is reminiscent of the $N=1$
pattern (\ref{2.5bxa}). (Indeed, the quantity $n^{2}$ is the second
Casimir eigenvalue $C_{2}(R)$ associated to the $U(1)$ 
irreducible representation $R$ labelled by the {\it single} integer $n$, and
$dimR={1}$ for $\forall{n}$.) More precisely, the two corresponding patterns
of the abelian contact interactions are equivalent, modulo certain
minor details irrelevant for the associated $D\geq{3}$ stringy sums.

To be more explicit, it is sufficient to restrict
our attention to the simplest situation when the support
$T(\{C_{k}\})$ (playing the role of the base-space for the $YM_{2}(T)$ theory
inherent in the l.h. side of eq. (\ref{0.5cxd})) does not contain {\it closed}
$2d$ submanifolds, i.e. closed 2-cycles\footnote{In other words,
$T(\{C_{k}\})$ can be embedded into a $2d$ manifold.
This restriction is always fulfilled, e.g., for the minimal area
worldsheets entering the $SC$ asymptotics (\ref{0.5bxx}) of $<W_{C}>$.}.
In this case, the partition function $Z(\{C_{k}\})$ in the l.h. side of
eq. (\ref{2.5bxh}) is equal to unity, while
the sum in the r.h. side of eq. (\ref{2.5bxh}) can be shown to
reduce to a single term for which the fusion-rule function $F(\{R_{q}\})$ is
equal to unity as well. On the other hand, recall that,
in the local limit (\ref{0.3cb}), the worldsheet's weight (\ref{0.1}) assumes
the form of eq. (\ref{2.5bxa}) where one is to reidentify 
$n^{2}=C_{2}(n)$. Furthermore, any admissible
$\{n_{q}\}$-assignement, induced by a particular immersion-map (\ref{0.4}),
satisfies the same fusion-rule constraints that are built into the $U(1)$
character-expansion (\ref{2.5bxh}) with $\{R_{q}\}\equiv{\{n_{q}\}}$.
Altogether, for the considered topology
of the supports $T(\{C_{k}\})$, one concludes that in the abelian case there
exists {\it the one-to-one correspondence} between the l.h side of eq.
(\ref{0.5cxd}) and the local limit (\ref{0.3cb}) of the associated
$T$-restriction of the $N=1$ stringy sum (\ref{0.1bbe})/(\ref{0.1}).

Finally, presuming the validity of the condition (\ref{0.1eeb}), let us
evaluate the leading boundary effects encoded in the weight (\ref{0.1})
for generic smearing function (\ref{0.1bb}). To simplify the consideration,
we concentrate on the simplest option of the $N=1$ Ansatz
(\ref{0.1bbe})/(\ref{0.1}) for a $D=2$ average $<W_{C}>$. For a given
worldsheet $\tilde{M}$ corresponding to a particular $\{n_{q}\}$-assignement,
the support
$C'$ (in ${\bf R^{2}}$) of the contour $C$ can be devided into the segments
${\mathcal{C}}'_{p}$ (of the length $\bar{L}_{p}$) which separate the $2d$
domains of the worldsheet's support $T(C)$ covered respectively $n^{+}_{p}$
and $n^{-}_{p}$ times. Then, a computation yields that the quasi-locality of
the weight (\ref{0.1}) results in the boundary effects which, in the leading
order of the short-distance expansion (akin to the one of eq.
(\ref{0.3bbq})), are described by the term
\be
exp\left(-m_{0}(\lambda)
\sum_{p}~(n^{+}_{p}+n^{-}_{p})^{2}~\bar{L}_{p}\right)~,
\label{2.5bxl}
\ee
where $m_{0}(\lambda)$ is given by eq. (\ref{0.3bbl}) of Appendix A.

One observes that $(n^{+}_{p}+n^{-}_{p})^{2}$ is the second Casimir
eigenvalue\footnote{Remark that it matches with the $C_{2}(R)$-scaling of the
perimeter term in the pattern (\ref{2.12bl}) of the $U(N)$ average
$<W_{C'}^{R}>$ in the extreme $SC$ limit (\ref{0.1eed}).}
of the $U(1)$ representation corresponding to the direct product of the
associated representations labelled by $n^{+}_{p}$ and $n^{-}_{p}$.
On the other hand, in the case of the
sheer Nambu-Goto pattern (\ref{2.5bb}), the exponent of eq. (\ref{2.5bxl})
would be traded for $\sum_{p}|m_{p}|\bar{L}_{p}$, where the integer $|m_{p}|$
counts how many times the contour $C$ covers the $p$th segment
${\mathcal{C}}'_{p}$ of the support $C'$ of $C$.
Therefore, unless $|n^{+}_{p}+n^{-}_{p}|=|m_{p}|$ for $\forall{p}$,
the boundary effects (encoded in the $D=2$ pattern (\ref{0.1})) are sensitive
to the quasi-contact interactions
(responsible also for the deviation of the factor (\ref{2.5bxb}) from unity)
which crucially depend on the choice of both the gauge group and the $YM_{D}$
action. Upon a reflection, it explains why in the regime
(\ref{0.1eea})/(\ref{0.1eec})
the necessary condition, for the Ansatz (\ref{0.1bbe})/(\ref{0.1}) to
describe properly the (leading) boundary effects within the $N\geq{2}$ $1/N$
$SC$ series, requires that the involved loops possess at most point-like
(self)intersections.

\subsection{The backtracking invariance of $<W_{C}>$.}

In conclusion, we clarify the two alternative prescriptions to implement
the invariance of the loop-averages $<W_{C}>$ with respect to zig-zag
backtrackings of the loop $C$. The short-cut prescription is provided by the
following
excessively simple rule (formalized by eq. (\ref{0.4ze}) of Appendix E)
that is predetermined by the basic duality-relation (\ref{0.5cxd}). Namely,
{\it the images $\tilde{M}(C)$ of the $\vartheta$-mappings (\ref{0.4})
(composed into the measure of the stringy representation (\ref{2.5}) of
$<W_{C}>$), being
kept intact in their {\bf interior}, should be coherently deformed on their
boundary $\partial\tilde{M}(C)=C$ so that any particular data of
backtrackings is introduced at a given nonbacktracking
reference-contour $C\in{\Upsilon_{nbt}}$.}

Although the above prescription is fully consistent, it looks somewhat
'external' to the $m_{0}=0$ option of the Nambu-Goto ($NG$) theory
(\ref{0.1bbe})/(\ref{2.5bb}) to
which the idealized Gauge String representation (\ref{2.5}) is equivalent for
nonbacktracking loops. Indeed, being extended from the subspace
$\Upsilon_{nbt}$ of nonbacktracking contours to the entire loop space
$\Upsilon$, the latter $NG$ theory {\it per se} does {\it not} refer to the
considered invariance which is, actually, {\it absent} (see Appendix E) in
the conventional $NG$ framework. Therefore, it is appropriate to discuss
how the backtracking symmetry of the average $<W_{C}>$ can be introduced as
resulting from certain symmetry of the judiciously reformulated Gauge String
action. To explain it in the simplest setting, let the
constraints (\ref{0.1eea})/(\ref{0.1eec}) be
satisfied so that, for contours without backtrackings, the 
Nambu-Goto pattern (\ref{0.1bbe})/(\ref{2.5bb}) correctly describes the
low-energy dynamics of the $D\geq{3}$ $YM_{D}$ system (\ref{1.1}). The
key-observation is that, as we already discussed,
the $m_{0}=0$ option of the $NG$ theory (\ref{0.1bbe})/(\ref{2.5bb}) is
equivalent to the local limit (\ref{0.1bbe})/(\ref{2.5bxa}) of the stringy
Ansatz (\ref{0.1bbe})/(\ref{0.1}) which, in turn, can be viewed as the
specific member of the general confining string's family \cite{PolyakCS}.

As a result, one can take
advantage of the fact that (compared to the Nambu-Goto action of eq.
(\ref{2.5bb})) the action in the exponent of eq. (\ref{0.1}) is invariant
under the {\it larger} group \cite{PolyakCS} of the reparametrizations 
which are now allowed to generate arbitrary worldsheet's foldings. The
central consequence of this extended worldsheet's symmetry is that, being
'projected' onto the group of the reparametrizations of the boundary contour
$C$, it results in the invariance of $<W_{C}>$ with respect to the contous's
foldings.
To elevate this classical symmetry into the full-fledged quantum invariance,
we observe that the l.h. side of the Ansatz (\ref{0.1bbe})/(\ref{0.1})
is {\it invariant} under certain deformation of the measure (in its r.h.
side) paired up with the backtracking invariance of the weight (\ref{0.1}).
As we explain in Appendix E, this deformation merely trades any given
worldsheet $\tilde{M}=\bar{\vartheta}(M)\in{{\mathcal{I}}(M,{\bf R^{D}})}$
for the whole equivalence class (i.e. {\it orbit}) of the surfaces. The
latter orbit is generated through attachments to $\tilde{M}$ all possible
'collapsed' (to bound a {\it zero} 3-volume) baby-universes which are
visualized as the corresponding worldsheet's foldings. In sum, there exists
the twofold modification
of the Nambu-Goto system (\ref{0.1bbe})/(\ref{2.5bb}) (i.e. when the weight
(\ref{2.5bb}) is substituted by the one of eq. (\ref{2.5bxa}), while the
measure is deformed as desribed above) that allows for the 'natural'
implementation of the required backtracking symmetry of $<W_{C}>$.

In culmination, we formulate the general {\it selection-rule} following
from the previous discussion. Consider a generic observable
${\mathcal{O}}_{j}$ in (open or closed string's sector of) the
Nambu-Goto theory (\ref{0.1bbe})/(\ref{2.5bb}) with $m_{0}=0$.
{\it The necessary condition, for such observable ${\mathcal{O}}_{j}$
to be associated to certain observable in the corresponding dual gauge theory
(\ref{1.1}), is that the above twofold modification of the pattern
(\ref{0.1bbe})/(\ref{2.5bb}), being possible, leaves the quantity
${\mathcal{O}}_{j}$ unchanged.} In this perspective, the backtracking
invariance of $<W_{C}>$ {\it together} with the invariance of the properly
introduced Gauge String action under the worldsheet's backtrackings can be
viewed as a sort of 'pseudogauge' symmetry. On the other hand, the stringy
action associated to either eq. (\ref{0.5c}) or eq. (\ref{2.5bb}) can be
viewed as arising {\it after} fixing of this 'pseudogauge' symmetry.

\section{The infrared equivalence with the Nambu-Goto Ansatz.}

It is time to put all the pieces together and
demonstrate that, in the $D\geq{3}$ regime (\ref{0.1eea})/(\ref{0.1eec}),
the $N\geq{1}$ Nambu-Goto Ansatz (\ref{0.1bbe})/(\ref{2.5bb}) appears as the
{\it reference-model~} that 'marks' the universality class which includes the
relevant Gauge String representation of the regularized $U(N)$ loop-averages.
Therefore, given a particular prescription to implement the $UV$ cut off,
the remaining regularization-dependence of the solution $<W_{C}>$ (of the Loop
equation) is localized\footnote{The considered 'localization' is
modulo the product of certain elementary factors
accumulated in the microscopic $1/\Lambda$-vicinity of the point-like
(self)intersections of the involved boundary contour(s).} within the pattern
of the coupling constants $\bar{\lambda}(\lambda)$ and $m_{0}(\lambda)$ of
the associated Nambu-Goto weight (\ref{2.5bb}).

For this purpose, we are to employ that {\it both} the analytical continuation
(\ref{2.5}) (describing the infinitely thin $YM$ vortices)
of the presumable Gauge String representation {\it and} the stringy sum
(\ref{0.1bbe})/(\ref{2.5bxa}) (where the weight (\ref{2.5bxa}) is the local
deformation (\ref{0.3b}) of the quasi-local pattern (\ref{0.1})) are exactly
equivalent to the $m_{0}=0$ option of the $NG$ theory
(\ref{0.1bbe})/(\ref{2.5bb}). In other words, with the exclusion
of the measure zero subset of the surfaces, the characteristic worldsheet's
configurations
within the two former sums are assigned with the $NG$ weight (\ref{2.5bb}).
In consequence, when the conditions (\ref{0.1eea}) are fulfilled, the
smearing (\ref{0.1}) of the local weight (\ref{2.5bxa}) simultaneously
provides with the admissible smearing of the 'bare' Gauge String weight
(\ref{0.5c}). The reason is that, in the regime (\ref{0.1eea}), the weight
(\ref{0.1}) can be traded for the curvature expansion (\ref{0.3bbq}) which,
for the {\it dense} variety of the worldsheets $\tilde{M}$, starts with
the leading term in the form of the $m_{0}=0$ Nambu-Goto pattern
(\ref{2.5bb}) rather than of the more general pattern\footnote{Remark that,
due to the effects of the quasi-locality, the deviation of the factor
(\ref{2.5bxb}) from unity is {\it observable}
for generic {\it finite} values of the ratio of the vortex width
and the characteristic amplitude of the fluctuations of the vortex.}
(\ref{2.5bxa}).
In turn, it implies that the data of the quasi-contact interactions (which
encodes the choice of the gauge group and the lagrangian for the
dual $YM_{D}$ theory) is unobservable within $<W_{C}>$ in the considered
$D\geq{3}$ regime (when the loop $C$ is {\it without} $1d$ selfintersections).

To implement this idea, we first derive the estimate (\ref{0.1eea}) for
the characteristic amplitude $\sqrt{<{\bf h}^{2}>}$ of the fluctuations of
the extended flux-tube. Then, we employ certain
general implications of the {\it Wilsonian} renormgroup in order to justify
the following. There is the pertinent variety (including the particular
choice (\ref{2.5bxd})) of the smearing functions (\ref{0.1bb}) so that, in the
regime (\ref{0.1eea})/(\ref{0.1eec}), the stringy systems
(\ref{0.1bbe})/(\ref{0.1}) and (\ref{0.1bbe})/(\ref{2.5bb}) possess
one and the same pattern of the low-energy theory.

\subsection{The estimate of the characteristic amplitude of the
fluctuations.}

The required estimate (\ref{0.1eea}) of the amplitude $\sqrt{<{\bf h}^{2}>}$
is fairly straightforward in the case when the amplitude is either much larger
or much smaller than the width of the vortex.
As we will see, in both cases\footnote{For a
given $\bar{\lambda}=\xi\lambda$, the peculiar logarithmic scaling
(\ref{0.1xeh}) is observable (as a profile of
the energy distribution) only in the regime when the height is sufficiently
larger than the width of the flux-tube.}, a simple computation (in the spirit
of \cite{Luscher}) provides with the equation
\be
\frac{<{\bf h}^{2}(\gamma)>}{(D-2)}~
\sim{~\frac{ln[A_{min}\Lambda^{2}]}{\xi\lambda\Lambda^{2}}}~~~~
\Longrightarrow{~~~~D_{H}\longrightarrow{+\infty}},
\label{0.1xeh}
\ee
where $A_{min}\equiv{A[\tilde{M}_{min}(C)]}$ is the minimal area corresponding
to the saddle-point worldsheet $\tilde{M}_{min}(C)$ while $\lambda$ and $\xi$
are defined in eqs. (\ref{1.1bx}) and (\ref{0.1ew}) respectively. (According
to the next section, in the regime (\ref{0.1eed}), the estimate (\ref{0.1xeh})
is modified through the substitution (\ref{2.12xc}).) As for
${\bf h}^{2}(\gamma)$, it is to be viewed as the squared distance between the
particular point $\gamma=(\gamma_{1},\gamma_{2})$ of a given worldsheet
$\tilde{M}(C)$ and the surface $\tilde{M}_{min}(C)$ (i.e.
the relative {\it height} of $\tilde{M}$ measured, at the point $\gamma$
sufficiently {\it far} from the boundary $C$, in
the direction normal to $\tilde{M}_{min}$) both spanned by the loop $C$ in
question. Then, taking into account that the width $\sqrt{<{\bf r}^{2}>}$ of
the stable flux-tube (associated to the $SC$ solution of the Loop equation
(\ref{1.11b})) is supposed to be $\sim{\Lambda^{-1}}$, we reproduce the first
of the previously asserted conditions (\ref{0.1eea}). Note also that
eq. (\ref{0.1xeh}) evidently implies that, in the large $N$ $SC$ regime
(\ref{0.1eec}), the Gauge String worldsheet
is characterized by the infinite Hausdorff dimension $D_{H}$.

Let us now return to the derivation of eq. (\ref{0.1xeh}) that, as a
by-product, will allow to present the semiclassical evaluation the physical
string tension $\sigma_{ph}(R)$ (see eqs. (\ref{2.12bxx}) and (\ref{0.7bxx}))
in the regime (\ref{0.1eea})/(\ref{0.1eec}). To this aim, we first
consider the simplest situation when the profile of the smearing function
${\mathcal{G}}({\bf z}^{2})$ is deformed in accordance with the local limit
(\ref{0.3cb}) when the flux-tube becomes infinitely thin. (The generalization
of the analysis to the case of the fat $YM$ vortex will be given in the end of
this subsection.) In the considered limit,
the equivalence theorem (stated in the begining of the section) ensures that
the stringy system (\ref{0.1bbe})/(\ref{0.1}) is dynamically reduced to the
Nambu-Goto one (\ref{0.1bbe})/(\ref{2.5bb}) with $m_{0}=0$ and
$\bar{\lambda}(\lambda)=\xi\lambda$. To make the required estimates for
sufficiently large $\bar{\lambda}=\xi\lambda$, one can employ the
reparametrization symmetry (\ref{0.4zh}) and introduce the
'semirelativistic' parametrization of the characteristic worldsheets
\be
x_{\mu}(\gamma)=y_{\mu}~~~if~~~\mu=1,2~~~~~~~;~~~~~~~
x_{j+2}(\gamma)=h_{j}(y)~~~if~~~j=1,...,D-2~,
\label{2.5bxv}
\ee
where the coordinates $y_{a},~a=1,2,$ parametrize the
minimal area surface $\tilde{M}_{min}(C)$ presumed to have disc's topology.
The position of $\tilde{M}_{min}(C)$ in ${\bf R^{D}}$ is encoded in
the metric $\hat{g}^{ab}(y)$ to be normalized by the condition
$det[\hat{g}(x)]=1$ so that ${A[\tilde{M}_{min}(C)]}=
\int_{\tilde{M}_{min}(C)}d^{2}y$. Therefore, within the semiclassical
approach one can employ the reduced pattern\footnote{The pattern
(\ref{2.5bxa}) is {\it not} applicable to those configurations of
$\tilde{M}(C)$ which, in the Hamiltonian framework,
correspond to the creation of closed 'sea'-strings from the vacuum.
At least for $\bar{\lambda}$ large enough, the latter creation processes
are infrared irrelevant.} of the area-functional
\be
\sigma_{0}A[\tilde{M}(C)]~\longrightarrow{~
\frac{\bar{\lambda}\Lambda^{2}}{2}}
\int\limits_{\tilde{M}_{min}(C)}~d^{2}y~\sqrt{1+\delta^{ij}\hat{g}^{ab}(y)
\partial_{a}h_{i}(y)\partial_{b}h_{j}(y)}~,
\label{2.5bxc}
\ee
where the height ${\bf h}(y)=\{h_{j}(y),~j=1,...,D-2\}$ has been introduced in eq.
(\ref{0.1xeh}), and $\tilde{M}_{min}(C)$ in what follows is
constrained to be flat: $\hat{g}^{ab}(y)\rightarrow{\delta^{ab}}$.

Evidently, the action (\ref{2.5bxc}) defines (together with the
ghost-determinants due to the fixing of the symmetry (\ref{0.4zh})) the
{\it two-dimensional} effective field theory in the finite box of the
2-volume $A_{min}(C)$ and with the {\it UV} cut off $\Lambda$. In the
quasi-classical limit $\bar{\lambda}>>1$, one can safely expand the square
root in eq. (\ref{2.5bxc}) that makes manifest the asymptotics
\be
\bar{\lambda}\rightarrow{\infty}~:~~~
\sqrt[n]{<(\partial_{a}{\bf h})^{2n}>}\sim{\frac{1}{\bar{\lambda}}}~~~
\Longrightarrow{~~~\sqrt{1+(\partial_{a}{\bf h})^{2}}\longrightarrow
{1+\frac{(\partial_{a}{\bf h})^{2}}{2}}}~,
\label{2.5bxm}
\ee
which, in the leading order (when the contribution of the ghost-determinants
can be neglected), reduces the system (\ref{2.5bxc}) to
the free $2d$ theory of the vector
field ${\bf h}(y)$ with the $(D-2)$ components $h_{j}$. This result can be
immediately utilized to justify that, in the regime (\ref{0.1eec}) (i.e. for
sufficiently large $\bar{\lambda}$), the estimate (\ref{0.1xeh}) is indeed
valid. The only subtlety is that one is to consider the variety of the large
contours $C$ so that
the corresponding minimal area $A_{min}(C)$ and the radius of curvature
${\mathcal{R}}(s)$ comply with the conditions (\ref{0.1eea}). Additionally,
we have to require that the considered contours $C$ are
{\it macroscopic} which can be formalized by the following twofold
constraint. Firstly, except for the $1/\Lambda$-vicinity of the
nontrivial (point-like) selfintersections of the boundary $C$, the distance
$|{\bf x}(s)-{\bf x}(s')|$ is much larger than the flux-tube's width
$\sqrt{<{\bf r^{2}}>}\sim{\Lambda^{-1}}$,
provided that the length of the corresponding segment of the
boundary $C=\partial \tilde{M}$ is much larger than $\Lambda^{-1}$:
\be
\int_{s'}^{s} dt~\sqrt{\dot{x}^{2}_{\mu}(t)}~>>\Lambda^{-1}~~~~~
\Longrightarrow~~~~~{|{\bf x}(s)-{\bf x}(s')|>>\Lambda^{-1}}~.
\label{0.1eel}
\ee
Secondly, we always presume that, once $C_{xx}=C_{xy}C_{yx}$ (where
${\bf x}(s)={\bf y}(s'),~s\neq{s'}$) satisfies the above
condition, then both $C_{xy}$ and $C_{yx}$ comply with this condition as well.

Next, employing the standard
proper-time regularization of the determinant of the $2d$ Laplacian (see eq.
(\ref{0.7bxx}) below), the semiclassical evaluation of the functional integral
(associated to the action (\ref{2.5bxc})) yields
\be
\zeta_{D}=\frac{D-2}{4\pi}
\label{0.7bcx}
\ee
for the entropy constant $\zeta_{D}$. The latter enters the corresponding
approximation for the physical string tension (\ref{0.1eec}) deduced,
in the regime (\ref{0.1eea})/(\ref{0.1eec}), from the loop-average
$<W_{C}^{f}>$ in the (anti)fundamental
representation of the $U(N)$ group in question. More generally, consider the
physical string tension $\sigma_{ph}(R)$ corresponding to the
average $<W_{C'}^{R}>$ in a given (anti)chiral $U(N)$ representation $R$.
Taking into account the condition (\ref{0.1eea}), in the leading order of the
$1/N$ expansion, one readily obtains within the same semiclassical approach
\be
\sigma^{(sc)}_{ph}(R)={|n(R)|
\left(\frac{\bar{\lambda}}{2}-\zeta_{D}\right)
\Lambda^{2}}~~~~~~~;~~~~~~~R\in{Y^{(N)}_{n}},
\label{2.12bxx}
\ee
where $n(R)\equiv{n}\sim{N^{0}}$ is the total number (see eq. (\ref{2.12bv})) of the
elementary $YM$-fluxes associated to $R$. The pattern (\ref{2.12bxx}) is
to be compared with the asymptotical $C_{2}(R)$-scaling (\ref{2.12bb}) of
$\sigma^{(sc)}_{ph}(R)$ taking place in the
complementary regime (\ref{0.1eed}).

Finally, let us show that, in the $SC$ regime (\ref{0.1eec}), the eq.
(\ref{0.1xeh}) remains valid for a generic profile of the smearing
function ${\mathcal{G}}({\bf z}^{2})$ satisfying eqs.
(\ref{0.1bb}) and (\ref{0.1ew}). Recall first that, once $\xi=O(1)$, the
normalization (\ref{0.1bb}) ensures that the width (\ref{0.3csb}) of
the $YM$ vortex is of order of $\Lambda^{-1}$. Next, we presume that
the width is much smaller than the amplitude of the fluctuations provided
the minimal area $A_{min}(C)$ is sufficiently large. To justify this
presumption, observe that the scaling (\ref{2.5bxm}) of
$\sqrt[n]{<({\bf h}^{2})^{n}>}$ evidently remains valid for
generic worldsheet weight (\ref{0.1}) constrained to support the
complementary scaling (\ref{0.1eec}) of $\sigma_{ph}$. Employing the same
parametrization\footnote{In this way, one has to neglect all the
selfintersections of $\tilde{M}(C)$ which, strictly speaking, are asigned in
the short-distance expansion of eq. (\ref{0.1}) with the weights additional to
the ones associated to eq. (\ref{0.1xem}). In the $D\geq{3}$ regime
(\ref{0.1eec}), employing the technique of Appendix A, it can
be shown that the latter extra weights, being infrared {\it irrelevant}, do
not alter the estimate (\ref{0.1xeh}).} of $\tilde{M}$ as in eq.
(\ref{2.5bxc}), the leading nontrivial order of the operator expansion
(\ref{0.3bbq}) is reduced in the limit $\bar{\lambda}\rightarrow{\infty}$
to the gaussian pattern
\be
\frac{{\lambda}\Lambda^{2}}{2}\int d^{2}y~
h_{j}(y)~\tilde{\mathcal{O}}(\Box)~
h^{j}(y)~~~~~;~~~~~
\tilde{\mathcal{O}}(t^{2})=
\left(t^{2}\tilde{\mathcal{G}}_{2}(t^{2})-
\tilde{\mathcal{O}}_{2}(t^{2})+\tilde{\mathcal{O}}_{2}(0)\right)
\Bigg|_{t^{2}>0}~>~0~,
\label{0.1xem}
\ee
where $\Box=-\partial_{b}\partial^{b}$ (with $\partial_{b}\equiv
{\partial/\partial y_{b}}$), while
$\tilde{\mathcal{G}}_{2}(q_{a}q^{a})=\int d^{2}z e^{iq_{b}z^{b}}
{\mathcal{G}}(z_{a}z^{a})$ and $\tilde{\mathcal{O}}_{2}(q_{a}q^{a})$ are the
$2d$ Fourier images of respectively the smearing function
${\mathcal{G}}(z_{a}z^{a})$ and its derivative
$\frac{d{\mathcal{G}}(r)}{dr}|_{r=z_{a}z^{a}}$ so that (in accordance with
eq. (\ref{0.1ew})) $\tilde{\mathcal{O}}(t^{2})\rightarrow{\xi t^{2}/2}$ when
$t^{2}\rightarrow{0}$.
Upon a reflection, in the $SC$ phase (\ref{0.1eec}), the Gauge String is
supposed to be stable with respect to the considered small fluctuations that
implies the positivity-constraint: $\tilde{\mathcal{O}}(t^{2})=
(\xi t^{2}/2+\sum_{n\geq{2}} b_{n} (t^{2}/\Lambda^{2})^{n})>0$
for $\forall{t^{2}}>{0}$. In turn, combining the latter stability with the
implications of the previous analysis, one justifies the required estimate
(\ref{0.1xeh}) since, in the expansion of $\tilde{\mathcal{O}}(t^{2})$,
the higher order $n\geq{2}$ terms in this case are
{\it irrelevant} for the considered logarithmic asymptotics of
$<{\bf h}^{2}>$. Also, in the
regime (\ref{0.1eea})/(\ref{0.1eec}), eq. (\ref{0.1xem}) can be
straightforwardly used to generalize the estimate (\ref{0.7bcx}):
\be
\zeta_{D}=\frac{D-2}{4\pi}\int\limits_{1}^{\infty}
\frac{d\tau}{\tau}\int\limits_{0}^{\infty} dt^{2}~
exp\left(-\frac{2\tau}{\xi} \tilde{\mathcal{O}}(t^{2})\right)~~~~~~,~~~~~~
\frac{\tilde{\mathcal{O}}(t^{2})}{dt^{2}}\Bigg|_{t^{2}=0}=\frac{\xi}{2}~,
\label{0.7bxx}
\ee
where the normalization of the exponent is fixed to match (in the limit
$\tilde{\mathcal{O}}(t^{2})=\xi t^{2}/2$) with the result which can be
obtained employing the Pauli-Villars regularization
used in the original computation of \cite{Luscher}.

\subsection{Identification of the proper universality class.}

Finally, let us demonstrate that, for any macroscopic contour $C$, the
conditions (\ref{0.1eea})/(\ref{0.1eec}) ensure the following important
{\it universality} of the low-energy results. Under the latter conditions,
for a large variety of the smearing functions ${\mathcal{G}}({\bf x}^{2})$,
the ($b=1$) stringy sum (\ref{0.1bbe})/(\ref{0.1})
is in the same universality class\footnote{Furthermore, judging from
the analysis of Section 8, it is reasonable to suppose that this is the only
universality class suitable to provide with the $\Lambda^{2}$-scaling
(\ref{0.1eec}) (of $\sigma_{ph}$) predetermined, in the large $N$ $SC$ phase,
by the pattern (\ref{1.11b}) of the Loop equation.} with the properly 
associated Nambu-Goto system (\ref{0.1bbe})/(\ref{2.5bb}). To support the
infrared equivalence with the Nambu-Goto theory, observe first
that in the regime (\ref{0.1eea}) the expansion (\ref{0.3bbq}) can be utilized
to consistently define (building on the stringy representation
(\ref{0.1bbe})/(\ref{0.1})) the effective 2-dimensional gravity on the
worldsheet $\tilde{M}$. In the infrared domain, the $p=0,~p=1$ and $p\geq{2}$
operators ${\mathcal{Q}}^{(k)}_{2p}$ in eq. (\ref{0.3bbq}) refer respectively
to the relevant, marginal and irrelevant operators associated to the former
operators via certain trick \cite{GFS,ExtrCurv/P} sketched below. In turn, it
allows to apply the general arguments of the {\it Wilsonian} renormgroup to
the effective theory (\ref{0.3bbq})/(\ref{0.1bbe}) in order to deduce that,
at least in the $SC$ phase (\ref{0.1eea})/(\ref{0.1eec}), the pattern of the
leading $p=0$ Nambu-Goto term correctly singles out the universality class in
question.

To sketch the purported arguments, recall first that the above mentioned trick
\cite{GFS,ExtrCurv/P} is aimed to circumvent the {\it nonpolynomial}
dependence (on the derivatives
$\partial x_{\mu}(\gamma)/\partial\gamma_{a}\equiv{\partial_{a}}$ with respect
to the coordinates $x_{\mu}(\gamma)$ maintaining the parametrization
(\ref{0.4zh}) of the surface $\tilde{M}$) of the
determinant $det[\hat{g}(\gamma)]=p^{2}_{\mu\nu}/2$
of the induced metric $\hat{g}_{ab}(\gamma)$ entering, see Appendix A, the
operators ${\mathcal{Q}}^{(k)}_{2p}(\gamma)$. For this purpose,
one is to introduce the auxiliary metric-tensor $h_{ab}(\gamma)$ to be
identified with $\hat{g}_{ab}(\gamma)$ through the Lagrange multipliers
$\varphi^{ab}$. Employing the conformal gauge $h_{ab}(\gamma)=
\rho(\gamma)\delta_{ab}$, it allows to trade $(det[\hat{g}(\gamma)])^{1/2}$
in ${\mathcal{Q}}^{(k)}_{2p}(\gamma)$ for the
{\it dimensionless} auxiliary scalar field $\rho(\gamma)$.
Next, consider the $SC$ phase (\ref{0.1eec}), where $\varphi^{ab}$
is supposed to exhibit a nontrivial condensate: $<\varphi^{\alpha\beta}>=
\tilde{\varphi}\delta^{ab}$ with $\tilde{\varphi}\neq{0}$.
As a result, for the computation of the loop-averages in the large-area
regime (\ref{0.1eea}), the expansion (\ref{0.3bbq}) can be traded
(generalizing the analysis of \cite{ExtrCurv/P}) for the
$\int_{\tilde{M}}d^{2}\gamma$-integral of certain {\it polynomial} (with the
$\rho(\gamma)$-{\it dependent} coefficients) expressed in terms of the
derivatives $\partial_{a_{1}}...\partial_{a_{k}}x_{\mu}(\gamma)$. In
particular, provided that $N,\lambda\rightarrow{\infty}$ while the conditions
(\ref{0.1eea}),(\ref{0.1eec}) are fulfilled, for any macroscopic contour $C$
the leading asymptotics of the $b=1$ stringy sum (\ref{0.1bbe})/(\ref{0.1})
is given by the sheer $m_{0}=0,~\bar{\lambda}(\lambda)=\xi\lambda$
Nambu-Goto weight (\ref{2.5bb}) assigned to the minimal-area surface
$\tilde{M}(C)$ with $\chi=1$.

More generally, given any macroscopic contour $C$, the pattern of the
renormgroup flow implies the following. For sufficiently large $\lambda$,
there is such a variety of the smearing functions
${\mathcal{G}}({\bf x}^{2})$ (including those for which all the moments
$K_{n}[{\mathcal{G}}]=O(1)$) that the
stringy systems (\ref{0.1bbe}), based respectively on the weights (\ref{0.1})
and (\ref{2.5bb}), merge in the regime (\ref{0.1eea})/(\ref{0.1eec}) provided
the judicious choice of the parameters $\bar{\lambda}(\lambda),~
m_{0}(\lambda)$. In particular, the second of the conditions (\ref{0.1eea})
reassures that (modulo the subleading numerical factor depending on the number
of the point-like (self)intersections of the loops) the loop-averages
$<W_{C}>$ in both of the systems can be
computed with the help of one and the same low-energy theory with the new
{\it UV} cut off $\Lambda'$; ${\mathcal{R}}^{-1}<<\Lambda'<<\Lambda$. As
previously, we presume that the boundary contour $C$ is devoid of
any $1d$ selfintersections (including zig-zag backtrackings).

The above renormgroup arguments are most straightforward to apply in the
$D\geq{5}$ case. Indeed, as the conditions (\ref{0.1eea})
ensure that, for any macroscopic contour $C$, the $YM$ vortex behaves (in its
interior) as if it is infinitely
thin, one can neglect all the selfintersections in the interior of the
flux-tube worldsheets. This is justified by the
theorem \cite{StabMap} that the selfintersecting $2d$ surfaces, entering the
measure of eq. (\ref{0.1bbe}), span the
subspace which in $D\geq{5}$ is of measure zero in the relevant space
${\mathcal{I}}(M,{\bf R^{D}})$ generated by the smooth immersions
(\ref{0.4}). In $D=3$ and $4$, one is to account for the extra contribution
to the short-distance expansion of the worldsheet weight (\ref{0.1}) which is
associated respectively to $1d$ and point-like stable selfintersections. A
more careful analysis shows then that the dynamical effect of the latter
selfintersections in $D\geq{3}$ is infrared-irrelevant at least as far as the
constraints (\ref{0.1eea})/(\ref{0.1eel}) are satisfied.
Altogether, it substantiates the applicability of the arguments based on the
notion of the effective $2d$ theory of gravity (e.g. of the vector field
${\bf h}(x)$ entering eq. (\ref{2.5bxc})) living on the worldsheet
$\tilde{M}$ as on the base-space.

\subsubsection{Infrared reduction of the $N$-dependence of the loop-averages.}

It is noteworthy that, compared to the $SC$ asymptotics (\ref{0.5bxx}) where
the $N$-dependent pattern of the quasi-contact interactions is observable,
in the (multiloop generalization of the) regime (\ref{0.1eea})
the $N$-dependence of the $U(N)$ loop-averages
$N^{b}<\prod_{k=1}^{b} W_{C_{k}}>$ is substantially reduced\footnote{
This reduction can be confronted with the hypothesis \cite{PolyakCS} that
all gauge theories are described by the same universal string theory, with
the dependence on the gauge group being reproduced through the factor 
(equal to $N^{\chi}$ in the $U(N)$ case) sensitive, in contradistinction to
the $\lambda$-dependence of the $J[..]$-factor (\ref{0.5c}), only to the
topology of the surface $\tilde{M}$.}. Once the macroscopic
contours $C_{k}$ do {\it not} possess $1d$ (self)intersections, 
in the considered regime this dependence is reproduced by the purely
{\it topological~} 't Hooft $N^{\chi}$-factor built into the weight
(\ref{2.5bb}) of the infrared equivalent Nambu-Goto theory. (As for the $SU(N)$ case, there appears the
additional source of the $1/N$-dependence which is formalized via the
substitution (\ref{0.1xaa}).)

Actually, the latter reduction
can be viewed as a particular consequence of the well-known
{\it information-loss} that generically takes place when the physical data
is restricted to its low-energy sector. Namely, the remaining block of eq.
(\ref{2.5bb}) is {\it universal}: modulo the particular functional dependence
of the parameters $\bar{\lambda}(\lambda)$ and $m_{0}(\lambda)$ on
$\lambda$, this block is common for a large variety of the actions of the
dual $U(N)$ gauge theory. (Complementary. there is a large variety of the
quasi-local weights, given by the substitution of the bilocal pattern in the
exponent of eq. (\ref{0.1}) by a generic area-functional (\ref{1.9adc}),
that result in the stringy systems belonging to the same universality class
'labelled' by the Nambu-Goto theory (\ref{2.5bb}).) Furthermore, combining
the previous analysis with the conclusions of \cite{Dub3}, we expect the
following large $N$ universality of the infrared properties of the $D\geq{3}$
$YM_{D}$ systems. Namely, the $N\rightarrow{\infty}$ limit of the proposed
implementation (\ref{0.1eec}) of the Nambu-Goto theory
(\ref{0.1bbe})/ (\ref{2.5bb}) is
presumed to provide with the low-energy description which is {\it universal}
(modulo the particular dependence of $\bar{\lambda}(\lambda)$ and
$m_{0}(\lambda)$ on $\lambda$) for {\it any} $D\geq{3}$ $U(\infty)$ or
$SU(\infty)$ pure gauge system (endowed with an arbitrary polynomial, in
terms of $F_{\mu\nu}$, lagrangian)
once the free energy of the latter exhibits the $N^{2}$-scaling.

\section{The loop-averages in the $SC$ regime (\ref{0.1eed}).}

In the extreme $SC$ limit $\lambda>>1$, the area-parameter
$ln[A_{min}\Lambda^{2}]$ can be used to vary the $D\geq{3}$ pattern of the
average $<W^{R}_{C'}>$ trading the regime (\ref{0.1eea}) for the opposite
regime (\ref{0.1eed}) where the specifics of the particular gauge theory's
action (\ref{1.1}) becomes most observable. In order to substantiate the
dimensional reduction (\ref{0.5bxx}), we first derive the modification
(\ref{2.12xc}) of the estimate (\ref{0.1xeh}). For simplicity, we
restrict the derivation to the simplest case of the $U(1)$
gauge theory (described through the $N=1$ $SC$ expansion
(\ref{0.1bbe})/(\ref{0.1})), presuming that the appropriately
formulated general conclusions are common for all $N\geq{1}$.

\subsection{The collective dynamics of the fat composed $YM$ vortex.}

Let us consider the $U(1)$ average $<W^{R}_{C'}>=<W^{(n)}_{C'}>$ in the
representation $R$ labelled by an integer number $n\in{\bf Z}$. The associated
$N=1$ stringy representation (\ref{0.1bbe})/(\ref{0.1}) of $<W^{(n)}_{C'}>$
implies that the worldsheet
$\tilde{M}(C)$ has the boundary $C=\cup^{n}_{k=1}C'_{k}$ (where $C'_{k}$ is
the $k$th copy of $C'$) resulting from the $n$-times winding around the base
loop $C'$. Therefore, $\tilde{M}(C)$ can be visualized as
the union $\cup^{n}_{k=1} \tilde{M}_{k}(C'_{k})$ of the $k$ 'elementary'
worldsheets $\tilde{M}_{k}(C'_{k})$ (each corresponding to the {\it unit}
$YM$ flux) spanned by the correspnoding $C'_{k}$. Altogether, we arrive at the
picture of $n$ {\it independently} fluctuating fat flux-tubes of the
width\footnote{It is the value of $\sqrt{<{\bf r}^{2}>}$, rather than
the averaged height (\ref{0.1xeh}), that determines the characteristic
$D\geq{3}$ distribution of the energy density of the {\it YM} vortex in the regime
(\ref{0.1eed}). As a result, the logarithmic scaling (\ref{0.1xeh}) is
essentially {\it unobservable} in this case.}
$\sqrt{<{\bf r}^{2}>}\sim{\Lambda^{-1}}$
which is much larger than the characteristic amplitude $\sqrt{<{\bf h}^{2}>}$
of their fluctuations, once the conditions (\ref{0.1eed}) are fulfilled.
In consequence, the quasi-contact interactions between the fat elementary
$YM$ vortices can {\it not} be neglected even for $[A_{min}\Lambda^{2}]>>1$.
The purpose of this subsection is to show that, from the infrared viewpoint
(i.e. at the distance-scales $>>\Lambda^{-1}$), these interactions
ensure the remarkable reduction of the number of the independent
low-energy stringy excitations. Namely, the considered conglomerate of the $n$ fat
elementary $YM$ vortices behaves, in the infrared, as a {\it single} composed
flux-tube which, fluctuating with the amplitude given by the modification
(\ref{2.12xc}) of eq. (\ref{0.1xeh}), supports the physical string tension
(\ref{2.12bb}).

To this aim, one is to introduce first the $n$ different heights
${\bf h}_{k}(y)=\{h_{j;k}(y)|~j=1,..,D-2;~k=1,..,n\}$ associated, according to
the parametrization (\ref{2.5bxv}), to each of the previously defined 
worldsheets $\tilde{M}_{k}(C'_{k})$. Then, the simple algebraic manipulations
varify that, for small fluctuations, the pattern of the $N=1$ weight
(\ref{0.1}) is reduced to the quadratic (with respect to ${\bf h}_{k}(y)$)
form given by the twofold modification of eq. (\ref{0.1xem}). In addition to
the $N=1$ option of the substitution (\ref{2.12xc}), the single variable
${\bf h}(y)$ is superseded by the {\it collective coordinate}
$(\sum^{n}_{k=1}{\bf h}_{k}(y))/n$. The latter can be interpreted as the
coordinate which, defining the auxiliary surface $\tilde{M}_{+}(C')$, refers
to the 'center of mass' of the $n$ worldsheets $\tilde{M}_{k}(C'_{k})$. The
key-point is that, due to the $S(n)$-invariance with respect to the
arbitrary permutations ${\bf h}_{k}(y)\rightarrow{{\bf h}_{\sigma(k)}(y)},~
\sigma\in{S(n)}$, the considered quadratic form is 
{\it independent} of all the $n-1$ 'relative' coordinates
${\bf h}_{k}(y)-{\bf h}_{k-1}(y)$.

This observation can
be strengthened further to reveal that, within the short-distance expansion
(\ref{0.3bbq}), the relative fields ${\bf h}_{k}(y)-{\bf h}_{q}(y)$ are
supposed to be {\it massive} owing to the presence of the
potential-like terms $({\bf h}_{k}(y)-{\bf h}_{q}(y))^{m},~m\geq{2}$.
Therefore, these degrees of freedom are supposed to be irrelevant at the
distances $>>\Lambda^{-1}$. Consequently, as long as the
constraints (\ref{0.1eed}) are satisfied, the residual infrared dynamics of
the 'center of mass' effective excitation $\tilde{M}_{+}(C')$ in the leading
order reduces to the one described by the $N=1$ variant of the quasi-local
weight (\ref{2.12bl}) where the support $T_{min}$ should be changed for the
collective coordinate $\tilde{M}_{+}(C')$ to be integrated over as in eq.
(\ref{0.1bbe}). To justify the consistency of the
arguments, one is to observe that
the fat flux-tube, assigned with the weight (\ref{2.12bl}), indeed fluctuates
with the amplitude given by the modification (\ref{2.12xc}) of the estimate
(\ref{0.1xeh}). In the limit $ln[A_{min}\Lambda^{2}]>>1$, this amplitude is
supposed to be much {\it larger} than the characteristic amplitude of 
the relative fluctuations $|{\bf h}_{k}(y)-{\bf h}_{k-1}(y)|$ which is
expected to be $\sim{1/\bar{\lambda}}$.

\subsection{The $YM_{D}\rightarrow{YM_{2}}(T_{min})$ dimensional reduction.}

Now we are in a position to deduce the prescription (\ref{0.5bxx}) for the
leading asymptotics of the average $<W_{C}>$ in the $D$-dimensional
$YM_{D}$ theory (\ref{1.1}) evaluated in the extreme $SC$ limit
(\ref{0.1eed}) accompanied by the auxiliary condition (\ref{0.1eeb}).
To begin with, from the above discussion it is clear that in the considered
limit, the $D\geq{3}$ sum of the $YM$ vortices is localized on the (moduli
space\footnote{Remark that, through the data of the singularities (excluded
from the immersion-space ${\mathcal{I}}(M,{\bf R^{D}})$) of the map
(\ref{0.4}), a given minimal area worldsheet $\tilde{M}_{min}(C)$ may have
additional 'quantum numbers' \cite{Dub3} apart from its area, genus and the
support in ${\bf R^{D}}$.}  of the) saddle-point worldsheet(s)
$\tilde{M}_{min}(C)$. The latter, implementing the absolute minimum of the
area-functional $A[\tilde{M}(C)]$, are presumed to possess a common support
$T=T_{min}(C)$. Let us further constrain that,
at any point of $T_{min}(C)$ which does not belong to the
$1/\Lambda$-vicinity of possible $0$- or $1$-dimensional selfintersections
of $T_{min}(C)$, the line in the normal direction either does not have
the second intersection with $T_{min}(C)$ or the second intersection takes
place at a distance $>>\Lambda^{-1}$. Then, combining the
condition (\ref{0.1eeb}) with the general pattern (\ref{0.3bbq}) of the
short-distance expansion, one arrives at the folowing important universality.
Namely, the smearing\footnote{For example, eq. (\ref{2.12bl}) presents
the smearing regularization of the local pattern (\ref{0.3csd}).} of the l.h.
side of eq. (\ref{0.5cxd})
becomes essentially unobservable modulo the subleading contribution of
the perimeter type akin to the one of eq. (\ref{2.5bxl}). (Furthermore, the
latter perimeter-dependent terms can be eliminated, up to a constant factor
depending on the number of point-like selfintersections of the loop, by the
proper choice of the regularization of the Loop equation which will ensure
$m_{0}(\lambda)=0$.) In other words, the flux-tube in its
interior can be treated as infinitely thin which facilitates the
{\it regularization-independence} of the leading asymptotics of the
$D\geq{3}$ average $<W_{C}>$ modulo the choice of auxiliary parameter $\xi$
in the expression (\ref{0.4bc}) for $g^{2}_{YM_{2}}$.

As a result, the corresponding leading contribution of the fat $YM$
flux-tube(s) is correctly reproduced by the r.h. side of eq. (\ref{0.5cxd})
associated to the support $T_{min}(C)$ (of the minimal area worldsheet(s)
$\tilde{M}_{min}(C)$) which, for simplicity, is presumed to be unique for
a given $C$. In turn, it is supposed to justify the
dimensionally reduced $SC$ asymptotics\footnote{A somewhat similar dimensional
reduction $YM_{D}\rightarrow{YM_{2}}$ is argued to take place in the heuristic
model \cite{AmbOles} (based on certain picture of the stochastic vacuum)
proposed to approximate the description of the $YM_{D}$ theory (\ref{1.1})
but in the standard phase when
$\lambda\rightarrow{0}$.} (\ref{0.5bxx}) of $<W_{C}>$,
once $\lambda,~({\mathcal{R}}(s)\Lambda)^{2}>>N^{2}$ so that
one can retain only the leading order of {\it both} the
$1/\lambda$-expansion {\it and} the curvature expansion (\ref{0.3bbq})
yet summing up the relevant $1/N^{2}$-subseries. (In the regime (\ref{0.1eed})
with $\lambda,~({\mathcal{R}}(s)\Lambda)^{2}\leq{N^{2}}$, one is to retain in
the r.h. side of eq. (\ref{0.5bxx}) only the leading $O(N)$ asymptotics of the
average $<W_{C}>|_{YM_{2}(T_{min})}$.)
In particular, when the general regularization prescription (\ref{0.1bb}) is
fixed, the parameter $\xi$ (entering e.g. eq. (\ref{0.4bc})) is to be
identified with the zeroth moment (\ref{0.1ew}) of the properly
introduced smearing function ${\mathcal{G}}(..)$.

On the one hand, the prescription (\ref{0.5bxx}) implies that, even in the
regime (\ref{0.1eed}), there is a subspace of the full loop space where the
$SC$ asymptotics (\ref{0.1eed}) of the average $<W_{C}>|_{YM_{D}}$ is still
reproduced by the
($m_{0}=0$ option of the) Nambu-Goto pattern (\ref{2.5bb}) applied to the
minimal area worldsheet. In turn, the simple pattern (\ref{0.1}) (applied to
$\tilde{M}_{min}(C)$) remains applicable when the saddle-point flux-tube
is given by the elementary $YM$ vortex with {\it unobservable} (within the
short-distance expansion (\ref{0.3bbq}), provided the latter is convergent at
least asymptotically) selfoverlapping. Then, the
dimensional reduction (\ref{0.5bxx}) equates (modulo the $\xi$-rescaling) the
$SC$ limit (\ref{0.1ew}) of the physical string
tension in the $YM_{D}$ system (\ref{1.1}) with the string
tension \cite{Kaz&Kost,Bralic} in the $D=2$ $YM_{2}$
theory (\ref{1.1}).

On the other hand, to implement in the  simplest setting the deviation of the
prescription (\ref{0.5bxx}) from the $m_{0}=0$ Nambu-Goto pattern
(\ref{2.5bb}), one is to consider the $D\geq{3}$
average $<W^{R}_{C'}>$ in a generic, not necessarily (anti)fundamental
representation $R$ belonging the $U(N)$ {\it (anti)chiral}\footnote{
Recall that, in the latter sector, the $U(N)$  character $\chi_{R}(V)$ is
composed of the products \cite{Dr&Zub} of the traces $tr[(V)^{k}]$ only.
The traces  $tr[(V^{-1})^{k}]$, associated to the holonomies winding in the
{\it opposite} direction, are {\it not} involved.
The choice of a chiral or antichiral irrep $R\in{Y^{(N)}_{n}}$
implies that, in eq. (\ref{2.12bv}), either $n_{i}\geq{0}$ or $n_{i}\leq{0}$
for $\forall{i}$. Consequently, all the $|n(R)|$ elementary fluxes,
composed into the fat collective $YM$ vortex associated to $<W^{R}_{C'}>$,
possess one and the same orientation.} sector. Conventionally,
we define the physical string tension $\sigma_{ph}(R)$ from the large
area asymptotics (to be considered {\it after}
the one of eq. (\ref{0.1eed})) of $ln[<W^{R}_{C'}>]$. Evaluating
$<W^{R}_{C'}>_{YM_{2}(T)}$ on a nonselfintersecting disc bounded by $C$,
one concludes from the pattern (\ref{0.3csd}) that in the regime
(\ref{0.1eed}) the tension 
\be
\sigma^{(sc)}_{ph}(R)={\left(\frac{\bar{\lambda}C_{2}(R)}{2N}-
\zeta_{D}\right) \Lambda^{2}}~~~~~~~,~~~~~~~R\in{Y^{(N)}_{n}}~,
\label{2.12bb}
\ee
asymptotically scales with respect to $R$ as the corresponding
eigenvalue $C_{2}(R)$ of the {\it second Casimir} operator of the $U(N)$
group in question:
\be
C_{2}(R)=\sum_{i=1}^{N}n_{i}\cdot(n_{i}+N-2i+1)~~~~~~~;~~~~~~
n(R)=\sum_{i=1}^{N} n_{i}~.
\label{2.12bv}
\ee
As for the $N$ integers $n_{i}$, parametrizing a generic $U(N)$ representation
$R$, they satisfy the condition: $n_{1}\geq{n_{2}}\geq...\geq{n_{N}}$ so that
$n\equiv{|n(R)|}$ measures the total amount of the elementary
$U(N)$ fluxes, and $C_{2}(R)\sim{N}$ once $|n(R)|\sim{N^{0}}$. Note also that,
for completeness, in the semiclassical result (\ref{2.12bb}) (to be compared
with eq. (\ref{2.12bxx})) we have also included the next-to-leading term
associated to the entropy of the fat flux-tube, parametrized by $R$, composed
of $|n(R)|$ elementary vortices of unit flux.

Finally, in the $SC$ limit (\ref{0.1eed}), the interactions between the latter
elementary vortices is predetermined by the previously considered amplitude
$<W^{R}_{C'}>_{YM_{2}(T)}$. To reveal the type of these interactions,
one is to take into accout that (owing to the constraint $n_{k}\geq{n_{k+1}}$)
the pattern (\ref{2.12bv}) ensures the inequality
\be
C_{2}(R)~\geq{~|n(R)|}~~~~~~,~~~~~~\forall{R}~,
\label{2.12bs}
\ee
where $n(R)$ is given by eq. (\ref{2.12bv}) and, for the subset of chiral
$U(N)$ representations, the lowest bound $C_{2}(R)=n(R)$ is achieved when
$n_{i}=1$ for $\forall{i}=1,2,...,N$. Therefore, apart from the latter
exceptional configuration (and its antichiral analogue with $n_{i}=-1$ for
$\forall{i}$), the {\it averaged} force between the elementary flux-tubes is
the repulsion that reminds of the picture associated to the type-II (dual)
abelian superconductor.

\section{Peculiarity of the dynamical regime (\ref{0.1eea}).}

In the regime (\ref{0.1eea})/(\ref{0.1eec}), the $D\geq{3}$ Nambu-Goto representation
(\ref{0.1bbe})/(\ref{2.5bb}) is presumed to be further reformulated in
compliance with the general spirit of the noncritical Polyakov's string
\cite{GFS}. Therefore, for macroscopic loops (satisfying, apart from the
$1/\Lambda$-vicinity of the selfintersections, eq. (\ref{0.1eel})), the
'microscopic'
$D\geq{3}$ Nambu-Goto system (\ref{0.1bbe})/(\ref{2.5bb}) is to be traded for
the 'macroscopic' low-energy theory with certain auxiliary worldsheet's fields
including the Liouville one. The important novelty is that the
{\it unconventional}
scaling (\ref{0.1eec}) of $\sigma_{ph}$ implies certain features of the Gauge
String dynamics which, in $D\geq{3}$, are expected to be essentially
different compared to the standard Polyakov's noncritical string theory
\cite{GFS}.

\subsection{Implications of the unconventional scaling (\ref{0.1eec}).}

To begin with, the straightforward consequence of equation
(\ref{0.1eec}) is that, generically, $\sigma_{ph}$ is of order of
$\Lambda^{2}$ unless $\bar{\lambda}(\lambda)\rightarrow{2\zeta_{D}}$.
The latter condition is in sharp contrast with the standard
scaling
\be
{\sigma_{ph}}/{\Lambda^{2}}~\longrightarrow{~+0}~,
\label{0.7bcb}
\ee
when the physical string tension is tuned to be much less than $\Lambda^{2}$.
As we already stated in the Introduction, the seemingly
bizarre scaling (\ref{0.1eec}) could be expected in advance, being
foreshadowed by the interpretation of $\Lambda$ as the confinement-scale
in the effective low-energy theory of the $D=4$ weakly-coupled
gauge theory (\ref{1.1}). Complementary, in view of eq. (\ref{0.1eec}),
the dynamics of the stringy system (\ref{0.1bbe})/(\ref{2.5bb}) is supposed to
be {\it devoid} of the notorious branched-polymer
'pathology' inherent in the 'fundamental' $D\geq{3}$ Nambu-Goto/Polyakov
string associated to the regime (\ref{0.7bcb}).

To see how the scaling (\ref{0.1eec}) improves the situation, let us
first recall the expected source of the above 'pathology' in the
conventional case (\ref{0.7bcb}). To begin with, choose an intermediate scale
$\Lambda'$ (with $\sqrt{\sigma_{ph}}<<\Lambda'<<\Lambda$) and consider the
effective low-energy theory (of the conventional Nambu-Goto string) with
$\Lambda'$ as the new {\it UV} cut off. Then, at the momentum-scales less than
$\Lambda'$, the dominant contribution to the worldsheet's entropy is
received from the graph-like surfaces visualized through the outgrowth of the
baby-universes reminiscent of fat graphs with the characteristic transverse
size of order of $1/\Lambda'$. Therefore, the considered effective theory is
presumed to fit in the pattern of the branched polymer
graphs filling the base-space ${\bf R^{D}}$.

On the other hand, it is easy to demonstrate that the $\Lambda^{2}$-scaling
(\ref{0.1eec}) of $\sigma_{ph}$ implies the severe suppression of the
considered outgrowth of the baby-universes. To this aim,
let us first note that (due to the {\it UV} cut off $\Lambda$) a tree-like
worldsheet's configuration of length $L$ costs the area at least of order of
$L\Lambda^{-1}$ where the well-developed tree-structure implies
$L>>{\Lambda^{-1}}$. In the branched polymer phase, the characteristic
worldsheet $\tilde{M}(C)$ would look like the minimal area surface
$\tilde{M}_{min}(C)$ with the attached multiple tree-like tentacles. For
large $A_{min}(C)\Lambda^{2}$, the area-density of these attachments is
supposed to be of order of $\Lambda^{-2}$. Therefore, the ratio of the
characteristic area $<A[\tilde{M}(C)]>$ to the minimal one is expected to be
much larger than unity. The point is that the latter estimate is {\it not}
consistent with the scaling (\ref{0.1eec}). To reveal the
mismatch, one can employ the pattern (\ref{2.5bxc}) that yields\footnote{It is
noteworthy that eq. (\ref{0.1xek}), being consistent with the the estimate
(\ref{0.1xeh}), is {\it not} in conflict with the infinite Hausdorff
dimension of the characteristic worldsheets.}
\be
\left({<A[\tilde{M}(C)]-A[\tilde{M}_{min}(C)]>}/{A[\tilde{M}_{min}(C)]}\right)
~\sim{~{1}/{\bar{\lambda}}}~,
\label{0.1xek}
\ee
which can not be much larger than unity in the $SC$ phase (\ref{0.1eec}) that
qualitatively justifies the lack of the 'branched polymer' outgrowth in
question.

At this step, it is appropriate to remark that the
considered mechanism of the suppression (of the outgrowth) is {\it not} in
conflict with the fact that the $3\leq{D}\leq{26}$ regime
(\ref{0.1eea})/(\ref{0.1eec}) of the Nambu-Goto theory
(\ref{0.1bbe})/(\ref{2.5bb}) is as much nontrivial as the conventional regime
(\ref{0.7bcb}) in the critical dimension $D=26$. Furthermore,
having presumably avoided the branched polymer phase, 
we can expect the absence of the tachion for $3\leq{D}\leq{26}$. Consequenly,
it implies that the mass $m^{(0)}_{gl}$ of the lowest glueball state, being
{\it positive} at least when $3\leq{D}\leq{25}$, is of order
(or even larger than) the inverse flux-tube's width $\sim{\Lambda}$.
Altogether, apart from the extreme {\it SC} limit
$\lambda\rightarrow{\infty}$
(where additional peculiarities may be anticipated), the physical spectrum
associated to the regime (\ref{0.1eec}) of the stringy system
(\ref{0.1bbe})/(\ref{2.5bb}) is expected to fit in
the pattern of the Regge trajectories similar to the one in glueball sector of
the realistic {\it QCD}. In particular, the trajectories are supposed
to contain the glueball states of arbitrary large masses
$m^{2}_{gl}>>\sigma_{ph}\sim{\Lambda^{2}}$. Actually, owing to the weight
pattern (\ref{2.5bb}), the latter condition
requires only that the characteristic 'size' $l_{gl}$ of the considered
states is much larger than the inverse $UV$ cut-off: $l_{gl}\Lambda>>1$.
Therefore, the asserted growth of the masses $m^{2}_{gl}$ is tantamount to the
constraint $l_{gl}\sqrt{\sigma_{ph}}>>1$ that is generically expected for
the highly excited string-like bound states.

Finally, there is {\it no} difficulty to implement the scaling (\ref{0.1eec})
at the level of the semiclassical evaluation (see e.g.
\cite{Fr&Tseyt}) of the Nambu-Goto averages (\ref{0.1bbe})/(\ref{2.5bb})
which is the adequate framework for obtaining a good approximation for
$<W_{C}>$ in the considered regime (\ref{0.1eec}). In particular, this
computation reproduces the value (\ref{0.7bcx}) of the entropy
constant $\zeta_{D}$ entering eqs. (\ref{0.1eec}) and (\ref{2.12bxx}).
To implement the regime (\ref{0.1eec}) on the full-fledged quantum level, the
problem arises because any direct 'brute force' {\it UV}
cut off $\Lambda$ (for the transverse string fluctuations)
by itself breaks the continuous symmetry under the worldsheet's
reparametrizations (\ref{0.4zh}). In consequence, thus introduced cut off
would produce spurious violation of the residual worldsheet's conformal
invariance (associated to the famous conformal anomaly \cite{GFS}).

Upon a reflection, the known practical prescriptions \cite{KPZ,DDK}, to
circumvent an explicit introduction of the $UV$ cut off, work only in the
conventional regime (\ref{0.7bcb}). Formally, the way out could be to
introduce the proper counterterms to restore the conformal invariance.
As this is rather unwieldy to perform explicitly, the better
alternative is to make up for
an {\it effective} $UV$ cut off defined in a conformally invariant manner.
(In particular, the manifest conformal symmetry of the regularized theory is
crucial for the determination of the coefficient, proportional to the
{\it central charge} associated to the conformal group, in
front of the so-called Luscher term \cite{Luscher}.)
To this aim, the general idea is suggested by the concept of the 
Pauli-Villars regularization. We are to design such a renormalizable string
theory, endowed with the new $UV$ cut off $\bar{\Lambda}$ (to be sent to
infinity), that has the following property. Approaching the Nambu-Goto
system (\ref{0.1bbe})/(\ref{2.5bb}) at the momentum scales less than
a fixed one $\Lambda=O(\sqrt{\sigma_{ph}})$, the theory in question should
{\it not} support propagating stringy degrees of freedom at the
momentum scales much larger than $\Lambda$. As for the 
scale $\Lambda^{2}$, being fixed compared to the physical string tension
$\sigma_{ph}$, it is constrained to vanish (i.e.
$\sigma_{ph}/\bar{\Lambda}^{2}\rightarrow{0}$) in the units of
$\bar{\Lambda}^{2}$.

\section{Including nontrivial selfintersections of the contour.}

Now, we are in a position to explicitly justify one of the major results of
the paper. To formulate the latter result, consider the subspace
$\Upsilon'$ (of the full loop space $\Upsilon$) which is comprised of
{\it macroscopic} loops (defined prior to eq. (\ref{0.1eel})) with at most
{\it point-like} selfintersections, i.e. {\it without}
selfintersections (e.g. backtrackings) on $1d$ submanifolds.
There is such a regularization of the Loop equation (\ref{1.11b}) that yields,
on the subspace\footnote{As it is clear from eq. (\ref{2.5bxl}), in the
presence of macroscopic $1d$
selfintersections (distinct from backtrackings) of
the loops involved, the Ansatz (\ref{0.1bbe})/(\ref{0.1}) properly represents
the infrared regime (\ref{0.1eea})/(\ref{0.1eec}) of the large $1/N$
$SC$ expansion (in the $YM_{D}$ theory (\ref{1.1})) only as far as the
area-dependent contributions are concerned. Also,
unless the conditions (\ref{0.1eea})/(\ref{0.1eec}) are satisfied,
the factorized form of the r.h. side of eq. (\ref{1.11b}) is in conflict with
quasi-local pattern (\ref{0.1}) implying the observability of 
the $\varrho[\tilde{M}(\{C_{k}\})]\neq{1}$ abelian factor (\ref{2.5bxb}).}
$\Upsilon'$, the solution in the form of the Ansatz
(\ref{0.1bbe})/(\ref{0.1}), provided the conditions
(\ref{0.1eea})/(\ref{0.1eec}) and certain additional constraint
(\ref{1.3zp}) (on the smearing function (\ref{0.1bb})) are satisfied.

The proof of the latter statement is built on the following idea.
To begin with, it is clear that the Ansatz (\ref{0.1bbe})/(\ref{0.1})
satisfies the (regularized variant of) the {\it linear} loop equation
(\ref{0.9za}) irrespectively whether or not the contour $C$ selfintersects.
Thus, the problem is to find such a regularization of the r.h. side of the
nonlinear Loop equation (\ref{1.11b}) so that, for a generic point-like
selfintersection, the latter can be reduced in the regime
(\ref{0.1eea})/(\ref{0.1eec}) to the regularized r.h. side of (the
$N\rightarrow{\infty}$ limit of) eq.
(\ref{0.9za}) on the entire subspace $\Upsilon'$. This is indeed possible to
accomplish owing to certain useful identity (\ref{0.9zz}).

\subsection{The problem to circumvent.}

Before we implement this strategy, it is instructive to reveal certain
important problem which resolution will require to impose the constraint
(\ref{0.1eec}) on $\sigma_{ph}$. For this purpose, it is sufficient to
simplify the discussion and restrict our attention to the {\it simple}
point-like selfintersections of $C$ where only {\it two} line-segments of the
boundary intersect. Then, take a
loop $C\in{\Upsilon'}$ with a simple nontrivial selfintersection at
${\bf x}(s)={{\bf y}(s')},~s'\neq{s}$. In this case, the r.h. side of eq.
(\ref{0.9za}) obtains {\it two} different nonvanishing contributions
both associated to ${\bf x}(s)$.
The first one, ${\bf x}(s)={{\bf x}(\bar{s})},~\bar{s}\rightarrow{s}$, as
previously corresponds to the trivial selfintersection that properly 'goes
through' the Loop equation (\ref{1.11b}). The apparent mismatch, between
the considered Ansatz and the {\it unregularized} pattern of eq.
(\ref{1.11b}), arises from the second term due to the additional contribution
of the nontrivial selfintersection at ${\bf x}(s)={{\bf y}(s')},~s'\neq{s}$.
For this term, the ratio of the right and the left hand sides of eq.
(\ref{1.11b}) is given by the factor
\be
\tau(C_{xx}|{\bf y})=\lim_{N\rightarrow{\infty}}
\frac{<W_{C_{xy}}><W_{C_{yx}}>}{<W_{C_{xx}}>}~,
\label{0.9aa}
\ee
where the averages are to be evaluated through the Ansatz
(\ref{0.1bbe})/(\ref{0.1}). The problem, which in the regime
(\ref{0.1eea})/(\ref{0.1eec}) will be circumvented by the appropriate
regularization of eq. (\ref{1.11b}), lies in the fact that the $D\geq{3}$
factor (\ref{0.9aa}) is generically {\it not} equal to unity.
The reason is twofold. First, even in the strictly local limit (\ref{0.3cb})
(when the stringy sum (\ref{0.1bbe}) is invariant under the substitution of
the weight (\ref{0.1}) by the $m_{0}=0,~\bar{\lambda}(\lambda)=\xi\lambda$
option of the Nambu-Goto weight (\ref{2.5bb})), the constraint
$\tau(C_{xx}|{\bf y})=1$ can {\it not} be generically satisfied if $D\geq{3}$.
Additionally, if the flux-tube width (\ref{0.3csb}) is not infinitesimal
compared to the amplitude of the fluctuations, the previously discussed
quasi-contact interactions (of the elementary $YM$ vortices) also
contribute, in the {\it abelian} manner, into the deviation of
$\tau(C_{xx}|{\bf y})$ from unity for $D\geq{2}$. Altogether,
the Ansatz (\ref{0.1bbe})/(\ref{0.1}) runs in conflict with the factorized
structure of the r.h. side of the $D\geq{3}$ Loop equation (\ref{1.11b}),
{\it prior} to the regularization of the latter.

In the standard regime (\ref{0.7bcb}), within the $1/N$ expansion, the
considered mismatch can be shown to persist even {\it after} a generic
regularization of the Loop equation (\ref{1.11b}) so that the considered
Ansatz is {\it not} consistent with the latter equation. The situation
improves drastically when the unconventional scaling (\ref{0.1eec}) of
$\sigma_{ph}$ is combined with the conditions (\ref{0.1eea}) which, in
particular, ensure the essential unobservability of the quasi-contact
interactions. Then, there exists an admissible regularization of eq.
(\ref{1.11b}) that reconciles the latter equation with the residual deviation
of the factor (\ref{0.9aa}) from unity.

\subsection{The proper regularization of the Loop equation.}

Let us proceed with a few generalities which are
relevant for the construction of such gauge-invariant regularization of
(the r.h. side of) the  Loop equation (\ref{1.11b}) that is consistent
with the Ansatz (\ref{0.1bbe})/(\ref{0.1}) in the regime
(\ref{0.1eea})/(\ref{0.1eec}). To begin with, in the presence
of a nontrivial selfintersection of a given contour $C$, the smearing-trick
(\ref{0.1bb}) can not be directly applied because of the two reasons:
due to the deviation of the factor (\ref{0.9aa}) from unity {\it and} the
nonlinearity of eq. (\ref{1.11b}). Complementary, the standard prescription,
to split the point of the nontrivial selfintersection, results formally in the
Wilson averages of {\it open} paths $<W_{C_{xy}}>_{\infty},
<W_{C_{yx}}>_{\infty}$ (with ${\bf x}\neq{\bf y}$) which would
be in conflict with the manifest gauge invariance. To properly
combine the smearing (\ref{0.1bb}) with the point-splitting, the general idea
is rather straightforward (see e.g. \cite{MigdRep}). In addition to the
ordered exponents along the open paths $C_{xy},C_{yx}$, the proper
regularization of eq. (\ref{1.11b}) should
include the ordered exponents
\be
{\mathcal{B}}(\Gamma_{xy})={\mathcal{P}}~e^{i\oint_{\Gamma_{xy}} dz^{\mu}
A^{a}_{\mu}(z)T^{a}}~~~~~~;~~~~~~
{\mathcal{B}}(\Gamma_{yx})=({\mathcal{B}}(\Gamma_{xy}))^{-1}~,
\label{1.3zj}
\ee
along two auxiliary (nonselfintersecting) paths
$\Gamma_{yx},~\Gamma_{xy}=\Gamma^{-1}_{yx}$ connecting ${\bf x}(s)$ and
${\bf y}(s')$. In consequence, the r.h. side of the regularized
eq. (\ref{1.11b}) can be expressed as some linear functional (to be
identified with r.h. side of eq. (\ref{0.9zz})) of the product
$<W_{C_{xy}\Gamma_{yx}}>_{\infty}<W_{C_{yx}\Gamma_{xy}}>_{\infty}$ that
supersedes $<W_{C_{xy}}><W_{C_{yx}}>$ in the numerator of the ratio
(\ref{0.9aa}).

Next, one of the advantages of the regime (\ref{0.1eea})/(\ref{0.1eec})
is that, dealing with any particular {\it macroscopic}\footnote{The definition
of macroscopic contours is given prior to eq. (\ref{0.1eel}).} contour $C$,
we can consistently distinguish between the trivial and nontrivial
selfintersections of the latter. The point is that, in the considered regime,
it is instructive first to design the
two somewhat distinct regularization prescriptions for the latter two types
of the selfintersections and then formulate certain condition (see eq.
(\ref{1.3zp})) that ensures the proper matching between the prescriptions.
As for the former type, it is convenient to retain the simple
regularization (\ref{0.1bb}). To handle nontrivial selfintersections,
our strategy is to make use of the ambiguity in the introduction of the
additional ordered exponents (\ref{1.3zj}): one is
to fix both the particular domain of integration over the positions of the
(open) contours $\Gamma_{xy}$ and the specific weight of the
integration over $\Gamma_{xy}$.

In the $SC$ phase (\ref{0.1eec}), to facilitate the required linearization of
the regularized Loop equation (\ref{1.11b}), this ambiguity is
to be fixed employing the following property of the $b=1,~\chi=1$ functional
integral (\ref{0.1bbe})/(\ref{0.1}). Let the point of a nontrivial
selfintersection (at ${\bf x}(s)={\bf y}(s')$) of a macroscopic contour
$C_{xx}=C_{xy}C_{yx}$ be resolved so that $|{\bf x}(s)-{\bf y}(s')|=
O(\Lambda^{-1})$. The auxiliary identity asserts that, for
any such $C_{xx}=C_{xy}C_{yx}$ satisfying the constraints (\ref{0.1eea}),
there is such effective action $S(\Gamma_{xy}|{\mathcal{G}})$ that
\be
<W_{C_{xx}}>_{\infty}~=
\int\limits_{\Gamma_{xy}\in{{\bf X^{D-1}}}}
{\mathcal{D}}z_{\mu}(t)~e^{-S(\Gamma_{xy}|{\mathcal{G}})}
<W_{C_{xy}\Gamma_{yx}}>_{\infty}<W_{C_{yx}\Gamma_{xy}}>_{\infty}~,
\label{0.9zz}
\ee
where the functional integral runs over the (open, when
${\bf x}(s)\neq{{\bf y}(s')}$) paths $\Gamma_{xy}=\Gamma^{-1}_{yx}$.
Presumed to be {\it nonselfintersecting} in its interior, each path is
parametrized, 
\be
\Gamma_{xy}\in{{\bf X^{D-1}}}~:~~~\{~z_{\mu}(t)~~|~~
z_{\mu}(0)=x_{\mu}(s)~,~z_{\mu}(1)=y_{\mu}(s')~\}~,
\label{0.9zza}
\ee
by the trajectory $z_{\mu}(t)$ embedded into some $D-1$ dimensional
subspace\footnote{The precise choice of ${\bf X^{D-1}}$, that fixes a part of
the regularization freedom, will not be necessary for our present
purposes.} ${\bf X^{D-1}}$ (diffeomorphic to ${\bf R^{D-1}}$) of the
base-space ${\bf R^{D}}$.

To explain the meaning of the relation (\ref{0.9zz}), let us first explain
the pattern of $z_{\mu}(t)$. For this purpose, one is to isolate the
cross-section $\tilde{M}(C)\cap {\bf X^{D-1}}$ of a particular
worldsheet $\tilde{M}(C)$ (entering the measure (\ref{0.1bbe})) with the
subspace ${\bf X^{D-1}}$ containing any two given points ${\bf x}(s)$ and
${\bf y}(s')$ of the macroscopic boundary loop $C\equiv{C_{xx}}$. Then, for
the dense subvariety of $\tilde{M}(C)\in{{\mathcal{I}}(M,{\bf R^{D}})}$, one
can uniquely separate a nonselfintersecting
contour\footnote{Moreover, the interior of the path $\Gamma_{xy}$ generically
does {\it not} intersect the contours $C_{xy}$ and $C_{yx}$.}, to be
identified with $z_{\mu}(t)$, from the connected component of the resulting
cross-section. As a result, in compliance with the pattern of the relation
(\ref{0.9zz}), in the regime (\ref{0.1eea}) we are to identify
$S(\Gamma_{xy}|{\mathcal{G}})$ with the {\it effective action} for the
cross-section paths (\ref{0.9zza}). This action is to be
deduced (see Appendix D.3 for more comments) from the full pattern
(\ref{0.1bbe})/(\ref{0.1}) via the {\it partial} integration over all the
worldsheets $\tilde{M}(C_{xy}\Gamma_{yx}), \tilde{M}(C_{yx}\Gamma_{xy})$
spaned by the fixed (for any given $\Gamma_{xy}$) loops $C_{xy}\Gamma_{yx}$
and $C_{yx}\Gamma_{xy}$.

Let us now utilize the identity (\ref{0.9zz}) as a building block of the
appropriate regularization scheme for the r.h. side of the Loop equation
(\ref{1.11b}). First of all, as it is clear from the structure of the Ansatz
(\ref{0.1bbe})/(\ref{0.1}), the following consequence of the unconventional
$\Lambda^{2}$-scaling (\ref{0.1eec}) of $\sigma_{ph}$ is crucial for our
purposes. Namely, once $|{\bf x}(s)-{\bf y}(s')|$ is microscopic, the subspace
${\bf X^{D-1}}$ can be chosen in such a way that {\it the
characteristic length $L[\Gamma_{xy}]$ of thus defined
contour $z_{\mu}(t)$ is {\it microscopic} as well}:
\be
|{\bf x}(s)-{\bf y}(s')|~\leq~{\Lambda^{-1}}~~~~\Longrightarrow~~~~{
<L[\Gamma_{xy}]>~=~<\int_{0}^{1} dt \sqrt{\dot{z}^{2}_{\mu}(t)}}>~
\sim~{\Lambda^{-1}}~.
\label{0.9czl}
\ee
Turning to the construction of the regularization (of eq. (\ref{1.11b}))
consistent with the Ansatz (\ref{0.1bbe})/(\ref{0.1}), we start by
acting on the average $<W_{C_{xx}}>_{\infty}$ by the Loop operator
$\hat{\mathcal{L}}_{\nu}$. The result can be represented as the linear
functional of $<W_{C_{xx}}>_{\infty}$ given by the ($N\rightarrow{\infty}$
limit of the) r.h. side of eq.
(\ref{0.9za}) {\it regularized} according to eq. (\ref{0.1bb}).
Next, we consider the simplest case of a contour $C_{xx}$
with a single\footnote{Multiple simple point-like selfintersections can be
treated through the straightforward generalization of the considered
algorithm. As for nonsimple point-like selfintersections (i.e. when more than two
line-segments intersect at a given point), they can be treated as the limiting
case of the simple ones.} simple point-like selfintersection at ${\bf x}(s)$.
Then, for any given pair
${\bf x}(s),~{\bf y}(s')$ of the points entering the regularized r.h. side of
 eq. (\ref{0.9za}), one is to transform $<W_{C_{xx}}>_{\infty}$ in
compliance with the identity (\ref{0.9zz}). In sum, the Ansatz
(\ref{0.1bbe})/(\ref{0.1}) satisfies (for macroscopic loops) the regularized
variant of the Loop equation so that the r.h. side
of eq. (\ref{1.11b}) is traded for
\be
\tilde{g}^{2}\oint\limits_{C_{xx}} dy_{\nu}(s')~
\Lambda^{D}{\mathcal{G}}(\Lambda^{2}({\bf x}(s)-{\bf y}(s'))^{2})
\int\limits_{\Gamma_{xy}} {\mathcal{D}}z_{\mu}~
e^{-S(\Gamma_{xy}|{\mathcal{G}})}
<W_{C_{xy}\Gamma_{yx}}>_{\infty}<W_{C_{yx}\Gamma_{xy}}>_{\infty}
\label{0.9zze}
\ee
which synthesizes eqs. (\ref{0.9zz}) and (\ref{0.1bb}). Note also that, as it
is clear from the smearing pattern of eq. (\ref{0.9zze}), the contours with
(point-like) selfintersection at ${\bf x}(s)={\bf y}(s')$ have to be
considered on equal footing together with nonselfintersecting contours
$C_{xx}=C_{xy}C_{yx}$ with $|{\bf x}(s)-{{\bf y}(s')}|\leq{\Lambda^{-1}}$.

Next, the matching between the prescriptions (\ref{0.1bb}) and
(\ref{0.9zze}) (applied respectively to the trivial and nontrivial
selfintersections of $C$) requires to ensure that the integrand in eq.
(\ref{0.9zze}) respects certain normalization condition inherited from
the normalization of the $\delta_{D}$-function in the r.h. side of eq.
(\ref{1.11b}). As a result, complementary to eq. (\ref{0.1bb}), one is to
impose the constraint
\be
{\mathcal{Y}[{\mathcal{G}}]}=\int d^{D}z~\frac{\delta_{ab}}{N^{2}}
<{\bf 0}|~\hat{\mathcal{E}}^{ab}[0|{\mathcal{G}}]~|{\bf z}>=1~,
\label{1.3zp}
\ee
where the operator $\hat{\mathcal{E}}^{ab}[0|{\mathcal{G}}]$, which
regularizes $\delta^{ab}\hat{1}$, is to be introduced (see eq. (\ref{1.3zb})
of Appendix D) rewriting eq. (\ref{0.9zze}) in the form
\cite{MigdRep,MakRev} conventional for the weak-coupling analysis.

Finally, so far we have been able to derive that, in the {\it SC} phase
(\ref{0.1eea})/(\ref{0.1eec}), the solution of the regularized Loop equation is consistent
with the stringy representation (\ref{0.1bbe})/(\ref{0.1}) for the $b=1,~h=0$
option of the 't Hooft factor $N^{2-2h-b}$. From the pattern of the latter
factor, it is manifest that, to justify it for $\forall{h}$ and $\forall{b}$,
one is to consider the entire Dyson-Schwinger chain of the loop equations
within the the $1/N$ expansion. This is done\footnote{The
conclusions of Appendix D are to be compared with the general expectation
(see e.g. \cite{MigdRep}) that if a free (i.e. of genus zero) stringy ansatz
satisfies the Loop equation (\ref{1.11b}) then the same interacting string
complies, within the $1/N$ expansion, with the full Dyson-Schwinger chain of
the loop equations.} in Appendix D where the validity of the $N^{\chi}$-factor
is shown to be predetermined by the topological relations (\ref{0.9zzi})
which formalize the {\it coupling} between different
worldsheet's topologies taking place in the
presence of a nontrivial (self)intersection of the loops.
In particular, the consistency of the
$N^{\chi}$-dependence of the weight (\ref{0.1}) with the chain of the
regularized $U(N)$ loop equations (\ref{1.11bz}) requires to impose the
following restriction (which strengthens the selection of the macroscopic
contours $C$ without $1d$ selfintersections that is required in the case of
the $N\rightarrow{\infty}$ eq. (\ref{1.11b})). For $b\geq{2}$ and large
$N\neq{\infty}$, both the characteristic size (understood in the sense of
eq. (\ref{0.1eel})) of each loop, presumed to be devoid of $1d$
selfintersections, and the minimal distance between any two different contours
(entering eq. (\ref{1.11bz})) should be of order of $N^{\alpha}$ with some
$\alpha>0$. This restriction is aimed to
suppress as $e^{-\beta N}$ (with some $\beta>0$) all kinds of the quasi-local
interactions (between the elementary flux-tubes) which abelian pattern,
encoded in eq. (\ref{0.1}), is in conflict with the $N\geq{2}$ $U(N)$
Dyson-Schwinger chain.

Let us also note
that, in view of the discussion in \cite{Dub3}, the $SU(N)$ counterpart of the
weight (\ref{0.1}) is deduced through the substitution
\be
U(N)~\longrightarrow{~SU(N)}~~~~~\Longrightarrow{~~~~~
\lambda_{U(N)}~\longrightarrow{~(1-1/N^{2})\cdot\lambda_{SU(N)}}}
\label{0.1xaa}
\ee
(where $\lambda_{SU(N)}=(g_{SU(N)})^{2}N\Lambda^{D-4}$) that can be
justified directly from the $SU(N)$ loop equations.

\section{Conclusions.}

To match with the regularization of the $YM_{D}$ theory (\ref{1.1}), we
introduce the smearing regularizations of the 'bare' Gauge String weight
(\ref{0.5c}) that allows to work with the {\it fat} $YM$ vortices of a
nonzero width $\sim{\Lambda^{-1}}$, where $\Lambda$ is the $UV$ cut off.
Given the appropriate smearing, the proposed in \cite{Dub3} stringy
construction is shown to provide with the {\it confining} solution
of the Dyson-Schwinger chain of the loop equations
in the large $N$ strong coupling phase (\ref{1.1bx}) of the
regularized\footnote{In particular, we resolve
the long standing problem of finding such class of the regularizations of the
loop equations that can be explicitly translated into the tractable
regularizations of the stringy solution of the latter equations. It is
noteworthy that, due to the constraint (\ref{0.1bb}), there is a variety of
the regularizations (including the 'most natural' one (\ref{0.3cb}) resulting
in the {\it infinitely thin} $YM$ vortex) of the Gauge String weight which do
{\it not} correspond to any regularization of the dual
$YM_{D}$ theory (\ref{1.1}).} $D\geq{3}$ $U(N)$ gauge theory (\ref{1.1}).
Also, the concise prescription (formalized by eq. (\ref{0.4ze})) to implement the
backtracking invariance of the average $<W_{C}>$ is found.
Complementary, with the help of the vocabulary \cite{MigdRep} that relates
the gauge theory's and the loop-space's operators, the proposal of \cite{Dub3}
can be strengthened further to present (at least at the formal level)
the algorithm to reconstruct the 'image' of the field-theoretic
$F_{\mu\nu}({\bf x})$-correlators (in the {\it regularized} $YM_{D}$ theory)
on the stringy side.

The land-mark of the considered $SC$ phase\footnote{This phase can be
reinterpreted as the continuum counterpart (maintaining the manifest $O(D)$
invariance of the Euclidean space ${\bf R^{D}}$) of the well-known $SC$
regime \cite{Wils} in the {\it lattice} gauge theories.} is that the
{\it infrared stable} effective excitations, implementing the $1/N$ $SC$
expansion of the regularized gauge system (\ref{1.1}), are the {\it fat
microscopic} flux-tubes rather than point-like gluonic excitations. As the
semiclassical analysis of the string fluctuations (leading, in particular, to the estimate
(\ref{0.7bxx}) of the entropy constant $\xi_{D}$) is supposed to correctly
reproduce the major qualitative features of the dynamics, the mechanism of
confinement in certain sense 'trivializes' within the proposed framework.
(More precisely, as the gluonic
excitations are {\it not} adequate to describe the large distance physics, the
nontriviality is hidden in the formal resummation relating the $1/N$ $WC$
and the $1/N$ $SC$ series in the $YM_{D}$ theory (\ref{1.1}).) Indeed, within
the $SC$ series ({\it avoiding} the spurious infrared divergences characteristic for
the individual Feynman graphs of the $WC$ expansion), the area-law
asymptotics of the average $<W_{C}>$ emerges as naturally as the ordinary
Coulomb interaction (between two heavy electric charges) appears in the
standard $QED$.

Next, the duality-correspondence (\ref{0.5cxd}) implies
that, in addition to the selfenergy (described, in the infrared domain, by
the Nambu-Goto pattern (\ref{2.5bb})), the worldsheet's weight encodes
certain (quasi-)contact interactions between the (self)overlapping elementary
vortices assigned with the unit $YM$-flux. In the leading order
(reproduced by eq. (\ref{0.5c})) of the derivative expansion (\ref{0.3bbq}),
these short-range interactions depend only on the choice of the gauge group
and the lagrangian of the dual gauge theory, while in the subleading
orders the dependence on the details of the regularization takes place. 
Nevertheless, we demonstrate that there are two dynamical regimes where the
pattern of the Wilson loop-averages depends on the regularization prescription
only through the dependence of a few relevant coupling constants on the bare
$YM_{D}$ coupling (\ref{1.1bx}).

The characteristic feature of the first regime (\ref{0.1eea})/(\ref{0.1eec})
is that the flux-tube width is much smaller than the amplitude of the
string fluctuations. In consequence,
for the loops without $1d$ (self)intersections, it entails the infrared
unobservability of the quasi-contact interactions
so that (for a large variety of the actions of the dual $U(N)$ gauge theory)
the corresponding implementation of the $U(N)$ Gauge String is in the same
universality class with the Nambu-Goto theory (\ref{0.1bbe})/(\ref{2.5bb}).
(Furthermore, we sort out those observables of the Nambu-Goto system
which can be mapped onto the properly associated observables in the dual
gauge theory (\ref{1.1}).) The important novel feature arises due to the
$\Lambda^{2}$-scaling (\ref{0.1eec}) of $\sigma_{ph}$ which, in the $SC$
phase (\ref{1.1bx}), is mandatory
at least for sufficiently large $N$. As a result, the stringy system
(\ref{0.1bbe})/(\ref{2.5bb}) is supposed to avoid the collapse into the
branched polymer phase. As for the formalism itself, in the considered
regime the Loop equation (\ref{1.11b}) possesses, for nonselfintersecting
loops, the particular solution (\ref{0.1bbe})/(\ref{0.1}) which satisfies the
simple eq. (\ref{1.11bb}). On the ${\bf \Psi}$-space of the area-functionals
(\ref{1.9adc}), the general solution (\ref{0.1bbe})/(\ref{1.9add}) of eq.
(\ref{0.9za}) is also found and reinterpreted as a specific member of the
'confining strings' family \cite{PolyakCS}.

On the contrary, in the extreme $SC$ limit (\ref{0.1eed}), the vortex width
is much larger than the amplitude of the string fluctuations. Therefore,
the full set of the loop-averages does encode the data of the quasi-contact
interactions and, consequently, reflects the choice of the particular action
for the associated $U(N)$ gauge theory (\ref{1.1}) in question. Due to the
relevant condition (\ref{0.1eeb}), the details of
the regularization are irrelevant, and the results are represented via the
universal local weight (\ref{0.5c}) assigned to the minimal area worldsheets.
It is observable, for example, in the
asymptotic Casimir scaling (\ref{2.12bb}) of $\sigma_{ph}(R)$ for any given
(anti)chiral $U(N)$ representation $R$. More generally, in the leading order
of the short-distance expansion (\ref{0.3bbq}), the $SC$ asymptotics of the
loop-averages in
the $D\geq{3}$ $YM_{D}$ theory (\ref{1.1}) is reduced to the one in the
l.h. side of eq. (\ref{0.5cxd}) evaluated in the corresponding
{\it dimensionally} reduced $D=2$ $YM_{2}$ system (\ref{1.1})/(\ref{0.4bc})
defined on the support $T$ of the minimal area surfaces. It is
noteworthy that, starting with the regime (\ref{0.1eed}) and than further
increasing $A_{min}(C)\Lambda^{2}$ (or, alternatively, decreasing
the value of $\lambda$), the asymptotic $C_{2}(R)$-scaling (\ref{2.12bb}) of
$\sigma_{ph}(R)$ is superseded by the $|n(R)|$-scaling (\ref{2.12bxx}).
                       
Actually, the previously discussed solution
(\ref{0.1bbe})/(\ref{0.1}) has the twofold utilization in the regime
(\ref{0.1eed}) as well. For $N=1$, the considered stringy sum
presents the general solution of the loop equations in the $SC$ phase of the
regularized $D\geq{2}$ $U(1)$ gauge theory (\ref{1.1bb}) (enriched with
the monopoles in $D\geq{3}$). In fact, this is valid {\it irrespectively} of
the ratio between the vortex width and the amplitude of the fluctuations.
As for $N\geq{2}$, the pattern (\ref{0.1bbe})/(\ref{1.9add}) properly
reproduces, (being applied to the saddle-point worldsheet
$\tilde{M}_{min}(\{C_{k}\})$, the extreme $SC$ asymptotics (\ref{0.1eed}) of
$<\prod_{k}W_{C_{k}}>$ when the minimal-area flux-tube is represented by the
elementary $YM$ vortices (of unit flux) with unobservable\footnote{For our
present purposes, it is sufficient to require the unobservablity within the
short-distance expansion (\ref{0.3bbq}) constrained to be at least
asymptotically convergent.} {\it (self)overlapping}.

Finally, the ultimate goal of our project is to develope an approach which
would make accessible the analysis of the low-energy phenomena in the $D=4$
$YM_{4}$ theory (\ref{1.1}) considered in the standard $WC$ phase with the
vanishing (according to the perturbative renormgroup flow) bare coupling
constant $\lambda\rightarrow{0}$. Unfortunately, the direct continuation of
the proposed stringy solution (of eq. (\ref{1.11b})) into the $WC$ phase,
being available, describes the metastable {\it microscopic} degrees of
freedom, as it is predetermined by the mandatory $\Lambda^{2}$-scaling
(\ref{0.1eec}) of $\sigma_{ph}$. Nevertheless, more subtle utilization of the
results obtained is expected to be possible in the $WC$ regime as well: the
derived stringy pattern
should be interpreted as describing the prototypes of the {\it macroscopic}
flux-tubes\footnote{In particular, we conjecture that in the $WC$ phase the
physical string tension $\sigma_{ph}(R)$ might exhibit a 'crossover', when
$A_{min}(C)\Lambda^{2}$ is varied, similar to the one interpolating between
the two patterns (\ref{2.12bb}) and (\ref{2.12bxx}).} which are supposed to
ensure that confinement is inherited by the $WC$ phase from the $SC$ one. In
other words, the extended $YM$ vortices are presumed to enter {\it not}
directly the microscopic $\lambda\rightarrow{0}$ gauge system (\ref{1.1})
itself but rather its effective {\it low-energy} theory considered at the
presumed confinement-scale. The latter is to be qualitatively identified with
the mass of the lowest glueball
excitation (that motivates the $\Lambda^{2}$-scaling (\ref{0.1eec}) of
$\sigma_{ph}$). At this scale, the
effective theory is expected to be in the $SC$ phase which, in particular,
manifests itself by the {\it absence} of the essentially logarithmic (with
respect to the $UV$ cut off $\bar{\Lambda}$) {\it Wilsonian} renormgroup flow
of the running coupling constant $\lambda(\bar{\Lambda})$ in front of the
$F^{2}_{\mu\nu}$-operator. (Roughly speaking, it can be reinterpreted as a
kind of {\it freezing} of $\lambda(\bar{\Lambda})$.) Thus, the major problem
is to find a technique which, being able to circumvent the 'brute force'
Wilsonian renormgroup reduction of the $UV$ cut off, should incorporate
the proposed formalism as the building block in the solution of the Loop
equation in the $\lambda\rightarrow{0}$ phase of the $YM_{D}$ theory
(\ref{1.1}). The work in this direction is in progress.

In conclusion, let us also remark that the regime (\ref{0.1eed}) is
reminiscent of the one which independently appeared in another recent
framework \cite{Witt2} (see also \cite{Gr&Ooguri}).
Being based on the conjecture \cite{Maldac} about the so-called
{\it AdS/CFT correspondence}, this approach follows the somewhat
alternative route towards certain stringy description of the pure gauge
theories. It is interesting that, in the extreme $D=4$ {\it SC} regime
$\lambda>>1$, the purported $\lambda\Lambda^{2}$-scaling \cite{Witt2} of the
physical string tension is consistent with the prediction \cite{Dub3} of our
formalism. Accidentally, the large $N$ asymptotics of the loop-averages
(entering the r.h. side of eq. (\ref{0.5bxx})) in the $D=2$ $YM_{2}$ theory
(\ref{1.1}) is discussed in \cite{Gr&Ooguri} as providing with the tentative
pattern of the $D=4$ averages which may arise from the $AdS/CFT$ duality.
Also, let us note the preliminary results
\cite{LE/SUSY} aimed at the justification of the latter correspondence
directly from a variant of the Loop equation\footnote{For
completeness, we remark that the formalism of the loop equations has been
recently applied, see \cite{LE/NCYM}, in the context of the noncommutative
gauge theories.} arising in the ${\mathcal{N}}=4$ $SUSY$ $YM_{4}$ theory.
One might expect that a deeper understanding of the interrelation between
the two approaches could pay back to both of them.

\begin{center}
{\bf Acknowledgements.}
\end{center}

The author is grateful to Yu.Makeenko for a discussion at the final stage of
the work. This project is partially supported by CRDF grant RP1-2108.

\app{The short-distance expansion of the weight (\ref{0.1}).}

The required operator expansion (\ref{0.3bbq}) of the (exponent of the
worldsheet's) weight (\ref{0.1}) is to be performed similarly to the more
conventional multipole (or derivative) expansions. First, we introduce the
convenient parametrization in terms of
\be
\gamma_{-}=(\gamma-\gamma')~~~,~~~\gamma_{+}=y\gamma+(1-y)\gamma'~~~~~;~~~~~
d^{2}\gamma d^{2}\gamma'=d^{2}\gamma_{+}d^{2}\gamma_{-}~,
\label{A.3b}
\ee
where $\gamma,\gamma'$ parametrize respectively the worldsheet's coordinates
${\bf x}(\gamma)$ and ${\bf y}(\gamma')\equiv{{\bf x}(\gamma')}$ (entering
eq. (\ref{0.1})), while the auxiliary parameter $0\leq{y}\leq{1}$ reflects
the ambiguity in the choice of
the 'center of mass' coordinate $\gamma_{+}$. Then, for any given
$\gamma_{+}$, one is to arrange for the Taylor expansion (of all
the structures involved into the bilocal exponent of eq.
(\ref{0.1})) in terms of the relative coordinate $\gamma_{-}$. Finally,
integrating out $\gamma_{-}$, the resulting 'effective action' is formulated
as the expansion (\ref{0.3bbq}) in terms of the integrals of the {\it local}
'scalar' operators ${\mathcal{Q}}^{(k)}_{2p}(\gamma_{+})$ which are composed
of the $\gamma_{+}$-dependent tensors of the worldsheet's curvature
(and torsion).

Before we briefly discuss the general pattern (\ref{0.3bbq}), let us first
apply this algorithm to derive the leading and the relevant next-to-leading
terms of the considered expansion.
Concerning the leading term (yielding the local limit (\ref{2.5bxa}) of
(\ref{0.1})) given by eq. (\ref{2.5bxa}), it is deduced as follows.
Keeping the leading $(\gamma_{-})^{0}$-contribution in
\be
p_{\mu\nu}(\gamma)=p_{\mu\nu}(\gamma_{+})+(1-y)\cdot
(\partial_{a}p_{\mu\nu}(\gamma_{+}))\gamma^{a}_{-}+
\frac{(1-y)^{2}}{2}(\partial_{a}\partial_{b}p_{\mu\nu}(\gamma_{+}))
\gamma^{a}_{-}\gamma^{b}_{-}+..~,
\label{A.3c}
\ee
\be
p_{\mu\nu}(\gamma')=p_{\mu\nu}(\gamma_{+})-y\cdot
(\partial_{a}p_{\mu\nu}(\gamma_{+}))\gamma^{a}_{-}+
\frac{y^{2}}{2}(\partial_{a}\partial_{b}p_{\mu\nu}(\gamma_{+}))
\gamma^{a}_{-}\gamma^{b}_{-}+..~,
\label{A.3d}
\ee
(where $p_{\mu\nu}(\gamma)$ is defined by eq. (\ref{0.2c})),
in the argument of the smearing function
\be
({\bf x}(\gamma)-{\bf x}(\gamma'))^{2}=
\left(\partial_{\alpha}{\bf x}(\gamma_{+})\gamma_{-}^{\alpha}+
\frac{(1-2y)}{2}\partial_{\alpha}\partial_{\beta}{\bf x}(\gamma_{+})
\gamma_{-}^{\alpha}\gamma_{-}^{\beta}+O((\gamma_{-})^{3})\right)^{2}
\label{A.3}
\ee
one is to neglect all the $O((\gamma_{-})^{3})$-terms retaining
the quadratic form
\be
{\mathcal{G}}(\Lambda^{2}({\bf x}(\gamma)-{\bf x}(\gamma'))^{2})
~\longrightarrow{~{\mathcal{G}}(\Lambda^{2}
\hat{g}_{\alpha\beta}(\gamma_{+})\gamma_{-}^{\alpha}\gamma_{-}^{\beta})}
\label{A.4}
\ee
which depends on the induced worldsheet's metric
\be
\hat{g}^{\alpha\beta}(\gamma)=
\frac{\partial x_{\mu}}{\partial\gamma_{\alpha}}
\frac{\partial x_{\mu}}{\partial\gamma_{\beta}}~.
\label{0.3c}
\ee
In consequence, one recovers the pattern (\ref{2.5bxa}), with the bare string
tension (\ref{0.1ew}) being given by the product of $\lambda\Lambda^{2}/2$
times the (curved space representation of the) zeroth moment (\ref{0.3bbx})
\be
K_{0}=\int_{\tilde{{\mathcal{P}}}}~d^{2}z~ \sqrt{\hat{g}(\gamma_{+})}~
{\mathcal{G}}(
\hat{g}_{\alpha\beta}(\gamma_{+})z^{\alpha}z^{\beta})=\xi~,
\label{A.4b}
\ee
where $\hat{g}(\gamma_{+})$ is the determinant of the worldsheet metric
(\ref{0.3c}), and $\tilde{{\mathcal{P}}}$ is an arbitrary manifold
diffeomorphic to
the $2d$ plane. In compliance with the l.h. side of eq. (\ref{0.3b}), the
integral (\ref{A.4b}) (being, by construction, {\it independent} of the metric
$\hat{g}_{\alpha\beta}$) is constrained to be equal to $\xi$.

Turning to the next-to-leading terms\footnote{Similar in spirit
computations (employing somewhat different techniques) can found e.g. in
\cite{AES,Orland}.} of the operator expansion, from
the dimensional power-counting, they are given by a combination
\cite{ExtrCurv/P,ExtrCurv/K}
\be
\frac{\lambda}{2}\left(H^{(1)}_{2}\int d^{2}\gamma_{+}
\sqrt{\hat{g}(\gamma_{+})}~
{\mathcal{K}}^{i\alpha}_{\beta}(\gamma_{+})
{\mathcal{K}}^{i\beta}_{\alpha}(\gamma_{+})+
H^{(2)}_{2}\int d^{2}\gamma_{+} \sqrt{\hat{g}(\gamma_{+})}~
R(\gamma_{+})\right)~,
\label{A.1}
\ee
where $R(\gamma_{+})$ is the intrinsic scalar curvature (with respect to the
metric $\hat{g}_{\alpha\beta}(\gamma_{+})$), while
${\mathcal{K}}^{i}_{\alpha\beta}(\gamma_{+})$ is the second fundamental
form introduced via the standard geometrical relation
\be
\partial_{\alpha} \partial_{\beta}{\bf x}=\Gamma^{\sigma}_{\alpha\beta}
\partial_{\sigma}{\bf x}+{\mathcal{K}}^{i}_{\alpha\beta}{\bf n}_{i}~~~~~;~~~~~
({\bf n}_{i},{\bf n}_{j})=\delta_{ij}~~~,~~~
({\bf n}_{i},\partial_{\alpha}{\bf x})=0~.
\label{A.2}
\ee
In eq. (\ref{A.2}), $\Gamma^{\sigma}_{\alpha\beta}(\gamma)$ is the usual
Christoffel symbol associated to $\hat{g}_{\alpha\beta}(\gamma)$, and
${\bf n}_{i}\equiv{{\bf n}_{i}(\gamma)},~i=1,...,D-2$, are the vectors normal
(at the point $\gamma$) to the worldsheet parametrized by eq. (\ref{0.4zh}).

As the second term in eq. (\ref{A.1}) is the full derivative
yielding vanishing contribution in the case of the closed
string, we concentrate on the determination of the coefficient
$H^{(1)}_{2}$ in front of the remaining extrinsic curvature term.
There are the two sources of the contribution in question.
First, keeping only the leading low-energy limit (\ref{A.4}) of the smearing
function, one is to pick up those terms of the expansions
(\ref{A.3c}),(\ref{A.3d}) which result in the contribution
quadratic in $\gamma_{-}$. Employing eq. (\ref{A.2}) and the identity
\be
\int_{\tilde{{\mathcal{P}}}}~d^{2}z~\sqrt{\hat{g}}~
{\mathcal{G}}(\hat{g}_{\alpha\beta}z^{\alpha}z^{\beta})~
z^{a}z^{b}=\frac{K_{2}}{2}~\hat{g}^{ab}~,
\label{A.5}
\ee
one concludes that the corresponding part of $H^{(1)}_{2}$ is given by
$-K_{2}/8$, where $K_{2}$ is defined in eq.
(\ref{0.3bbx}). (In the process of the derivation, we have traded the pair of
cartesian antisymmetric tensors $\varepsilon^{ab}$
(with $\varepsilon^{12}=1$) for their covariant counterparts:
$E^{ab}=\varepsilon^{ab}/\sqrt{\hat{g}}$, so that $E^{ab}E^{cd}
\hat{g}_{ac}=\hat{g}^{bd}$.)

As for the second type of the contribution, retaining the leading
$(\gamma_{-})^{0}$-terms in eqs. (\ref{A.3c}) and (\ref{A.3d}),
from the expansion of the smearing function one is to pick the term
\be
{\mathcal{G}}'(\Lambda^{2}
\hat{g}_{\alpha\beta}(\gamma_{+})\gamma_{-}^{\alpha}\gamma_{-}^{\beta})
\cdot\left(\frac{1-2y}{2}\partial_{a}\partial_{b}{\bf x}(\gamma_{+})
\gamma_{-}^{a}\gamma_{-}^{b}\right)^{2}\Lambda^{2}
\label{A.6}
\ee
and employ the identity (where ${\mathcal{G}}'(t^{2})\equiv{\partial
{\mathcal{G}}(t^{2})/\partial t^{2}}$)
\be
\int_{\tilde{{\mathcal{P}}}}~d^{2}z~\sqrt{\hat{g}}~
{\mathcal{G}}'(\hat{g}_{\alpha\beta}z^{\alpha}z^{\beta})~
z^{a}z^{b}z^{c}z^{d}=-\frac{K_{2}}{4}~(\hat{g}^{ab}\hat{g}^{cd}+
\hat{g}^{ac}\hat{g}^{bd}+\hat{g}^{ad}\hat{g}^{bc})
\label{A.7}
\ee
together with the standard relation $({\mathcal{K}}^{i\alpha}_{\alpha})^{2}=
R+{\mathcal{K}}^{i\alpha}_{\beta}{\mathcal{K}}^{i\beta}_{\alpha}$.
In sum, collecting both of the contributions, one obtains
\be
\frac{\lambda}{2}~H^{(1)}_{2}=-
\lambda~\frac{K_{2}}{16}\cdot\left(1+3(1-2y)^{2}\right)<0~,
\label{A.8}
\ee
where, in the $D=4$ case, $K_{2}$ is fixed as in eq. (\ref{2.5bxd}). In turn,
it implies that,
given the regularization prescription (\ref{2.5bxd}), the positivity of the
expansion (\ref{0.3bbq}) in $D=4$ requires that the factor $\xi$ should be
larger than certain critical value $\xi_{cr}$.

Next, the above computations demonstrate that the generic area-functional
(\ref{1.9adc}) can be expanded according to the pattern of eq. (\ref{0.3bbq}).
In particular, for macroscopic contours, the domain of the integration in the
relevant $n=2p$ integrals (\ref{0.3bbx}) (defining the even moments $K_{2p},~
p\geq{0}$) can be extended up to the
infinite $2d$ plane for the points $\gamma_{+}$ in 
the interior of the worldsheet $\tilde{M}$ (i.e. sufficiently far, in the
units of $\Lambda^{-1}$, from the boundary of $\tilde{M}$). Note that,
in contradistinction to the previous approaches \cite{AES,Orland},
the moments $K_{2p}$
are originally represented by the expressions like
eq. (\ref{A.4b}) or (\ref{A.7}) which are manifestly $2d$ generally-covariant.
By construction, the latter integrals do {\it not} depend on
the choice of the metric $\hat{g}_{\alpha\beta}$ that, in eq. (\ref{0.3bbx}),
is restricted to be {\it flat}: $\hat{g}_{\alpha\beta}
\rightarrow{\delta_{\alpha\beta}}$.

\sapp{The leading boundary term.}

Finally, the part of the short-distance expansion of eq. (\ref{0.1}),
associated to the points ${\bf x}(\gamma)$ in the vicinity of the boundary of
the worldsheet $\tilde{M}$, deserve a special treatment. Here, the domain of
the integration in the moments $K_{n}$ can be extended at best up to a $2d$
half-plane rather than to the full $2d$ plane.
In the leading nontrivial order (of the expansion), this discrepancy results
in the boundary contribution\footnote{For macroscopic loops, the
next-to-leading term (\ref{2.5bxl}) is negligible compared to the leading
area-term in the exponent of eq. (\ref{2.5bxa}).} (\ref{2.5bxl}) that is
reduced to the second term of the exponent of the Nambu-Goto weight
(\ref{2.5bb}) when the boundary contours are devoid of the $1d$
selfintersections. To obtain the expression (see eq. (\ref{0.3bbl}) below) for
the effective mass $m_{0}(\lambda)$ entering eq. (\ref{2.5bxl}), the simplest
option is to consider $\tilde{M}(C)$ in the form of a flat rectangle of the
size $2l_{1}\times 2l_{2}$. Then, the exponent of the weight (\ref{0.1})
can be easily evaluated in the form
\be
\frac{\lambda}{2}
\int\limits_{-\infty}^{+\infty}\int\limits_{-\infty}^{+\infty}
\frac{dp_{1}dp_{2}}{(2\pi)^{2}}
~\tilde{\mathcal{G}}_{2}(p_{1}^{2}+p_{2}^{2})
\left(\frac{2sin(p_{1}\bar{l}_{1})}{p_{1}}\right)^{2}
\left(\frac{2sin(p_{2}\bar{l}_{2})}{p_{2}}\right)^{2}~,
\label{0.3bbk}
\ee
where $\bar{l}_{1}={l}_{1}\Lambda$, $\bar{l}_{2}={l}_{2}\Lambda$, and
$\tilde{\mathcal{G}}_{2}(p_{a}p^{a})$ is the two-dimensional Fourier image
(on a $2d$ plane) of ${\mathcal{G}}(z_{a}z^{a})$ which is presumed to be a
meramorphic function of $t^{2}=p_{a}p^{a}$. Assuming that
$\tilde{\mathcal{G}}_{2}(t^{2})=O(1/t^{4})$
when $|t|^{2}\rightarrow{\infty}$, the straightforward
computation yields\footnote{In the process of the derivation, we have
consistently neglected the exponentially suppressed (for $l_{1},l_{2}>>
\Lambda^{-1}$) terms $\sim{e^{-\mu l_{1}},e^{-\mu l_{2}}}$, where
$\pm i\mu$ (with $-t^{2}=\mu^{2}\sim{\Lambda^{2}}$) are the positions of the
two nearest, to the real axis, poles of $\tilde{\mathcal{G}}_{2}(t^{2})$.} for
the effective perimeter-mass
\be
m_{0}(\lambda)=-\lambda\Lambda~
\oint\limits_{\Gamma_{+}}~\frac{dt}{2\pi}~
\frac{\tilde{\mathcal{G}}_{2}(t^{2})}{(t+i\epsilon)^{2}}~,
\label{0.3bbl}
\ee
where $\epsilon\rightarrow{+0}$, and the contour integral runs along the path
$\Gamma_{+}$ which encircles the upper semiplane of the complex-valued
variable $t$.

In particular, in the local limit (\ref{0.3cb}), we obtain
$\tilde{{\mathcal{G}}}(0)=\xi$, while $\partial^{q}
\tilde{\mathcal{G}}_{2}(t^{2})/\partial t^{2q}\rightarrow{0}$ for $q\geq{1}$
so that one indeed recovers: $m_{0}(\lambda)\rightarrow{0}$.
Let us also note that, being combined with the natural positivity constraint
$\tilde{\mathcal{G}}_{2}(t^{2})>0$ (for $\forall{t^{2}}\geq{0}$), eq.
(\ref{0.3bbl}) implies that $m_{0}(\lambda)<0$.

\app{The zero modes of $\hat{\mathcal{L}}_{\nu}$ and the KR representation.}

Before we analyse the zero-mode equation (\ref{1.9add}), let us first recall
a few relevant details concerning the loop calculus. Recall first the meaning
of the area- and the path-derivatives composed into the Loop operator
(\ref{1.6bf}). As for the area-derivative $\delta/\delta
\sigma_{\mu\nu}({\bf x}(s))$, it is associated to the variation of the base-loop
$C\equiv{C_{xx}}\rightarrow{C_{xx}\tilde{C}_{xx}}$ with
respect to creation of the infinitesimal auxiliary loop $\tilde{C}_{xx}$
attached to $C_{xx}$ at the point ${\bf x}(s)$. By definition, the loop
$\tilde{C}_{xx}$, being projected onto the $\mu\nu$-plane, spans the 
surface-element of the (minimal) area (\ref{0.2c}). As for the
path-derivative $\partial_{\mu}^{{\bf x}(s)}\equiv{\delta/\delta x_{\mu}(s)}$,
it is associated to the variation \cite{MigdRep} of the functional (resulting
after the application of $\delta/\delta \sigma_{\mu\nu}({\bf x}(s))$ to
$<W_{C}>$)
with respect to the infinitesimal deformation $C_{xx}\rightarrow
{\Gamma_{zx}C_{xx}\Gamma_{xz}}$ (with $\Gamma_{xz}=\Gamma^{-1}_{zx}$) of the
contour $C\equiv{C_{xx}}$. Here, the deformed loop
$\Gamma_{zx}C_{xx}\Gamma_{xz}$ is obtained from $C_{xx}$ creating, at
${\bf x}(s)$, a backtracking appendix to ${\bf z}$ along the path
$\Gamma_{xz}$ so that $z_{\mu}-x_{\mu}(s) \equiv{\delta x_{\mu}(s)}
\rightarrow{0}$ for $\forall{\mu}=1,...,D$.

In sum, on the ${\bf \Psi}$-space (\ref{1.9adc}), one easily deduces
(employing, in particular, eq. (\ref{1.9adb})) that the action of the Loop operator
$\hat{\mathcal{L}}_{\nu}({\bf x}(s))$ considerably simplifies and can be
represented by the corresponding differential operator
\be
\hat{\mathcal{L}}_{\nu}({\bf x}(s))~
ln\left(w_{2}[\tilde{M}(C)]\right)=
\frac{\partial}{\partial x_{\mu}(s)}~
\frac{\delta_{f}}{\delta p_{\mu\nu}({\bf x}(s))}~
ln\left(w_{2}[\tilde{M}(C)]\right)~,
\label{1.11bxf}
\ee
where $p_{\mu\nu}({\bf x})$ is the area-element (\ref{0.2c}), and
the prescription (\ref{0.4ze}) (introduced in Appendix E to
implement the backtracking invariance of $<W_{C}>$) is implied.

Given eq. (\ref{1.11bxf}), the zero-mode equation (\ref{1.9add}) can be
reformulated on the ${\bf \Psi}$-space as the set of the {\it linear}
equations yielding the corresponding constraints separately for each
$n$th order tensor $\bar{\bf \mathcal{S}}^{(n)}_{\{\mu_{k}\nu_{k}\}}(..)$
entering $-ln(w^{(0)}_{2}[..])$. In the
$n=2$ case, one thus obtains the condition stated (after eq. (\ref{0.1xx}))
in Section 2.3. To properly reinterprete the full set of the constraints,
it is helpful to view (in the spirit of \cite{Kalb&Ramond,PolyakCS}) the set
of the tensor-functions $\bar{\bf \mathcal{S}}^{(n)}_{\{\mu_{k}\nu_{k}\}}(..)$
as the {\it connected}\footnote{This requirement ensures the appropriate
cluster decomposition (akin to the one of \cite{MigdRep}) that is necessary
to maintain the selfconsistency of the gauge
theory represented through the set of the Wilson loop-averages.
This condition can be shown to exclude a variety of the spurious solutions of
the zero-mode equation (\ref{1.9add}).} correlators,
\be
\bar{\bf \mathcal{S}}^{(n)}_{\{\mu_{k}\nu_{k}\}}
(\{{\bf x}_{i}-{\bf x}_{j})\})=
<<B_{\mu_{1}\nu_{1}}({\bf x}_{1})B_{\mu_{2}\nu_{2}}({\bf x}_{2})...
B_{\mu_{n}\nu_{n}}({\bf x}_{n})>>~,
\label{1.6zxa}
\ee
in some theory of the tensor Kalb-Ramond (KR) field $B_{\mu\nu}({\bf x})$
defined in the base-space ${\bf R^{D}}$. Then, taking into account eq.
(\ref{1.11bxf}), the zero-mode equation (\ref{1.9add}) implies that
\be
\partial_{\mu}B_{\mu\nu}({\bf x})=0~.
\label{1.6zxb}
\ee
To visualize this constraint, let us take for simplicity the $D=4$ case when
the
Kalb-Ramond tensor can be canonically decomposed into the so-called exact and
co-exact 2-forms $F_{\mu\nu}$ and $H_{\mu\nu}$,
\be
B_{\mu\nu}=F_{\mu\nu}+\frac{1}{2}\epsilon_{\mu\nu\rho\sigma}
H_{\rho\sigma}~~~~~;~~~~~F_{\mu\nu}({\bf x})=
\partial_{\mu}\wedge A_{\nu}({\bf x})~~,~~H_{\mu\nu}({\bf x})=
\partial_{\mu}\wedge C_{\nu}({\bf x})~,
\label{1.6zxe}
\ee
which are 
conventionally expressed in terms of the normal and dual gauge potentials
$A_{\nu}({\bf x})$ and $C_{\nu}({\bf x})$ respectively so that the ordinary
$U(1)$ gauge symmetry is {\it enlarged} to the product
$U_{e}(1)\otimes U_{m}(1)$ of the 'electric' and 'magnetic' factors. Then, the condition
(\ref{1.6zxb}) (being $C_{\rho}$-{\it independent}) is transformed into
the constraint that $\partial_{\mu}\wedge A_{\nu}({\bf x})$ vanishes for
$\forall{\mu,\nu}$, i.e.
$A_{\nu}({\bf x})=\partial_{\nu}f({\bf x})$ with any $f({\bf x})$. Finally,
to make contact with \cite{PolyakCS}, we note the following. On the
${\bf \Psi}$-space of the area-functionals (\ref{1.9adc}), the general
$D=4$ solution of eq. (\ref{1.11bfb}) can be reproduced (for a given Euler
character $\chi$ of $\tilde{M}_{\chi}(C)$)
in the abelian $KR$ theory with the action
\be
\int\limits_{\bf R^{4}}d^{4}x \left(\frac{1}{4g^{2}}B^{2}_{\mu\nu}({\bf x})+
{\mathcal{W}}(\partial_{\mu}{^{*}B}_{\mu\nu}({\bf x}))\right)+
\frac{i}{2}\int\limits_{\tilde{M}_{\chi}}d\sigma_{\mu\nu}({\bf y})
B_{\mu\nu}({\bf y})~,
\label{1.6zxc}
\ee
where $d\sigma_{\mu\nu}({\bf x})$ is given by
eq. (\ref{0.2c}), while ${\mathcal{W}}(..)$ is an
arbitrary\footnote{To interprete this ambiguity, one is to observe that, in
any $U_{e}(1)\otimes U_{m}(1)$ gauge theory
with the monopoles, the full set of the gauge-invariant observables is
{\it larger} than the set of the Wilson loop averages (\ref{1.2}).
Additionally, we should fix the averages including the {\it dual} Wilson
loops that measure the interaction between the monopoles. The latter loops
are to be deduced from (the $N=1$ option of) eq.
(\ref{1.2}) trading $A_{\rho}({\bf x})$ for its dual counterpart
$C_{\rho}({\bf x})$ introduced through the resummation (\ref{1.1bd}):
in the r.h. side of eq. (\ref{1.1bd}) one is to resolve the Bianchi
identities via the decomposition (\ref{1.6zxe}), i.e. $\partial_{\mu}
(\partial^{\mu}\wedge C_{\nu}({\bf x}))=k_{\nu}({\bf x})$.} scalar function
(of the combination\footnote{As ${^{*}B}_{\mu\nu}
\equiv{\epsilon_{\mu\nu\rho\sigma}
B_{\rho\sigma}/2}$, the vector $\partial_{\mu}{^{*}B_{\mu\nu}}$ is expressed
in terms of the dual potential $C_{\rho}$ only.}
$\partial_{\mu}{^{*}B_{\mu\nu}}$) presumed to be consistent with the
existence of the corresponding functional integral over $B_{\mu\nu}({\bf x})$.
In particular, the case of
${\mathcal{W}}(..)=0$ reproduces the ($N=1$ option of the) weight (\ref{0.1}).

Next, to make contact between eq. (\ref{1.6zxc}) and the action
(\ref{1.1bb}) of the abelian gauge theory enriched with the monopoles, let us
consider the $SC$ series representing the partition function $Z_{U(1)}$
of the compactified $U(1)$ gauge theory associated to
eq. (\ref{1.6zxc}). To begin with, the logarithm of $Z_{U(1)}$ is
proportional to the partition function of the {\it closed} string. Therefore,
in eq. (\ref{1.6zxc}) one is to sum over all $\bar{\vartheta}$-immersions
(\ref{0.3b}) into ${\bf R^{4}}$ (spanning the space
${\mathcal{I}}(M,{\bf R^{4}})$), where $\tilde{M}=
\bar{\vartheta}(\tilde{M})\equiv{\bar{\vartheta}}$ is {\it not} necessarily
connected worldsheet (of an arbitrary Euler character (\ref{0.9zzk})) without
boundaries. Then, the short-cut way between eqs. (\ref{1.6zxc}) and
(\ref{1.1bb}) is provided\footnote{In the simplest setting, eq. (\ref{1.1bd})
is to be applied to the local limit (\ref{0.3cb}) of $Z_{U(1)}$ (when
the action ${\mathcal{W}}(..)$ can be {\it locally} expressed in terms of the co-exact
$H_{\mu\nu}$-form (\ref{1.6zxe}) as long as the gauge
$\partial_{\mu} C_{\mu}({\bf x})=0$ is fixed) that allows to reproduce the
limit (\ref{0.1bbe})/(\ref{2.5bxa}) of the zero flux-tube's width of the stringy
representation (\ref{0.1bbe})/(\ref{0.1}).} by the {\it formal} resummation
prescription (reminiscent of the one in \cite{PolyakCS})
\be
\int\limits_{{\mathcal{I}}(M,{\bf R^{4}})}{\mathcal{D}}\bar{\vartheta}
e^{-\frac{i}{2}\int\limits_{\bar{\vartheta}}
d\sigma_{\mu\nu}({\bf y})B_{\mu\nu}({\bf y})}~\longleftrightarrow{~
\int{\mathcal{D}}k_{\mu}({\bf x}){\mathcal{D}}\tilde{C}_{\mu}({\bf x})
e^{-i\int\limits_{{\bf R^{4}}} d^{4}x~\tilde{C}_{\nu}({\bf x})(
\partial_{\mu}{^{*}B_{\mu\nu}}({\bf x})-k_{\nu}({\bf x}))}},
\label{1.1bd}
\ee
to be substantiated in the end of this Appendix. In eq. (\ref{1.1bd}),
the functional integral over the monopole's current $k_{\mu}({\bf x})$
symbolizes the averaging over the canonical multiloop ensembles akin to the
one in eq. (\ref{1.1bb}). As for the auxiliary dual potential
$\tilde{C}_{\nu}({\bf x})$, it plays the role of the Lagrange multiplier that
enforces the Bianchi identities modified in the presence of the magnetic
charges quantized according to the Dirac rule.

Upon a reflection, the resummation-prescription (\ref{1.1bd}) implies that
the $B_{\mu\nu}$-pattern of the $g^{2}$-dependent part of the lagrangian
(\ref{1.1bb}) is to be {\it identified}, when one reintroduces the
$\delta_{D}({\bf x}-{\bf y})$-kernal removing the smearing (\ref{0.1bb}), with
the sum of the two terms (in the large round brakets) of the lagrangian
(\ref{1.6zxc}). Therefore, it requires to put ${\mathcal{W}}(..)=0$ in the
latter equation. In turn,
the fact, that the system (\ref{1.1bb}) supports the existence of the
flux-tubes of a {\it finite} width, implies (in compliance with the
well-known dual superconductivity hypothesis) that the
monopoles are in certain sense condensed. More precisely, the properly
introduced dual gauge field $C_{\nu}(\bf x)$ is associated to the
{\it massive} rather than massless degrees of freedom.

Finally, the somewhat heuristic derivation of the prescription (\ref{1.1bd})
(where both sides have to be coherently regularized at some $UV$ scale
$\Lambda$) is based on the following observation. In the spirit of the
arguments of \cite{PolyakCS}, the formal summation in the l.h. side of eq.
(\ref{1.1bd}) singles out only those configurations of the tensor Kalb-Ramond
field $B_{\mu\nu}({\bf y})$ for which the exponent (in the l.h side) does
{\it not} depend on the choice of the worldsheet
$\tilde{M}=\bar{\vartheta}(M)$ at all. The latter configurations are
precisely those that comply with the modified Bianchi identities
implemented by the r.h. side of eq. (\ref{1.1bd}). Actually, the considered
resummation can be understood as an analogue of the more familiar
Borel resummation. While the l.h. side of eq. (\ref{1.1bd}) yields the 
representation pertinent for the $SC$ phase (i.e. for sufficiently large
value of $\lambda=g^{2}$), the r.h. side of this equation provides with
pattern adequate in the $WC$ regime for sufficiently small $\lambda$.

\app{The reduction to take advantage of.}

The generic pattern of the type (\ref{0.1bbe}) is defined by
fixing the configuration space ${\mathcal{X}}$ of the mappings (\ref{0.4})
together with the weight $w[\tilde{M}],~\tilde{M}=\vartheta(M)$. The latter,
being presumed {\it finite} for any map $\vartheta\in{{\mathcal{X}}}$,
is further constrained to be {\it local} on the worldsheet $\tilde{M}$.
Let the two $D\geq{3}$ stringy representations (\ref{0.1bbe}) be determined
respectively by $\{{\mathcal{X}}_{a},w_{a}[\tilde{M}]\}$ and a
presumably simpler pair $\{{\mathcal{X}}_{b},w_{b}[\tilde{M}]\}$ so that
${\mathcal{X}}_{b}$ is {\it dense} in ${{\mathcal{X}}_{a}}$ (while
$w_{b}[\tilde{M}]$ can be formally continued onto the entire space
${\mathcal{X}}_{a}$). By construction of the functional
measure, the {\it equivalence} of the latter two representations
(\ref{0.1bbe}) takes place if the two weight-patterns\footnote{For example,
the pattern (\ref{2.5bxa}) vs.
the $m_{0}=0$ option of eq. (\ref{2.5bb}); or eq. (\ref{0.5c}) vs. the same
$m_{0}=0$ option of eq. (\ref{2.5bb}).} are
distinct only on a {\it measure zero} subspace of ${\mathcal{X}}_{a}$ (where
the relative complexity of $w_{a}[\tilde{M}]$ is supposed to show up.)

To present the rigorous formalization of this equivalence, assume first
that there exists a subspace ${\mathcal{X}}_{c}\in{{\mathcal{X}}_{b}}$ which
is {\it dense} in ${\mathcal{X}}_{b}$ (and, as a result, in
${\mathcal{X}}_{a}$), i.e.
\be
{\bf V}[{\mathcal{X}}_{c}]~/~{\bf V}[{\mathcal{X}}_{a}]=
{\bf V}[{\mathcal{X}}_{c}]~/~{\bf V}[{\mathcal{X}}_{b}]=1~,
\label{0.1xea}
\ee
where ${\bf V}[{\mathcal{X}}]$ is the properly normalized volume of the
space ${\mathcal{X}}$. In addition, we postulate
that the patterns of the weights $w_{a}[\tilde{M}]$ and $w_{b}[\tilde{M}]$
{\it coincide if} $\tilde{M}=\vartheta(M)$ {\it does} belong to
${\mathcal{X}}_{c}$:
\be
\vartheta(M)\in{\mathcal{X}}_{c}~~~~~\Longrightarrow{~~~~~
w_{a}[\vartheta(M)]=w_{b}[\vartheta(M)]}~.
\label{0.1xeb}
\ee
Then, the theorem is that the two functional integrals, fixed
respectively by the pairs $\{{\mathcal{X}}_{a},w_{a}[\tilde{M}]\}$ and
$\{{\mathcal{X}}_{b},w_{b}[\tilde{M}]\}$,
coincide
\be
\int\limits_{{\mathcal{X}}_{a}}d{\vartheta}~
w_{a}[\vartheta(M)]=\int\limits_{{\mathcal{X}}_{b}}d{\vartheta}~
w_{b}[\vartheta(M)]~.
\label{0.1xee}
\ee
When ${\partial {\vartheta}=\cup_{k}C_{k}}$, the above sums provide with
the two equivalent representations of the loop-averages
(\ref{0.1bbe}). In the case of the worldsheets
{\it without} boudaries (i.e. when ${\partial {\vartheta}=0}$), the identity
(\ref{0.1xee}) equates the two partition functions 
associated to the corresponding averages (\ref{0.1bbe}).

\sapp{Application to the Ansatz (\ref{0.1bbe})/(\ref{2.5bxa}).}

Let us apply the developed technology in order to prove that the $D\geq{3}$
representation (\ref{0.1bbe}) is invariant under the substitution of the
weight (\ref{2.5bxa}) by the $m_{0}=0$ option of the Nambu-Goto pattern
(\ref{2.5bb}). For this purpose, one is to identify $w_{a}[\tilde{M}]$ and
$w_{b}[\tilde{M}]$ respectively with eq. (\ref{2.5bxa}) and the $m_{0}=0$
reduction of eq. (\ref{2.5bb}), while
\be
{\mathcal{X}}_{a}={\mathcal{X}}_{b}={\mathcal{I}}(M,{\bf R^{D}})~~~~~~,~~~~~~
{\mathcal{X}}_{c}={\mathcal{I}}_{1}(M,{\bf R^{D}})~,
\label{0.1xef}
\ee
where ${\mathcal{I}}(M,{\bf R^{D}})$ stands for the space of the smooth
immersions (\ref{0.4}) (of a $2d$ manifold $M$ into ${\bf R^{D}}$) entering
the $D\geq{3}$ functional integral (\ref{0.1bbe}).
As for ${\mathcal{I}}_{d}(M,{\bf R^{D}})$,
it denotes the subspace of ${\mathcal{I}}(M,{\bf R^{D}})$ where the
corresponding worldsheet $\tilde{M}=\vartheta (M)$, in its  {\it interior}, is
either a nonselfintersecting $2d$ manifold
or a smooth surface with selfintersections on
submanifolds of dimension {\it less or equal to} $d=0,1,2$.
As a result, according to eq. (\ref{2.5bxb}), on
${\mathcal{I}}_{1}(M,{\bf R^{D}})$ the $m_{0}=0$ option of the pattern
(\ref{2.5bb}) {\it coincides} with the one of eq. (\ref{2.5bxa}).

Finally, in order to apply eq. (\ref{0.1xee}), as in \cite{Dub3} one is to
make use of the basic theorem \cite{StabMap} from the theory of immersions.
For $d\geq{4-D}$, the subspace ${\mathcal{I}}_{d}(M,{\bf R^{D}})$ is
{\it dense} in ${\mathcal{I}}(M,{\bf R^{D}})$:
\be
{\bf V}[{\mathcal{I}}_{d}(M,{\bf R^{D}})]~/~
{\bf V}[{\mathcal{I}}(M,{\bf R^{D}})]=1~~~~~if~~~~~d\geq{4-D}~.
\label{0.1xe}
\ee
and, in particular, the subspace ${\mathcal{I}}_{1}(M,{\bf R^{D}})$
is indeed {\it dense} in ${\mathcal{I}}(M,{\bf R^{D}})$ when $D\geq{3}$. It
completes the proof.

\sapp{Application to the Ansatz (\ref{2.5}).}

The same general formula (\ref{0.1xee}) implies the equivalence of the
idealized Gauge String representation (\ref{2.5}) to the $m_{0}=0$ option of
the Nambu-Goto Ansatz (\ref{0.1bbe})/(\ref{2.5bb}).
To apply eq. (\ref{0.1xee}), recall first that any differentiable immersion
is given \cite{StabMap} by a mapping (\ref{0.4}) which is constrained to be
locally one-to-one and in both sides $k\geq{1}$ times differentiable.
The $YM_{D}/String$ duality \cite{Dub3} prescribes that
$\Delta(M,{\bf R^{D}})$ presents a specific extension\footnote{The extension
is performed allowing for certain {\it singularities} which correspond to
the violation of the latter constraints on the mapping (\ref{0.4}).
Strictly speaking, the detailed pattern of $\Delta(M,{\bf R^{D}})$ (and of
the $J[..]$-factor in eq. (\ref{0.5c})) does depend on the choice of the gauge
group and the $YM_{D}$ lagrangian.} of the space
${\mathcal{I}}(M,{\bf R^{D}})$
of the smooth (i.e., infinitely differentiable) immersions. We are to take
advantage that the space ${\mathcal{I}}(M,{\bf R^{D}})$ is
{\it dense} in the option of $\Delta(M,{\bf R^{D}})$ associated to the
$D\geq{3}$ $U(N)$ gauge theory (\ref{1.1}):
\be
U(N)~:~~~~~{\bf V}[{\mathcal{I}}(M,{\bf R^{D}})]~/~
{\bf V}[\Delta(M,{\bf R^{D}})]=1~~~~~~~if~~~~~~D\geq{3}~,
\label{0.5ccs}
\ee
provided the boundary contours are either devoid of backtracking pieces or
absent at all. (The $SU(N)$ counterpart of eq. (\ref{0.5ccs}) is obtained
trading $\Delta(M,{\bf R^{D}})$ for its reduction
$\tilde{\Delta}(M,{\bf R^{D}})$ which is to be induced, see \cite{Dub3},
eliminating certain type of the singularities via the
redefinition (\ref{0.1xaa}) of the bare string tension.)
In the absence of the boundaries, the proof of the $D\geq{4}$ variant of eq.
(\ref{0.5ccs}) is given in \cite{Dub3}. Its extension to the case, when the
boundary loops are without backtrackings, is straightforward. As for the $D=3$
case, it takes a little more effort (to be reported elsewhere) to adapt the
arguments of \cite{Dub3} in order to justify the required identity.

Next, in consequence of eq. (\ref{0.5ccs}), one is to identify
$w_{a}[\tilde{M}]$ and $w_{b}[\tilde{M}]$
with respectively eq. (\ref{0.5c}) and the $m_{0}=0$ reduction of eq.
(\ref{2.5bb}) so that
\be
{\Delta}(M,{\bf R^{D}})={\mathcal{X}}_{a}~~~,~~~
{\mathcal{I}}(M,{\bf R^{D}})={\mathcal{X}}_{b}~~~,~~~
{\mathcal{I}}_{1}(M,{\bf R^{D}})={\mathcal{X}}_{c}~,
\label{0.5ccz}
\ee
where the subspace ${\mathcal{I}}_{1}(M,{\bf R^{D}})$ of
${\mathcal{I}}(M,{\bf R^{D}})$ is specified after
eq. (\ref{0.1xef}). To support the relevance of the latter identification,
we first note that the selfintersection factor $J[..]$ (entering eq.
(\ref{0.5c})) greatly simplifies
on ${\mathcal{I}}(M,{\bf R^{D}})$ as it is formalized by eq. (\ref{0.5ccx}).
Then, the key-observation is that, by construction of
$|{\mathcal{C}}_{\vartheta}|$, the {\it dense} (see eq. (\ref{0.1xe}))
$D\geq{3}$ subspace ${\mathcal{I}}_{1}(M,{\bf R^{D}})$ of ${\mathcal{I}}(M,{\bf R^{D}})$
is comprised of the worldsheets assigned with the $J[..]=1$
factor:
\be
|{\mathcal{C}}_{\vartheta}|=1~~~~~~~if~~~~~~~
\vartheta(M)\in{{\mathcal{I}}_{1}(M,{\bf R^{D}})}~,
\label{0.5cxc}
\ee
so that the weight (\ref{0.5c}) is reduced to the $m_{0}=0$ option of
the weight (\ref{2.5bb}) which, in turn, allows to
apply the basic equivalence-identity (\ref{0.1xee}). In sum, combining all
the pieces together, we conclude that, in $D\geq{3}$, the $m_{0}=0$ variant
of the Nambu-Goto Ansatz (\ref{0.1bbe})/(\ref{2.5bb}) and the idealized Gauge
String pattern (\ref{2.5}) are indeed equivalent for {\it nonbacktracking}
boundary contours. By the same token, the equivalence is maintained between
the closed string's sector of the latter Ansatz and the partition function
of the idealized {\it closed} Gauge String (corresponding to the analytical
continuation of the free energy of the $U(N)$ gauge theory (\ref{1.1})).

\app{Justifying the 't Hooft factor.}

To justify the 't Hooft factor $N^{\chi}$ in eq. (\ref{0.1}), our strategy
is as follows. First, we demonstrate that the $N=1$
reduction of the Ansatz (\ref{0.1bbe})/(\ref{0.1}) is consistent with the
entire Dyson-Schwinger chain of the regularized abelian loop equations. Then,
certain topological relations are revealed which play the crucial role
in the proof of the following matching. When the considered Ansatz is
restricted to the (multiloop generalization\footnote{
The generalization of the conditions (\ref{0.1eea})/(\ref{0.1eec}) has been
formulated in the very end of Section 8.} of the) regime
(\ref{0.1eea})/(\ref{0.1eec}), the $N^{\chi}$-factor allows to match
(within each order of the $1/N$ expansion) the overall degree of
$1/N$ in both sides of any given $U(N)$ loop equation (\ref{1.11bz}).
Combining the two previous steps, we then
sketch the proof that, under the above conditions, the Ansatz
(\ref{0.1bbe})/(\ref{0.1}) indeed 'goes through' the chain of the
regularized $U(N)$ loop equations.

Before we proceed with the implementation of this program, let us briefly
recall that a generic $n$th order loop equation in the
{\it U(N)} chain reads
$$
\tilde{g}^{-2}\cdot\hat{\mathcal{L}}_{\nu}({\bf x}(s_{q}))
<W_{C^{(1)}}W_{C^{(2)}}...W_{C^{(q)}_{xx}}...W_{C^{(n)}}>=
$$
\be
=\oint\limits_{C^{(q)}_{xx}} dy_{\nu}(s'_{q})~
\delta_{D}({\bf y}(s'_{q})-{\bf x}(s_{q}))
<W_{C^{(1)}}W_{C^{(2)}}..W_{C^{(q)}_{xy}}W_{C^{(q)}_{yx}}..W_{C^{(n)}}>+
\label{1.11bz}
\ee
$$
+\sum_{k\neq{q}}^{n}~\frac{1}{N^{2}}
\oint\limits_{C^{(k)}} dy_{\nu}(s'_{k})~
\delta_{D}({\bf y}(s'_{k})-{\bf x}(s_{q}))
<W_{C^{(1)}}..W_{C^{(q)}_{xx}C^{(k)}_{xx}}..W_{C^{(k-1)}}W_{C^{(k+1)}}..
W_{C^{(n)}}>
$$
relating $n$-loop average with the
$(n+1)$- and $(n-1)$-loop ones. Here, for any given $q=1,...,n,$ the Loop
operator (\ref{1.6bf}) is applied to the $q$th  
contour $C^{(q)}\equiv{C^{(q)}_{xx}}$ which is decomposable
(see the second line of eq. (\ref{1.11bz})) as $C^{(q)}_{xy}C^{(q)}_{yx}$ in
the same way as in eq. (\ref{1.11b}). Complementary, in the last line, the
$k$th contour $C^{(k)}$ (with $k\neq{q}$) is combined, if the
$\delta_{D}({\bf y}-{\bf x})$-function does allow, with the contour
$C^{(q)}$ gluing the two loops at the point ${\bf x}$
into the single loop $C^{(q)}_{xx}C^{(k)}_{xx}$.

\sapp{The full $U(1)$ solution.}

To handle the abelian variant of eqs. (\ref{1.11bz}), consider first the
simplest case when all the involved contours, being nonselfintersecting, do
{\it not} mutually intersect. For such loops, one easily observes that
the $N=1$ option of the Ansatz (\ref{0.1bbe})/(\ref{0.1}) goes through the
entire $U(1)$ chain of the regularized eqs. (\ref{1.11bz}) for any
$n\geq{1}$. This can be proved by a straightforward generalization of the
steps resulting in eq. (\ref{1.11bbx}). In particular, on the considered
subspace of the multiloop space,
both the $U(1)$ and the $U(N)$ chain  reduce to the set of the linear loop
equations. The latter can be deduced trading in eq. (\ref{1.11bbx}) the
surface $\tilde{M}(C)$ with a single connected boundary for the worldsheet 
$\tilde{M}(\{C^{(k)}\})$ with an arbitrary number of connected boundary
components so that both of the functional derivatives act on any given loop
$C^{(q)}$.

Finally, for a generic set of the contours, the
analysis\footnote{Being maintained {\it without} imposing the constraints
(\ref{0.1eec}),(\ref{0.1eea}) (or (\ref{0.1eed})), the formal consistency
of the $N=1$ Ansatz (\ref{0.1bbe})/(\ref{0.1}) with the $U(1)$ eq.
(\ref{1.11bz}) does {\it not} account for possible phase transitions. The
latter are expected to invalidate the considered $SC$
expansion as the faithful representation of the corresponding $D\geq{3}$
abelian gauge theory (with the monopoles), defined by the action (\ref{1.1bb})
in the $D=4$ case, in the $WC$ phase $g^{2}\rightarrow{0}$.} can be reduced to
the previous case due to the well-known property of the $U(1)$  Wilson loops.
Namely, if any two contours $C^{(1)},~C^{(2)}$ mutually intersect at ${\bf x}$
then $W_{C^{(1)}}W_{C^{(2)}}=W_{C^{(1)}_{xx}C^{(2)}_{xx}}$, where the new
single contour $C^{(1)}_{xx}C^{(2)}_{xx}$ is composed as in the third line of
eq. (\ref{1.11bz}).

\sapp{The coupling between the different worldsheet's topologies.}

Let us now turn to the topological relations that necessitate the 't Hooft
$N^{\chi}$-factor of the weight (\ref{0.1}). The guiding idea (underlying
the construction of the $D\geq{3}$ low-energy solution
(\ref{0.1bbe})/(\ref{0.1})) is that, in any given order of the
$1/N$-expansion, the correspondence between the worldsheet's topologies
(associated, see eq. (\ref{0.9zzi}), to each of the three lines of
eq. (\ref{1.11bz})) is $N$-{\it independent}, being thus the same as in the
abelian case. In this perspective, we note first that a part of the reason
behind the simplicity of the reduced $U(N)$ loop equation (\ref{0.9za}) lies
in the fact that the latter is {\it diagonal}\footnote{The decoupling of
different worldsheet's topologies immediately follows from the possibility to
trade eq. (\ref{0.9za}) for eq. (\ref{1.11bfb}) formulated directly for any
particular {\it single} surface $\tilde{M}_{\chi}(C)$.} with respect to the
genus $(1-\chi)/2$ of each particular worldsheet $\tilde{M}_{\chi}(C)$
entering the measure of the $b=1$ stringy sum (\ref{0.1bbe}).
On the other hand, the decoupling is {\it no} more valid when we consider the
contribution (into the r.h. side of eq. (\ref{1.11bz})) either from a
nontrivial selfintersection of a given contour or from an
intersection of any two different loops.

Then, as it is shown in the end of
this subappendix, in each particular order of the $1/N$
expansion, the following matching between the corresponding total Euler
characters (\ref{0.9zzk}) takes place irrespectively whether or not
the proper regularization of the loop equations is performed. Namely, the
variety of the (not necessarily connected) worldsheets with the total Euler
character $\chi$ in the first line of eq. (\ref{1.11bz}) must be associated,
through the action of $\hat{\mathcal{L}}_{\nu}({\bf x}(s))$, to the
corresponding variety of the worldsheets in the second line with the total
Euler character $\bar{\chi}$,
\be
\bar{\chi}-\chi=\bar{b}-b=1~~~~~~~;~~~~~~~\chi-\hat{\chi}=b-\hat{b}=1~,
\label{0.9zzi}
\ee
which is one unit larger than $\chi$ so that the {\it difference}
$\bar{\chi}-\chi$ is equal to the one between the numbers $\bar{b}$ and $b$ of
the connected boundary components in the second and in the first lines
respectively. By the same token\footnote{Both of the relations (\ref{0.9zzi})
can be reformulated as the {\it conservation} of the composed topological
number $(p-h-b)$, where $p,~h,$ and $b$ are introduced in eq.
(\ref{0.9zzk}).}, the Euler character $\hat{\chi}$, associated to the
third line, is one unit {\it less} than the character $\chi$ associated to
the first line that matches with the difference between the numbers $\hat{b}$
and $b$ of the boundaries respectively in the third and in the first lines of eq.
(\ref{1.11bz}). 

Upon a reflection, the relations (\ref{0.9zzi}) imply certain
restriction which is necessary to impose, for the entire
$1/N$ expansion (\ref{0.1bbe})/(\ref{0.1}) to 'go through' the regularized
loop equations (\ref{1.11bz}). One is to deal only with those
multiloop sets $\{C^{(k)}\}$ for which the (regularized) third line of 
eq. (\ref{1.11bz}) vanishes in any order of the $1/N$ series. To justify this
statement, we are to combine the relations (\ref{0.9zzi}) with the
representation of the averages $<\prod_{k}W_{C^{(k)}}>$ (of the Wilson loops
(\ref{1.2})) in the form of the $1/N$-expansion starting with the
${N^{0}}$-term. Altogether, it suggests to search for the $N$-dependence of
the low-energy weight $w_{2}[\tilde{M}_{\chi}(\{C^{(k)}\})]$ in the form
of the Ansatz $[f(N)]^{\chi}$.
Then, it is straightforward to verify that the chain of the loop-equations
(\ref{1.11bz}) unambiguously require to put $f(N)=N$. Indeed, in this case,
the overall powers $N^{\chi-b}$ and $N^{\bar{\chi}-\bar{b}}$
in the first and the second lines of eq. (\ref{1.11bz}) are precisely equal
for the considered (in the context of the first of the relations (\ref{0.9zzi}))
paired varieties of the worldsheets.

Next, when the third line of the regularized eq. (\ref{1.11bz}) can not be
discarded, the second of the relations (\ref{0.9zzi}) implies that the
the overall powers $N^{\chi-b}$ and $N^{\hat{\chi}-\hat{b}}/N^{2}$
in the first and the third lines do {\it not} match.
To clarify when this harmful contribution can be consistently
discarded, observe first that prior to the regularization the considered third
line identically vanishes once the involved contours $C^{(k)}$ do {\it not}
mutually intersect. Therefore, after the regularization, one is to require
that {\it both} the characteristic size of each (macroscopic) loop $C^{(k)}$
{\it and} the minimal distance between any two different loops
are of order of ${N^{\alpha}},~\alpha>0$. Indeed, due to the latter condition,
all sorts of the quasi-local interactions (between the elementary flux-tubes)
are suppress as $e^{-\beta N}$ with some $\beta>0$. This is necessary to
accomplish because, if not suppressed\footnote{In the $N\rightarrow{\infty}$
limit, it is sufficient to weaken the $N^{\alpha}$-scaling of all the
distances to the $|\phi|$-scaling with some $N$-independent
$|\phi|\rightarrow{\infty}$. In the case of a single loop $C$, it is
tantamount to the requirement (formulated
prior to eq. (\ref{0.1eel})) that the contour $C$ is macroscopic.}, the
abelian pattern of these interactions is in conflict\footnote{To say the
least, the above interactions  would produce spurious violation of the large
$N$ factorization of the multiloop averages $<\prod_{k}W_{C^{(k)}}>$.} with
the $N\geq{2}$ $U(N)$ loop equations (\ref{1.11bz}).

Returning to the derivation of the relations (\ref{0.9zzi}),
(from the previous abelian analysis) it is clear that the
second of the relations is topologically equivalent to the first one
which we now focus on. Furthermore, for simplicity,
we concentrate on the $n=1$ eq. (\ref{1.11bz}) that
(in the $U(N)$ case) can be deduced from the large $N$ eq. (\ref{1.11b})
through the substitution (\ref{0.9zzg}).
The generalization of the analysis to the $n\geq{2}$
case of (nonintersecting) loops is straightforward.

As for the $n=1$ option, it is instructive to prove the corresponding
relation (\ref{0.9zzi}) for the regularized variant of the associated loop
equation (\ref{1.11bz}) so that we first sketch how the large $N$ identity
(\ref{0.9zz}) can be adapted to the case at hand. For this purpose, take any
particular connected worldsheet
$\tilde{M}_{\chi}(C)\in{\mathcal{I}}(M_{\chi},{\bf R^{D}})$
(entering eq. (\ref{0.1bbe})) of the Euler character $\chi$.
Let a nontrivial selfintersection point (at ${\bf x}(s)={\bf y}(s')$ with
$s\neq{s'}$) of a macroscopic loop $C_{xx}=C_{xy}C_{yx}$ be resolved according
to the preliminary smearing (\ref{0.1bb}). To properly generalize eq.
(\ref{0.9zz}), as previously one is to consider the intersection of
$\tilde{M}_{\chi}(C)$
with some $(D-1)$-dimensional domain ${\bf X^{D-1}}$ including the points
${\bf x}(s)$ and ${\bf y}(s')$. The considered cross-section generically
allows to select the nonselfintersecting contour
(\ref{0.9zza}). Then, in the generalized variant of the identity (\ref{0.9zz})
written for generic values of $\chi$, the surface
$\tilde{M}_{\chi}(C_{xx})$ (in its l.h. side) with the single boundary
$C_{xx}$ is associated to the worldsheet
${\tilde{M}_{\bar{\chi}}(C_{xy}\Gamma_{yx};C_{yx}\Gamma_{xy})}$ (in the r.h.
side). The latter, being endowed with the two boundaries $C_{xy}\Gamma_{yx}$
and $C_{yx}\Gamma_{xy}$, has the {\it same} area $A[\tilde{M}_{\chi}(C_{xx})]=
A[\tilde{M}_{\bar{\chi}}(C_{xy}\Gamma_{yx};C_{yx}\Gamma_{xy})]$ but possesses
a different value of the Euler character $\bar{\chi}$ (to be determined
below). Moreover, for any macroscopic contour $C_{xx}$, the effective action
$S_{\chi}(\Gamma_{xy}|{\mathcal{G}})$ (introduced akin to eq.
(\ref{0.9zz})) is evidently {\it independent} of $\chi$ in the relevant limit
(\ref{0.1eea}) of the large minimal areas associated to the loops
$C_{xy}\Gamma_{yx}$ and $C_{yx}\Gamma_{xy}$ in question.

To substantiate eq. (\ref{0.9zzi}), there remains to prove that the total
Euler character $\bar{\chi}$ of the worldsheet
$\tilde{M}_{\bar{\chi}}(C_{xy}\Gamma_{yx};C_{yx}\Gamma_{xy})$ is one unit
{\it larger} than the one $\chi$ of $\tilde{M}_{\chi}(C_{xx})$. This fact
immediately follows from the observation that, when
being discretized, the former ({\it not} necessarily connected) worldsheet
can be obtained from the latter by adding four extra cells: the two sites
(0-cell), the two links (1-cells) and a single plaquette (2-cell).  According
to the definition of $\chi$, a given $k$-cell
contributes to the total Euler character as $(-1)^{k},~k=0,1,2$.
Altogether, it reproduces the difference (\ref{0.9zzi}) between $\bar{\chi}$
and $\chi$. To have an example of how it works, take the option when
the surface $\tilde{M}_{\bar{\chi}}(C_{xy}\Gamma_{yx};C_{yx}\Gamma_{xy})$
remains to be {\it connected}. It implies that the trajectory (\ref{0.9zza})
is supposed to cut exactly one (equivalence class of the) uncontractible 
closed path of $\tilde{M}_{\chi}(C_{xx})$. In turn, it ensures that 
the genus $\bar{h}=-\bar{\chi}/2$ is one unit {\it smaller}
than $h=(1-\chi)/2$ which is in exact agreement with eq. (\ref{0.9zzi}).

Finally, synthesizing the previous arguments with the analysis of Section 8, one can
show the following. Let the regularization prescription (\ref{0.9zze}) is
appropriately adapted (for nontrivial point-like selfintersection of the
loops) to the higher genus cases. Then, provided the $\Lambda^{2}$-scaling
(\ref{0.1eec}) of $\sigma_{ph}$ is fulfilled, the Ansatz
(\ref{0.1bbe})/(\ref{0.1}) is the genuine $1/N$ expansion of the
$U(N)$ solution for the entire Dyson-Schwinger chain of eqs. (\ref{1.11bz})
restricted to the subspace of the macroscopic contours satisfying the
conditions formulated in the very end of Section 8.

\sapp{The derivation of the constraint (\ref{1.3zp}).}

In order to derive the constraint (\ref{1.3zp}), let us recast the
presecription (\ref{0.9zze}) into the form matching
with the gauge-invariant regularization discussed in \cite{MigdRep,MakRev}.
Recall, that the latter is associated (see \cite{MakRev}) to the
so-called second order Dyson-Schwinger equation\footnote{Eq. (\ref{1.3za})
regularizes the modified representation \cite{LE/P} of the Loop equation
(\ref{1.11b}) which can be obtained from eq. (\ref{1.11b}) integrating both
sides along the contour $C$, i.e. performing the integral
$\oint_{C} dx_{\nu}$.}
$$
\left<\int d^{D}x~(\hat{D}^{ab}_{\mu}(A)F^{b}_{\mu\nu}({\bf x}))
\frac{\delta}{\delta A^{a}_{\nu}({\bf x})}W_{C}[A]\right>=
$$
\be
=\tilde{g}^{2}\left<\int d^{D}x~d^{D}y~
<{\bf y}|~\hat{\mathcal{E}}^{ab}[A]~|{\bf x}>
\frac{\delta}{\delta A^{a}_{\nu}({\bf y})}
\frac{\delta}{\delta A^{b}_{\nu}({\bf x})} W_{C}[A]\right>_{YM_{D}},
\label{1.3za}
\ee
where $W_{C}[A]\equiv{W_{C}}$ is the Wilson loop (\ref{1.2}), and
$<{\bf y}|\hat{\mathcal{E}}^{ab}[A]|{\bf x}>$ is the matrix element
which is to regularize $\delta^{ab}\delta_{D}({\bf x}-{\bf y})$.
To make contact with eq. (\ref{0.9zze}), we first decompose
\be
<{\bf y}|~\hat{\mathcal{E}}^{ab}[A|{\mathcal{G}}]~|{\bf x}>=
{\Lambda}^{D}{\mathcal{G}}({\Lambda}^{2}({\bf x}-{\bf y})^{2})
~<{\bf y}|~\hat{\mathcal{E}}_{1}^{ab}[A|{\mathcal{G}}]~|{\bf x}>~,
\label{1.3zb}
\ee
where the gauge field $A^{a}_{\nu}$ is supposed to enter the operator
$\hat{\mathcal{E}}_{1}^{ab}[A|{\mathcal{G}}]$ through the covariant
derivative $\hat{D}^{ab}_{\mu}(A)$ (in the adjoint representation), and we
have introduced an explicit
functional dependence on the smearing ${\mathcal{G}}$-function
(\ref{0.1bb}). Secondly, one is to identify 
that the path-integral representation of the matrix element in the r.h.
side of eq. (\ref{1.3zb}) is given by
\be
<{\bf y}|~\hat{\mathcal{E}}_{1}^{ab}[A|{\mathcal{G}}]~|{\bf x}>=
\int\limits_{\Gamma_{xy}\in{{\bf X^{D-1}}}}
{\mathcal{D}}z_{\mu}(t)~e^{-S(\Gamma_{xy}|{\mathcal{G}})}~
tr\left(T^{a}{\mathcal{B}}(\Gamma_{xy})T^{b}
{\mathcal{B}}(\Gamma_{yx})\right)~,
\label{1.3zc}
\ee
where ${\mathcal{B}}(\Gamma_{xy})$ is the path-ordered exponent
(\ref{1.3zj}) (with $T^{a}
\equiv{T^{a}_{ij}}$ standing for the $U(N)$ generator in the fundamental
representation with $tr(T^{a}T^{b})=\delta^{ab}$) so that one recognizes
the building block of eq. (\ref{0.9zze}) in the r.h. side of eq.
(\ref{1.3zc}).

Finally, for a generic smearing function, the normalization of the
r.h. side of eq. (\ref{1.3zb}) does {\it not} allow to interprete
$\hat{\mathcal{E}}^{ab}[0|{\mathcal{G}}]$ as the smearing of
$\delta^{ab}\hat{1}$ (where $\hat{1}$ acts in the coordinate space) as
it is implied in the conventional gauge-invariant regularization
\cite{MigdRep,MakRev}. Therefore, in order to match between the prescriptions
(\ref{0.9zze}) and (\ref{0.1bb}), one is to impose
the constraint (\ref{1.3zp}) complementary to the normalization (\ref{0.1bb}).

In conclusion, let us make a few comments concerning the relation
(\ref{0.9zz}) underlying the considered above regularization.
For a preliminary orientation, consider first the zero vortex-width limit
(\ref{0.3cb}) when the weight (\ref{0.1}) can be traded, within 
the stringy sum (\ref{0.1bbe}), for the Nambu-Goto weight
(\ref{2.5bb}) with $m_{0}=0,~\bar{\lambda}(\lambda)=\xi\lambda$. In this case,
the interpretation of $S(\Gamma_{xy}|{\mathcal{G}})$ (in the exponent of eq.
(\ref{0.9zz})) as the effective action for the cross-section paths
$\Gamma_{xy}$ is justified by
the following observation. The consistency of the latter interpretation
requires that $S(\Gamma_{xy}|{\mathcal{G}})$ has a {\it finite} limit when
the minimal area of both $\tilde{M}_{min}(C_{xy}\Gamma_{yx})$ and
$\tilde{M}_{min}(C_{yx}\Gamma_{xy})$ is sent to infinity (in the units of
$\Lambda^{-1}$). In turn, the required finiteness is predetermined by the fact
that the correlators of the normals
to the worldsheet $\tilde{M}(C_{xy}C_{yx})$ are supposed to be short-ranged
in the considered regime (\ref{0.1eec}). As a result, even when
$|{\bf x}-{\bf y}|\leq{\Lambda^{-1}}$, the implicit dependence of the action
$S(\Gamma_{xy}|{\mathcal{G}})$ on the {\it geometry}
of the contour $C=C_{xy}C_{yx}$ is reduced to the one on
the geometry in  the $1/\Lambda$ vicinity of ${\bf x}$ and
${\bf y}$.

As for the case, when the width of the $YM$ vortex is $\sim{\Lambda^{-1}}$,
the only essential difference with the previous consideration is
due to the presence of the quasi-contact interactions (of the abelian nature)
between the elementary flux-tubes. The consistency is then maintained owing to
the fact that, for any macroscopic contour $C$ in the regime (\ref{0.1eea}),
the contribution of the latter interactions is unobservable everywhere except
for the $1/\Lambda$ vicinity of the (point-like) selfintersections of $C$.
In consequence, by the same token as previously,
$S(\Gamma_{xy}|{\mathcal{G}})$ remains finite when 
$A[\tilde{M}_{min}(C_{xy}\Gamma_{yx})],~A[\tilde{M}_{min}(C_{yx}\Gamma_{xy})]
\rightarrow{\infty}$.

\app{The backtracking invariance of $<W_{C}>$.}

The basic symmetry of $<W_{C}>$, encoded in the duality-relation
(\ref{0.5cxd}), is the invariance with respect to the creation of arbitrary
zig-zag backtrackings (bounding zero 1-volume) of the
contour $C$. Our aim is to discuss the two alternative prescriptions
to implement the backtracking invariance of $<W_{C}>$ starting
from any particular contour $C\in{\Upsilon_{nbt}}$ devoid of backtrackings.

Before we deduce from eq. (\ref{0.5cxd}) the first prescription (introduced
in Section 4.4), let us first present the strict formalization of what we are
going to obtain. Actually, the required prescription amounts to the following
simple modification (given by eq. (\ref{0.4ze})) of the measure in eq.
(\ref{2.5}) (or (\ref{0.1bbe})) while the action associated to the weight
(\ref{0.5c}) (or, respectively, (\ref{2.5bb})) is kept intact. To begin with,
it is helpful to reinterprete the loop space $\Upsilon$ as the space of the
{\it orbits}\footnote{Thus, the $\Upsilon_{nbt}$ subspace (comprised of all
nonbacktracking contours) represents the proper {\it section} of $\Upsilon$ so
that $\Upsilon={\mathcal{O}}(\Upsilon_{nbt})$.} ${\mathcal{O}}(C)$, each being
generated via the appropriate attachment of the backtracking segments
from a particular nonbacktracking contour $C\in{\Upsilon_{nbt}}$. To introduce
the orbit structure, one is to associate to a given smooth immersion-map
$\bar{\vartheta}_{b}:~\breve{C}\rightarrow{C}$, resulting in
the boundary contour $C\equiv{C[\hat{1}]}$, the smooth composed (i.e.
inner-product) mapping
\be
C[\hat{1}]=\bar{\vartheta}_{b}(\breve{C})
\longrightarrow{\bar{\vartheta}_{b}(\breve{C})\circ{\upsilon}(\breve{C})}
\equiv{C[\upsilon]}~,
\label{0.4zd}
\ee
where the map ${\upsilon}(\breve{C})$ induces on a given {\it reference}-loop
$C[\hat{1}]\in{\Upsilon_{nbt}}$ a particular data of backtrackings.
Provided the backtracking segments possess the support on
$C[\hat{1}]\in{\Upsilon_{nbt}}$, the deformation (\ref{0.4zd}) can be
evidently represented as the reparametrization of the contour $C[\hat{1}]$,
\be
x_{\mu}(s)\rightarrow{x_{\mu}(f(s))}~~~;~~~
\{f(0)=0~,~f(1)=1~|~df(s)/ds>-\infty\}~,
\label{0.1zz}
\ee
originally introduced through the trajectory $x_{\mu}(s),~s\in{[0,1]}$.
Then, the foldings are associated to those points $s_{k}$ where
the derivative $df(s)/ds$ changes its {\it sign} (while the composite
boundary-mapping (\ref{0.4zd}) ceases to be an immersion).

Now we are ready to formalize the required prescription.
The Gauge String representation of $<W_{C[\upsilon]}>$ can be
deduced from the one of $<W_{C[\hat{1}]}>$ trading, in the measure of eq.
(\ref{2.5}) (or (\ref{0.1bbe})), the ${\vartheta}(M(\breve{C}))$- (or,
respectively, $\bar{\vartheta}(M(\breve{C}))$-) maps for the
modified mappings
\be
{\vartheta}(M(\breve{C}))
\longrightarrow{{\vartheta}(M(\breve{C}))\circ{\upsilon}(\breve{C})}
\equiv{\tilde{M}(C[\upsilon])}~
\label{0.4ze}
\ee
which, leaving the
{\it interior} ${\vartheta}(M(\breve{C}))/C$ of the surface intact
\be
\left({\vartheta}(M(\breve{C}))\circ{\upsilon}(\breve{C})\right)/
C[\upsilon]={\vartheta}(M(\breve{C}))/C[\hat{1}]~,
\label{1.6za}
\ee
create a given data of the foldings (of the boundary contour
$C\in{\Upsilon_{nbt}}$) in accordance with the $\upsilon$-pattern
(\ref{0.4zd}). The backtracking invariance of $<W_{C}>$ follows then from
the fact that all the ingredients of the 'action' in eq. (\ref{2.5}) (i.e.
the area $A[\tilde{M}]$, the Euler character $\chi$, and the selfintersection
factor $J[..]$ deduced from eq. (\ref{0.5cxd})) are invariant under
the transformation (\ref{0.4ze}).

Next, to obtain eq. (\ref{0.4ze}) from the duality-relation (\ref{0.5cxd}),
one is to observe that the prescription (\ref{0.4ze}) in $D\geq{3}$
effectively implements the following {\it tree-irreducibility}\footnote{Recall
that a tree is a graph without any closed 1-cycles.} constraint (to be deduced
below), {\it lacking} in the conventional Nambu-Goto paradigm.
The considered irreducibility implies that, being cut along an
arbitrary tree, any admissible worldsheet $\tilde{M}(\{C_{k}\})$ does
{\it not} split into a union of {\it disjoint} components.
In order to reveal why the considered constraint is crucial for
the backtracking invariance of $<W_{C}>$, one notes first that for
$\upsilon\neq{\hat{1}}$ the support (in ${\bf R^{D}}$) of the complement
$C[\upsilon]/C[\hat{1}]$ has topology of a (not necessarily connected) tree
graph. Then, in addition to the surfaces $\tilde{M}(C[\upsilon])$ complying
with eq. (\ref{0.4ze}), there is another relevant
contribution within the conventional Nambu-Goto measure: along
$C[\upsilon]/C[\hat{1}]$, one can glue to $\bar{\vartheta}(M(\breve{C}))$
arbitrary {\it closed} 'baby-universes'. The latter contribution, being of
nonzero measure in $D\geq{3}$, evidently violates both
the tree-irreducibility and the backtracking invariance. It is the
tree-irreducibility which in $D\geq{3}$ makes the Gauge String representation
rid of such 'baby-universes'.

Finally, it is straightforward to derive the considered irreducibility from
the basic relation (\ref{0.5cxd}). Indeed, the functional integral of the
$YM$ theory, defined (via, see \cite{Dub3}, the Heat-Kernal lattice gauge
system) on a taget-space $T$ irreducible with respect to the cut along a
particular tree ${\mathcal{T}}$, {\it factorizes}. The latter follows from
the possibility \cite{Dr&Zub} to choose such a gauge that all the
link-variables (in the corresponding Heat-Kernal system on $T$), associated
to the tree ${\mathcal{T}}$ in question, are constrained to unity.
As a result, the weight of the (conglomerates of the) tree-{\it reducible}
worldsheets properly factorizes as well so that the latter worldsheets
are to be interpreted as the {\it disconnected} contribution properly composed
of the tree-irreducible {\it connected} components. Therefore, the required
irreducibility constraint follows from the fact that, by construction, the
stringy representation of the average in the r.h. side of eq. (\ref{0.5cxd})
involves only {\it connected} worldsheets.

\sapp{Employing the larger symmetry of the string action (\ref{0.1}).}

Let us now turn to the complementary prescription to implement the
backtracking invariance of $<W_{C}>$, for simplicity restricting our attention
to the regime (\ref{0.1eea})/(\ref{0.1eec}) where the stringy pattern
(\ref{0.1bbe})/(\ref{0.1}) applies. To begin with, the logarithm of eq.
(\ref{0.1}) (akin to a generic polynomial functional (\ref{1.9adc}) of the
area-element $d\sigma_{\mu\nu}({\bf x})$) is evidently
invariant under the extention \cite{PolyakCS} of the group of the worldsheet's
diffeomorphisms (\ref{0.4zh}). The extension allows for a {\it nonpositive}
Jacobian (\ref{0.4zhh}), and zeros of the Jacobian ${\mathcal{J}}(\gamma)$
can possess support on closed curves which bound (on the worldsheet
$\tilde{M}$ in question) connected $2d$ domains where
${\mathcal{J}}(\gamma)<0$.

As a result, the extended reparametrizations (\ref{0.1zz}) of the boundary
loop (allowing for negative $df(s)/ds$) can be alternatively implemented
as the boundary restriction of the {\it smooth}\footnote{
This is in contrast to the previous prescription formalized
via rather {\it singular} reparametrization (\ref{0.4zh}) corresponding to
the mapping (\ref{0.4ze}).} reparametrization (\ref{0.4zh}) ({\it violating}
the ${\mathcal{J}}(\gamma)>0$ constraint) which involves the worldsheet's
{\it interior} as opposed to eq. (\ref{1.6za}).
The latter reparametrization associates to a given strict immersion
$\bar{\vartheta}(M(\breve{C}))$ the whole {\it orbit}
\be
\bar{\vartheta}(M(\breve{C}))
~\longrightarrow{~\bar{\vartheta}(M(\breve{C}))\circ{\omega}(M(\breve{C}))}
\equiv{\phi(M(\breve{C}))}~,
\label{0.4z}
\ee
generated by the auxiliary maps ${\omega}(M(\breve{C}))$. Each particular
${\omega}(M(\breve{C}))$, geometrically, is visualized as an attachment of a
given pattern of the worldsheet's $2d$ foldings with the support on the
surface $\tilde{M}(C)$ represented by the original immersion-map
$\bar{\vartheta}(M(\breve{C}))$. (In fact, $ln(w_{2}[\tilde{M}(C)])$ is
invariant under the generalization of (\ref{0.4z}) when the foldings are
substituted by arbitrary backtracking $2d$ segments (i.e. closed
baby-universes bounding zero 3-volume). The latter segments, assumed to
possess at least a 1-dimensional domain of intersection with
$\tilde{M}(C)$, do {\it not} necessarily have the entire
support on $\tilde{M}(C)$.)

To make contact with the extended group of the reparametrizations
(\ref{0.1zz}), eq. (\ref{0.4z}) should be compared with eq. (\ref{0.4ze}).
One concludes that the boundary restriction ${\omega}_{b}(\breve{C})$ of
${\omega}(M(\breve{C}))$ is to be identified with the corresponding mappings
$\upsilon(\breve{C})$.
The smoothness of the reparametrizarion functions $f_{\alpha}(\gamma)$ is now
achieved because the loop's backtrackings are {\it not} an isolated
1-dimensional 'defect' of the transformation (\ref{0.4zh}) but a part of the
2-dimensional worldsheet's folding associated to the domain of negative
Jacobian. It is also noteworthy that, for $\upsilon\neq{\hat{1}}$, the
topology of $\tilde{M}(C[\upsilon])=\phi(M(\breve{C}))$ does
{\it violate} the tree-irreducibility condition exactly
on the tree where the worldsheet's backtrackings (encoded in
${\omega}(M(\breve{C}))$) are attached.

Finally, in order to implement the backtracking invariance of $<W_{C}>$
(i.e. extended symmetry (\ref{0.1zz})) on the
quantum level, the measure in the Ansatz (\ref{0.1bbe}) is to be modified
according to the pattern (\ref{0.4z}):
\be
\int {\mathcal{D}}\bar{\vartheta}(M(\breve{C}))
~\longrightarrow{~\int {\mathcal{D}}\bar{\vartheta}(M(\breve{C}))
~\frac{d\omega(M(\breve{C}))}{{\mathcal{K}}_{\bar{\vartheta}}}}
\equiv{\int d\phi(M(\breve{C}))}~,
\label{0.4zz}
\ee
where ${\mathcal{K}}_{\bar{\vartheta}}$ stands for the proper
normalization-factor. As the string action (\ref{0.1}) is {\it independent} of
${\omega}(M(\breve{C}))$ (while the Wilson loops are invariant under the
reparametrizations (\ref{0.1zz})), the original representation
(\ref{0.1bbe})/(\ref{0.1}) is invariant under the extension (\ref{0.4zz})
of the measure.


\begin{thebibliography}{99}




\bibitem{Wils} K.Wilson, {\it Phys. Rev.} {\bf D10} (1974) 2445.
\bibitem{'t Hooft1} G.'t Hooft, {\it Nucl. Phys.} {\bf B72} (1974) 461.
\bibitem{MigdRep} A.Migdal, {\it Phys. Rep.} {\bf 102} (1983) 199.
\bibitem{GFS} A.Polyakov, {\it Gauge Fields and Strings},
 Harwood Academic Publishers, Chur, 1987.
\bibitem{Gr&Tayl} D.Gross, {\it Nucl. Phys.} {\bf B400} (1993) 161; \\
D.Gross and W.Taylor, {\it Nucl. Phys.} {\bf B400} (1993) 181;
{\it Nucl. Phys.} {\bf B403} (1993) 395.
\bibitem{Dub2} A.Dubin, {\it Nucl. Phys.} {\bf B584} (2000) 749.
\bibitem{Dub3} A.Dubin, {\it Nucl. Phys.} {\bf B582} (2000) 677.
\bibitem{MigdElf} A.Migdal, {\it Nucl. Phys.} {\bf B189} (1981) 253.
\bibitem{LE/MM} Yu. Makeenko and A.Migdal,
{\it Phys. Lett.} {\bf 88B} (1979) 135;\\
{\it Nucl. Phys.} {\bf B188} (1981) 269.
\bibitem{LE/P} A.Polyakov, {\it Nucl. Phys.} {\bf B164} (1981) 171.
\bibitem{PolyakCS} A.Polyakov, {\it Nucl. Phys.} {\bf B486} (1997) 23.
\bibitem{Wu&Yang} T.Wu and C.Yang, {\it Phys. Rev.} {\bf D12} (1975) 3845.
\bibitem{Kaz&Kost} V.Kazakov and I.Kostov, {\it Nucl. Phys.} {\bf B176}
(1980) 199;\\ V.Kazakov, {\it Nucl. Phys.} {\bf B179} (1981) 283.
\bibitem{Bralic} N.Bralic, {\it Phys. Rev.} {\bf D22} (1980) 3090.
\bibitem{ALaw} Yu.Makeenko and A.Migdal, {\it Phys. Lett.} {\bf 97B} (1980)
253.
\bibitem{Dr&Zub} J.Drouffe and J.Zuber, {\it Phys. Rep.} {\bf 102} (1983) 1.
\bibitem{Luscher} M.Luscher, K.Symanzik, and P.Weisz,
{\it Nucl. Phys.} {\bf B173} (1980) 365;\\
M.Luscher, {\it Nucl. Phys.} {\bf B180} (1981) 317.
\bibitem{StabMap} M.Golubitsky and V.Guilemin,
{\it Stable Mappings and Their Singularities},\\ Springer-Verlag, (1973).
\bibitem{Ol&Pet} P.Olesen and J.Petersen, {\it Nucl. Phys.} {\bf B181}
(1981) 157.
\bibitem{Gop&Mack} M.Gopfert and G.Mack, {\it Comm. Math. Phys.} {\bf 82}
(1981) 545;\\
J.Frohlich and T.Spenser, {\it Comm. Math. Phys.} {\bf 83}
(1982) 411.
\bibitem{Kalb&Ramond} M.Kalb and P.Ramond, {\it Phys. Rev.} {\bf D9} (1974)
2273;\\ Y.Nambu, {\it Phys. Rev.} {\bf D10} (1974) 4262.
\bibitem{Maldac} J.Maldacena,
{\it Adv. Theor. Math. Phys.} {\bf 2} (1998) 231;
{\it Phys. Rev. Lett.} {\bf 80} (1998) 4859.
\bibitem{Polyak2} A.Polyakov, {\it Nucl. Phys. Proc. Suppl.} {\bf 68}
(1998) 1.
\bibitem{Witt2} E.Witten,
{\it Adv. Theor. Math. Phys.} {\bf 2} (1998) 505. 
\bibitem{Gr&Ooguri} D.Gross and H.Ooguri, {\it Phys. Rev.} {\bf D58} (1998)
1060.
\bibitem{LE/SUSY} N.Drukker, D.Gross and H.Ooguri,
{\it Phys. Rev.} {\bf D60} (1999) 1250;\\
A.Polyakov and V.Rychkov, {\it Nucl. Phys.} {\bf B581} (2000) 116.
\bibitem{Fr&Tseyt} E.Fradkin and A.Tseytlin, {\it Annal. Phys.}
{\bf 143} (1982) 413.
\bibitem{ExtrCurv/P} A.Polyakov, {\it Nucl. Phys.} {\bf B268} (1986) 406.
\bibitem{ExtrCurv/K} H.Kleinert, {\it Phys. Lett.} {\bf 174B} (1986) 335. 
\bibitem{KPZ} V.Knizhnik, A.Polyakov and A.Zamolodchikov,
{\it Mod. Phys. Lett.} {\bf A3} (1988) 1213.
\bibitem{DDK} F.David, {\it Mod. Phys. Lett.} {\bf A3} (1988) 1651;\\
J.Distler and H.Kawai, {\it Nucl. Phys.} {\bf B321} (1989) 509.
\bibitem{PolyakLH} A.Polyakov, {\it Proceedings of Les Houches School},
1992, Elsevier.
\bibitem{MakRev} M.Halpern and Yu.Makeenko, {\it Phys. Lett.} {\bf 218B}
(1989) 230.
\bibitem{LE/NCYM} J.Ambjorn, Yu.Makeenko, J.Nishimura and R. Szabo,
{\it JHEP} {\bf 11} (1999) 029;\\
M. Abou-Zeid and H.Dorn, hep-th/0009231;
\bibitem{Orland} P.Orland, {\it Nucl. Phys.} {\bf B428} (1994) 221;\\
M.Sato and S.Yahikozawa, {\it Nucl. Phys.} {\bf B436} (1995) 100.
\bibitem{AES} D.Antonov, D.Ebert and Yu. Simonov, {\it Mod. Phys. Lett.}
{\bf A11} (1996) 1905;\\
M.Diamantini, F.Quevedo and C.Trugenberger, {\it Phys. Lett.} {\bf B396}
(1996) 115.
\bibitem{AmbOles} J.Ambjorn, P.Olesen and C.Petersen,
{\it Nucl. Phys.} {\bf B240} (1984) 533.

\end{thebibliography}
\enddocument